\documentclass[natbib,twocolumn]{svjour3}       
\smartqed  
\usepackage{graphicx}
\usepackage[dvipsnames]{xcolor}
\usepackage[hyphens]{url}
\usepackage[breaklinks, colorlinks, citecolor=MidnightBlue]{hyperref}
\usepackage{amssymb}
\usepackage{booktabs}
\usepackage{commath}
\usepackage{comment}
\usepackage{microtype}
\usepackage{txfonts}
\usepackage{aas_macros}
\usepackage{adjustbox}
\newcommand{\simgt}{\lower.5ex\hbox{$\; \buildrel > \over \sim \;$}}
\newcommand{\simlt}{\lower.5ex\hbox{$\; \buildrel < \over \sim \;$}}

\def\btheta{\mbox{\boldmath $\theta$}}

\newcommand{\cxo}{{\it Chandra}}

\newcommand{\planck}{{\it Planck}}
\newcommand{\xmm}{XMM-{\it Newton}}

\newcommand{\chandra}{{\it Chandra}}

\def\rhoc{\mathrel{\rho_{\rm crit}}}
\def\Msol{\mathrel{M_\odot}}

\def\h70Msol{\mathrel{h_{70}^{-1}M_\odot}}
\newcount\Comments  
\newcommand{\kolor}[2]{\ifnum\Comments=1\textcolor{#1}{#2}\fi}

\def \Mgas  {M_{{\rm gas}, \Delta}}
\def \fgas  {f_{\rm gas}}
\def \Mv  {M_{\Delta}}
\def \Rv  {R_{\Delta}}
\def \rhoc {\rho_{\rm c}}
\def\YX {Y_{\rm X}}
\def\TX {T_{\rm X}}
\def\LX {L_{\rm X}}
\def\YSZ {Y_{\rm SZ}}
\def \rhoDM {\rho_{\rm DM}}
\def \rhog {\rho_{\rm gas}}

\def\MY {$M_{500}$--$Y_{\rm X}$}

\begin{document}
\title{The galaxy cluster mass scale and its impact on cosmological constraints from the cluster population
}
\subtitle{ }

\titlerunning{The galaxy cluster mass scale and its cosmological impact}        

\author{G.W.~Pratt    \and
        		M.~Arnaud \and
		A.~Biviano \and
		D.~Eckert \and
		S.~Ettori \and
		D.~Nagai \and
		N.~Okabe \and
		T.H.~Reiprich
}


\institute{G.W. Pratt \at
              AIM, CEA, CNRS, Universit\'e Paris-Saclay, Universit\'e Paris Diderot, Sorbonne Paris Cit\'e, F-91191 Gif-sur-Yvette, France \\
              \email{\href{mailto:gabriel.pratt@cea.fr}{gabriel.pratt@cea.fr}}      
           \and
M. Arnaud \at
AIM, CEA, CNRS, Universit\'e Paris-Saclay, Universit\'e Paris Diderot, Sorbonne Paris Cit\'e, F-91191 Gif-sur-Yvette, France
\and
A. Biviano \at
INAF-Osservatorio Astronomico di Trieste, via G.B. Tiepolo 11, 34143, Trieste, Italy
\and
D. Eckert \at
Max-Planck-Institut f\"{u}r extraterrestrische Physik, Giessenbachstrasse 1, 85748 Garching, Germany 
\and
S. Ettori \at 
INAF-Osservatorio di Astrofisica e Scienza dello Spazio, via P. Gobetti 93/3, 40129 Bologna, Italy,
\and
D. Nagai \at
Department of Physics, Yale University, PO Box 208101, New Haven, CT, USA \\
Yale Center for Astronomy and Astrophysics, PO Box 208101, New Haven, CT, USA
\and
N. Okabe \at
Department of Physical Science, Hiroshima University, 1-3-1 Kagamiyama, Higashi-Hiroshima, Hiroshima 739-8526, Japan
\and
T. H. Reiprich \at
Argelander Institute for Astronomy, University of Bonn, Auf dem H\"ugel 71, 53121 Bonn, Germany
}

\date{Received: 31 December 2018 / Accepted: 14 February 2019}

\maketitle

\begin{abstract}

The total mass of a galaxy cluster is one of its most fundamental properties. Together with the redshift, the mass  links observation and theory, allowing us to use the cluster population to test models of structure formation and to constrain cosmological parameters. Building on the rich heritage from X-ray surveys, new results from  Sunyaev-Zeldovich and optical surveys have stimulated a resurgence of interest in cluster cosmology. These studies have generally found fewer clusters than predicted by the baseline {\it Planck} $\Lambda$CDM model, prompting a renewed effort on the part of the community to obtain a definitive measure of the true cluster mass scale. Here we review recent progress on this front. Our theoretical understanding continues to advance, with numerical simulations being the cornerstone of this effort. On the observational side, new, sophisticated techniques are being deployed in individual mass measurements and to account for selection biases in cluster surveys. We summarise the state of the art in cluster mass estimation methods and the systematic uncertainties and biases inherent in each approach, which are now well identified and understood, and explore how current uncertainties propagate into the cosmological parameter analysis.  We discuss the prospects for improvements to the measurement of the mass scale using upcoming multi-wavelength data, and the future use of the cluster population as a cosmological probe.


\end{abstract}

{
  \hypersetup{linkcolor=MidnightBlue}
  \tableofcontents
}

\section{Introduction}
\label{intro}

Clusters of galaxies represent the highest-density peaks of the matter distribution in the Universe. Forming at the intersection of cosmic filaments, they grow hierarchically through continuous accretion of material. Composed of dark matter (DM; 85\%), ionised hot gas in the intracluster medium (ICM; 12\%), and stars ($\sim 3\%$), their matter content reflects that of the Universe.
Their distribution in mass and redshift, and its evolution, allow us to probe both the physics of structure formation through gravitational collapse and the underlying cosmology in which this process takes place \citep[e.g.][]{all11,kra12}. Thus together with the redshift, the mass of a cluster is its most fundamental property. 

X-ray follow-up of objects in the {\it R\"ontgensatellit} (ROSAT) catalogues\footnote{\url{http://www.mpe.mpg.de/xray/wave/rosat/index.php}} allowed significant progress to be made on obtaining cosmological constraints from cluster number counts \cite[e.g.][]{bor01,rb02,Vikhlinin09b,man10} and baryon fraction \cite[e.g.][]{Allen08}. From the beginning, such studies consistently indicated a low matter density, with a mean normalized matter density $\Omega_{\rm m}\sim0.3$, and a matter fluctuation amplitude $\sigma_{8} \sim 0.7-0.8$.
However, while the first cosmological constraints from Sunyaev-Zeldovich (SZ) cluster number count studies broadly confirmed these findings \cite[][]{rei13,has13,PCXX2014}, the high statistical precision of the {\it Planck}\footnote{\url{https://www.cosmos.esa.int/web/planck}} Cosmic Microwave Background (CMB) measurements revealed an up to $\sim 2\sigma$ difference in the measurement of the key parameter $\sigma_8$ \citep{PCXX2014}. A number of physical effects have been advanced to explain this discrepancy, including invoking `new physics' (a massive neutrino component), but Occam's Razor would suggest that the simplest explanation lies in uncertainties in the cluster mass scale.

A number of different methods can be used to obtain individual cluster masses. The most commonly used are galaxy kinematics (the use of galaxy orbits as tracers of the underlying potential), X-ray and SZ observations (using the distribution of the ICM as a probe of the potential), and lensing (using distorsions of background galaxies to probe the intervening mass distribution). Each method has its inherent assumptions, and much work has gone into using numerical simulations to explore the possible biases that these assumptions might introduce into the final mass estimation. 

When cluster surveys are used to trace the growth of structure and samples are defined for use as cosmological probes, it is not possible to obtain individual masses for every object. Furthermore, one must understand the probability that a cluster of a given mass is detected with a given value of the survey observable  $\mathcal{O}$ (generally the X-ray or SZ signal, and more recently, the total optical richness), i.e. the relationship between $\mathcal{O}$ and the mass and the scatter about this relation. It is common practice to calibrate such a relationship for a limited number of objects, and then apply the resulting scaling law to the full sample. This approach has been successfully applied to a number of cluster samples. It requires accurate mass estimates of the calibration sample, and understanding of how the calibration and survey sample(s) map to the underlying population (i.e. knowledge of the sample selection function). While these uncertainties can be built into the marginalisation over cosmological parameters, tighter  parameter constraints go hand in hand with our understanding of these issues.

The mass scale is thus fundamental for the study of clusters. This review aims to take stock of the current status of cluster mass estimation methods and its impact on cosmological parameter estimation using the cluster population, and to address the prospects for future improvements.


\section{Theoretical insights from cosmological simulations} 

Cosmological simulations 
have been a workhorse for making predictions for the structure and shape of dark matter haloes for more than twenty years \citep[see e.g.][for reviews]{kra12,pla15}. Moreover, the abundance and clustering properties of dark matter haloes that form in the concordance cold dark matter (CDM) models are the standard against which observations are compared in order to derive cosmological constraints. Modern hydrodynamic simulations further provide insights into the effects of baryons on the dark matter halo properties, and on the internal structure of gas and stars within the dominant dark matter potential. In this Section we summarise a number of important insights that numerical simulations have provided for the interpretation of observational data.

The most commonly-used definition of mass, derived from theoretical studies but now used almost universally, is the three-dimensional mass enclosed within a given radius $R_{\Delta}$ inside which the mean interior density is $\Delta$ times the critical mass density, $\rho_{\rm c}(z)$, at the redshift of the cluster. Alternatively, one can use $\Delta$ times the mean mass density $\rho_{\rm m}(z)=\Omega_{\rm m}(z)\,\rho_{\rm c}(z)$. The standard notation expresses these quantities as 
\begin{eqnarray}
 M_{\Delta \rm c}&=&\frac{4\pi}{3}\, \Delta \rho_{\rm c}\,(z)\,  R_{\Delta
  \rm c}^3, \nonumber \\
 M_{\Delta \rm m}&=&\frac{4\pi}{3}\, \Delta \rho_{\rm m}\,(z)\,  R_{\Delta \rm m}^3.
 \label{eq:delta}
\end{eqnarray}
One sometimes simply uses $M_\Delta$ and $R_\Delta$ for the former case. Commonly-used values of $\Delta$ in observational studies include  2500 (corresponding to the central parts of the halo), 500 (roughly equivalent to the virialised region that is accessible to the current generation of X-ray telescopes), and 200 (corresponding approximately to the `virial' radius).


\subsection{Dark matter density profiles}

\subsubsection{NFW model}

The mass and internal structure of galaxy clusters reflect the properties of primordial density perturbations and the nature of the dark matter. In the standard hierarchical CDM scenario of cosmic structure formation, numerical simulations predict that dark matter haloes spanning a wide mass range can be well described by a universal mass density profile \citep{NFW96,NFW97}. The so-called Navarro-Frenk-White (NFW) profile is expressed in the form:
\begin{equation}
\rho_{\rm NFW}(R)=\frac{\rho_s}{(R/R_s)(1+R/R_s)^2},
\label{eq:rho_nfw}
\end{equation}
where $\rho_s$ is the central density parameter and $R_s$ is the scale radius that divides the two distinct regimes of asymptotic mass density slopes $\rho\propto R^{-1}$ and $R^{-3}$. 

The NFW profile is fully specified by two parameters: $M_\Delta$ and the halo concentration $c_\Delta=R_\Delta/R_s$. The three-dimensional spherical mass, $M_\Delta$, enclosed by the radius, $R_\Delta$, is given by
\begin{equation}
M_{\rm NFW}(<R_\Delta)=\frac{4\pi \rho_s R_\Delta^3}{c_\Delta^3}\left[
\ln(1+c_\Delta)-\frac{c_\Delta}{1+c_\Delta}\right].
\label{eq:MNFW}
\end{equation}
As the central density reflects the mean density of the Universe at the time of formation, haloes with increasing mass are
expected to have lower mass concentration at a given redshift \citep[e.g.][]{NFW97,Bullock01,Gao04b,Dolag04,Duffy08,Stanek10,Klypin11,Bhattacharya13,Meneghetti14,Ludlow14,Diemer15}.

Numerical simulations usually describe the relation between the mass and the NFW concentration (i.e. the $c-M$ relation) for simulated haloes using a power-law function \cite[e.g.][]{Bhattacharya13, Meneghetti14, Diemer15}. This relation exhibits large intrinsic scatter for a given halo mass owing to  the wide distribution in formation times \citep[e.g.][]{neto07} and  the evidence that not all systems are well described by a NFW model \citep[e.g.][]{jin00}. 
Recently, \citet{Diemer15} have proposed that a seven-parameter, double power-law functional form computed by peak height and the slope of the linear matter power spectrum can describe concentrations in the fiducial $\Lambda$CDM cosmology with $5\%$ accuracy. 

Although non-baryonic dark matter exceeds baryonic matter by a factor of $\Omega_{\rm DM}/\Omega_{\rm b} \approx 6$ on average, the gravitational field in the central regions of galaxies is dominated by stars.  In the hierarchical galaxy formation model the stars are formed in the condensations of cooling baryons in the halo centre.  As the baryons condense in the centre, they pull the dark matter particles inward thereby increasing their density in the central region. The response of dark matter to baryonic infall has traditionally been calculated using the model of adiabatic contraction \citep{eggen62}, which has also been tested and/or calibrated numerically using both idealised \citep{ryden_gunn87,blumenthal86} and cosmological simulations \citep{gnedin04, rudd08, Duffy08, velliscig14,shir18}.


\subsubsection{Einasto model}

Recent high-resolution N-body simulations \citep[e.g.][]{Navarro04,Gao12} indicate that an Einasto profile \citep{Einasto65} better describes the spherically averaged mass density profile for dark matter haloes than the NFW profile. The Einasto profile has the form:
\begin{eqnarray}
 \rho_{\rm Einasto}=\rho_{-2} \exp\left(-\frac{2}{\alpha}\left[\left(\frac{R}{R_{-2}}\right)^\alpha-1\right]\right) \label{eq:Einasto}
\end{eqnarray}
where $\alpha$ is a shape parameter that describes the degree of curvature
of the profile, and $\rho_{-2}$ and $R_{-2}$ are a mass density and a
scale radius at which the logarithmic slope is $-2$, respectively.
The NFW model corresponds to a $\alpha\sim0.18$ case for the Einasto
profile. The spherical mass enclosed within $R_\Delta$ is given by 
\begin{eqnarray}
 & & M_\Delta= 4\pi \rho_{-2}\, R_{-2}^3 \frac{1}{\alpha}
  \left(\frac{2}{\alpha}\right)^{-3/\alpha} \nonumber \\
& & \qquad \times   \exp\left(2/\alpha \right)\left[ \Gamma\left(\frac{3}{\alpha}\right)-\Gamma\left(\frac{3}{\alpha}, \frac{2}{\alpha}\left(\frac{R_\Delta}{R_{-2}}\right)^\alpha\right)\right],
\end{eqnarray}
where $\Gamma(x)$ and $\Gamma(a,x)$ are the gamma function and the upper
incomplete gamma function, respectively.
The Einasto profile is specified by the three parameters $M_\Delta$,
$c_\Delta=R_\Delta/R_{-2}$, and $\alpha$. 


\subsubsection{Sparsity}
An alternative to a parameterised description of the dark matter profile is the use of the halo sparsity, $s_{\Delta}$. This quantity measures the ratio of halo masses at two different overdensities:
\begin{equation}
s_{\Delta_1\, \Delta_2} = M_{\Delta_1} / M_{\Delta_2}
\end{equation}
 and has been recently proposed as new cosmological probe for galaxy clusters \citep{balmes14,corasaniti18}. If the halo follows a NFW profile, the sparsity and concentration are directly related. However, halo sparsity has the key feature that the ensemble average value at a given redshift exhibits much smaller scatter than that of the mass concentration, and does not require any modelling of the mass density profile, but only the mass measurements within two overdensities. It is thus also an attractive quantity for comparison with observations.


\subsection{The shape and distribution of dark matter and gas}

Although the above discussion assumes spherical symmetry, the CDM model predicts that cluster-size dark matter haloes are generally triaxial and are elongated along the direction of their most recent major mergers \citep[e.g.][]{thomas98,jing_suto02,hopkins05,kasun_evrard05,bett07,gottloeber_yepes07}. The degree of triaxiality is correlated with the halo formation time \citep[e.g.][]{allgood06}, suggesting that at a given epoch more massive haloes are more triaxial. For the same reason, triaxiality is sensitive to the linear structure growth function and is higher in cosmological models in which haloes form more recently \citep{maccio08}. Furthermore, inclusion of baryons in simulations modifies the shapes of cluster-size dark matter haloes, causing them to become rounder due to gas dissipation associated with galaxy formation processes \citep[e.g.][]{kazantzidis04}.

\begin{figure*}
\hfill\resizebox{\hsize}{!}{\includegraphics[width=0.465\textwidth]{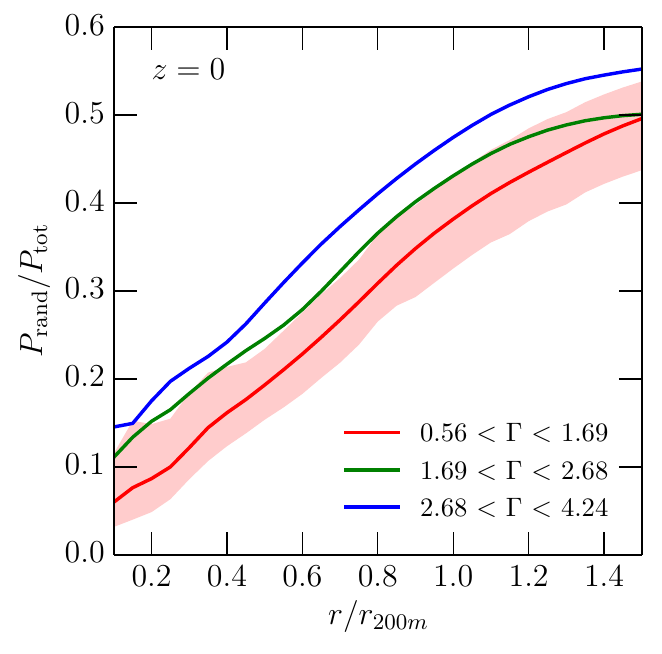}
\hfill
\includegraphics[width=0.40\textwidth]{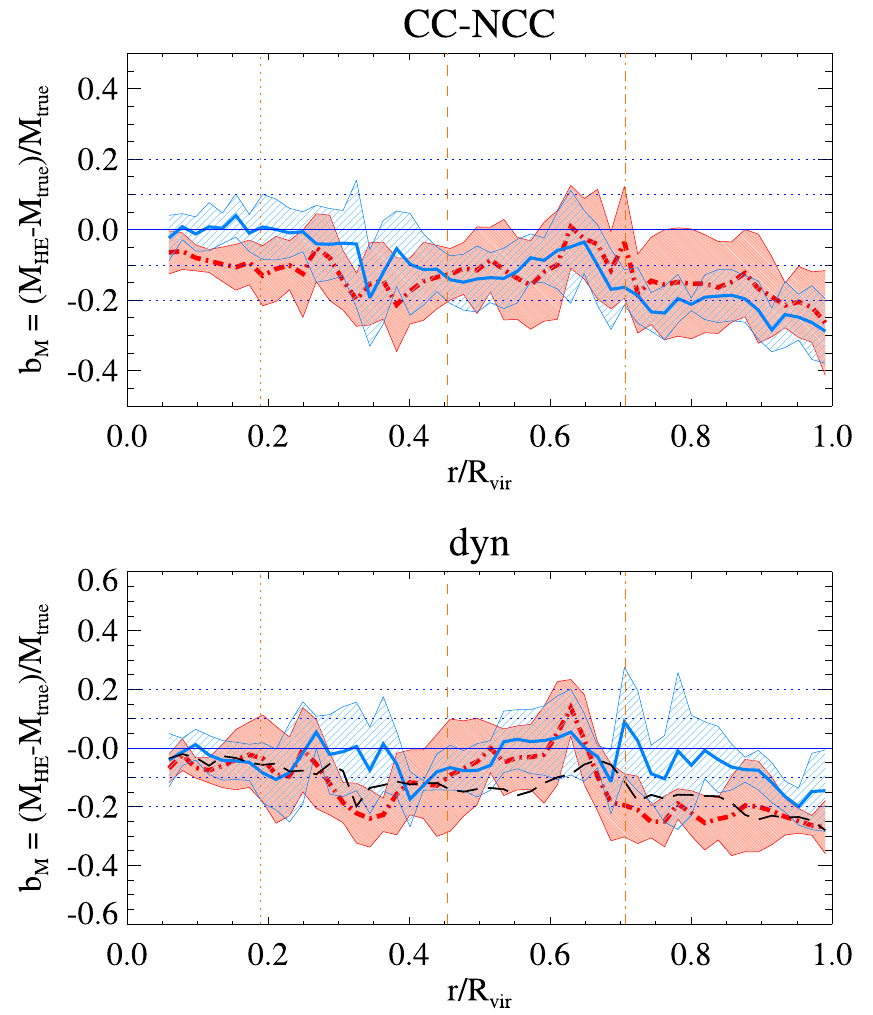}}
\caption{\emph{Left:} Random-to-thermal pressure ratio for simulated clusters from the $\Omega_{500}$ simulation \citep[reproduced from][]{Nelson14}. The various curves show the median non-thermal pressure profiles sorted by a proxy for  the mass accretion rate $\Gamma$ defined as the difference in mass between $z=0.5$ and $z=0$ \citep{Diemer14}. \emph{Right:} Hydrostatic mass bias $b_{M}=(M_{\rm HSE}-M_{\rm true})/M_{\rm true}$ as a function of radius for a sample of 29 clusters simulated with the SPH code \texttt{GADGET-3} \citep[reproduced from][]{biffi16}. The two panels show the radial profiles of $b_M$ sorted by the system core state, where CC indicates cool core and NCC denotes non-cool core objects (top), and dynamical state (bottom). The shaded areas show the dispersion around the median.}
\label{fig:hse_sim}
\end{figure*}

\citet{lau11} showed that gas traces the shape of the underlying potential rather well outside the core, as expected if the gas were in hydrostatic equilibrium (HE hereafter) in the cluster potential, but that the gas and potential shapes differ significantly at smaller radii. These simulations further suggest that with radiative cooling, star formation and stellar feedback (CSF) intracluster gas outside the cluster core ($R\gtrsim 0.1\,R_{500}$) is more spherical compared to non-radiative simulations, while in the core the gas in the CSF runs is more triaxial and has a distinctly oblate shape. The latter reflects the ongoing cooling of gas, which settles into a thick oblate ellipsoid as it loses thermal energy. In the CSF runs, the difference reflects the fact that gas is partly rotationally supported. In non-radiative simulations the difference between gas and potential shape at small radii is due to random gas motions, which make the gas distribution more spherical than the equipotential surfaces. Results are similar for unrelaxed clusters but with considerable scatter. In both CSF and non-radiative runs, the gravitational potential was found to be  much more spherical than DM. 

Stochastic feedback from a central active galactic nucleus (AGN) will also heat and redistribute the gas in the core regions \citep[e.g.][]{lebrun14,tru18}. Due to their shallower potential wells, such feedback has a stronger effect on the gas distribution of lower mass systems, leading to a radial and mass dependent modification of the gas content in the core regions.


\subsection{ICM energy budget and departures from equilibrium} 
\label{sec:icm_energy}

The deep potential well of galaxy clusters compresses the collapsing baryons (consisting mostly of pristine hydrogen and helium with densities of $\sim 10^{3}$ particle cm$^{-3}$), and heats them to temperatures of $10^7$ K ($\sim 1$ keV) and above. Given its high temperature, the ICM emits in X-rays principally via thermal Bremsstrahlung, with a continuum emission typically following 
$\epsilon \propto n_{\rm gas}^2 T^{1/2}$. Inverse Compton scattering of CMB photons by ICM electrons produces the SZ effect that is observed in millimetric bands \citep{sun72}. The SZ signal is proportional to the electron pressure integrated along the line-of-sight. 

The spatial distribution and thermodynamical properties of the ICM depend on the underlying dark matter potential and the merging history of a cluster. From a general point of view, the dynamics of an inviscid collisional gas follows the Euler equation,
\begin{equation}
\frac{\partial \mathbf{v}}{\partial t}+(\mathbf{v}\cdot\nabla)\, \mathbf{v}+\frac{1}{\rho}\nabla P = -\nabla\Phi
\label{eq:euler}
\end{equation}
where $\mathbf{v}$ denotes the three-dimensional velocity field, $P$, $\rho$ are the gas pressure and density, and $\Phi$ the cluster gravitational potential. After few sound crossing times of the order of $10^9$ years, the ICM is expected to reach HE, and the kinetic energy thermalises, such that the pressure support should be dominated by the thermal pressure ($P\approx P_{\rm th}$). The velocity field becomes negligible and the Euler equation reduces to the HE equation,
\begin{equation}
\frac{1}{\rho} \frac{d P_{\rm th}}{d r} = -\frac{GM(<R)}{R^2}, 
\label{eq:hee}
\end{equation}
\noindent where $G$ is the gravitational constant. Under this assumption, the mass profile can be reconstructed from the radial profiles of ICM thermodynamic quantities (see Sect. \ref{sec:X-ray}). 

\begin{figure*}
\includegraphics[width=0.6\textwidth]{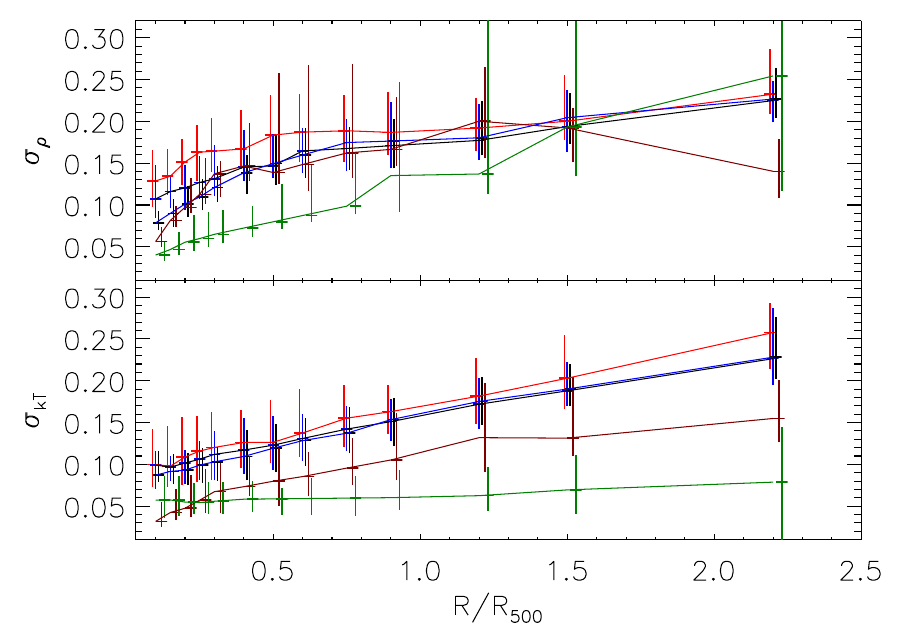}
\hfill
\includegraphics[width=0.4\textwidth]{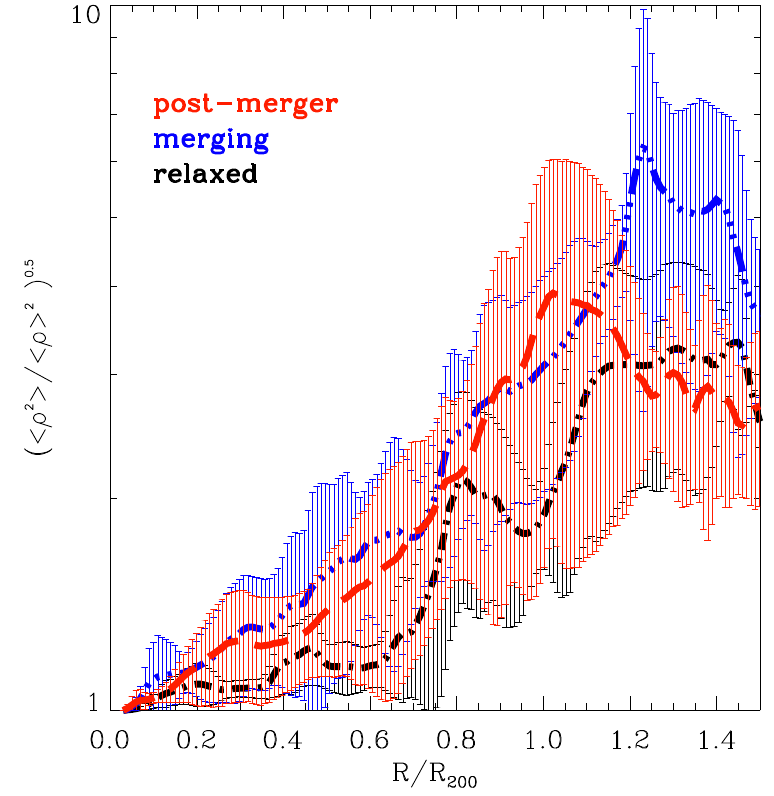}
\caption{\emph{Left:} A measure of the predicted inhomogeneity in the gas density and temperature distributions. The Figure shows the average width of the distribution of gas density (top) and temperature (bottom) within spherical shells for five different simulation setups \citep[reproduced from][]{rasia14}. The colours refer to the following runs: non-radiative SPH (red), non-radiative AMR (brown), SPH with cooling and star formation (blue), AMR with cooling and star formation (green), and SPH with AGN feedback (black). \emph{Right:} Radial profiles of gas clumping factor $C=\left(\langle \rho^2\rangle/\langle\rho\rangle^2\right)^{1/2}$ estimated from non-radiative AMR simulations of a set of massive clusters \citep[reproduced from][]{vazza13}. The simulated systems are sorted into relaxed (black), merging (blue), and post-merger (red) categories. }
\label{fig:inhomogeneities}
\end{figure*}

However, residual random gas motions can produce a non-negligible contribution to balance the gravitational field, which causes an underestimation of the true mass when the energy is assumed to be fully thermalised. The total pressure balancing gravity can be described as the sum of thermal pressure and random kinetic pressure,
\begin{equation}
P_{\rm tot} \simeq P_{\rm th}+\frac{1}{3}\, \rho\, \sigma_v^2,
\end{equation}
where $\sigma_v$ denotes the velocity dispersion of isotropically moving gas particles (i.e. turbulent motions). More generally, by integrating the Eqn.~\ref{eq:euler} over a radial shell, the total enclosed mass within the radius $R$ can be written as,
\begin{equation}
M(<R) = M_{\rm therm} + M_{\rm rand} + M_{\rm rot} + M_{\rm cross}+ M_{\rm stream} +M_{\rm accel},
\label{eq:mhse_tot}
\end{equation} 
where the expressions for all six terms are given in \citet{Lau13}. The first term in the equation is the hydrostatic mass (see Eqn.~\ref{eq:mhe}). The second and third terms indicate the pressure support induced by random and rotational gas motions, respectively. The fourth and fifth terms describe the contribution from cross and stream motions; the final term is the acceleration. 

Each of these additional terms will introduce corrections to the HE assumption and need to be properly understood in order to derive accurate masses from the hydrostatic method. Given the difficulty of directly measuring gas motions in the ICM\footnote{Existing experimental constraints and future prospects on gas motions in the ICM are discussed in detail in another chapter of this series Simionescu et al. (2019).}, numerical simulations have been widely exploited to set constraints on the relative importance of each of these terms \citep{rasia06,Nagai07,Vazza09,Lau09,nelson12,battaglia13,Suto13,rasia14,Nelson14,Nelson14b,biffi16,shi15,shi16}. Most studies consistently predict that random residual gas motions (i.e. turbulence) dominate the required correction, independent of the dynamical state. The amplitude of the turbulent pressure support, however, varies from cluster-to-cluster, with predictions in the range of $10-30\%$ at $R_{500}$ depending on the mass accretion histories of clusters \citep{Nelson14,shi15,shi16}. Bulk motions and acceleration provide an important contribution only in merging clusters. 
Note that the acceleration term is very small within the virialised regions of galaxy clusters, but becomes a non-negligible and irreducible mass bias in merging clusters or the outskirts of all clusters \citep{Lau13,Suto13,Nelson14b}.

Figure~\ref{fig:hse_sim} shows the radial profiles of non-thermal pressure and hydrostatic mass bias from two different sets of simulations \citep{Nelson14,biffi16}. Both studies predict a trend of increasing non-thermal-to-thermal pressure ratio with radius, and hydrostatic mass biases ranging from $<5\%$ in the core to $\sim30\%$ at $R_{500}$. Both studies also find a dependence of the predicted hydrostatic mass bias on the cluster dynamical state and accretion rate, the non-thermal pressure contribution being on average higher in highly accreting systems. The relatively low-values of the non-thermal pressure derived from the X-COP data (in Sect.~\ref{sec:fgasxcop}) are consistent with the expectation that relaxed clusters have the lower level of non-thermal pressure support. Future work should focus on detailed understanding of the nature of gas flows in the density-stratified ICM in cluster outskirts \citep{shi18,vazza18}.


\subsection{The presence of gas inhomogeneities}
\label{sec:icm_inhom}

In practice, interpretation of X-ray measurements of the thermodynamical properties of the ICM may be complicated by the presence of structure and inhomogeneities in the gas temperature and density distributions \citep{Mazzotta04,vikhlinin06b}.
Unfortunately, numerical simulations have not yet converged on what the typical level of temperature inhomogeneities in the ICM should be, as the result appears to depend substantially on the adopted physical and computational setup. The two main hydrodynamical solvers in numerical simulations of clusters are Smoothed Particle Hydrodynamics (SPH) and Adaptive Mesh Refinement (AMR). \citet{rasia14} compared the predicted level of temperature anisotropies in five sets of numerical simulations featuring both SPH and AMR hydrodynamical solvers, and for different implementations of baryonic physics (non-radiative, cooling and star formation, and AGN feedback). The predicted  level of temperature inhomogeneity  ranges from 5 to 25\% (see Fig. \ref{fig:inhomogeneities}). 

Similarly, the gas density determined from X-ray observations of the ICM may be biased by the presence of inhomogeneities in the gas distribution of the ICM. Over-dense regions exhibit an enhanced X-ray signal because of the $\rho^2$ dependence of the emissivity, which boosts the estimated gas density towards high values \citep{mathiesen99}. The overestimation of the gas density is usually quantified by the clumping factor $C=\left(\langle \rho^2\rangle/\langle\rho\rangle^2\right)^{1/2}\geq1$. Numerical studies predict that the clumping factor $C$ should increase from values close to 1 in the central regions to $1.2-1.3$ around $R_{200}$ \citep{nagai07b,vazza13,roncarelli13,zhu13,planelles17}, with substantial scatter from one system to another. As an example, the right-hand panel of Fig. \ref{fig:inhomogeneities} shows the radial profiles of the clumping factor in a set of 20 massive clusters simulated with the AMR code {\sc Enzo} and sorted according to their dynamical state, showing that the ICM in merging systems is on average more clumpy than in relaxed objects \citep{vazza13}. The HE equation in the presence of clumping should be modified by the gradient of the clumping factor \citep{roncarelli13}, neglect of which can cause biases of $\sim5\%$ on the reconstructed masses. Note that the effect of clumping on the gas fraction is expected to be larger, as it biases simultaneously the gas mass high and the hydrostatic mass low. The corresponding values of $f_{\rm gas}$ can be overestimated by $\sim10\%$ at $R_{500}$ \citep{eckert+15}.


\subsection{Baryon budget}
\label{sec:depletion}

\subsubsection{Total baryonic content}

\begin{figure*}[t]
\includegraphics[width=0.48\textwidth]{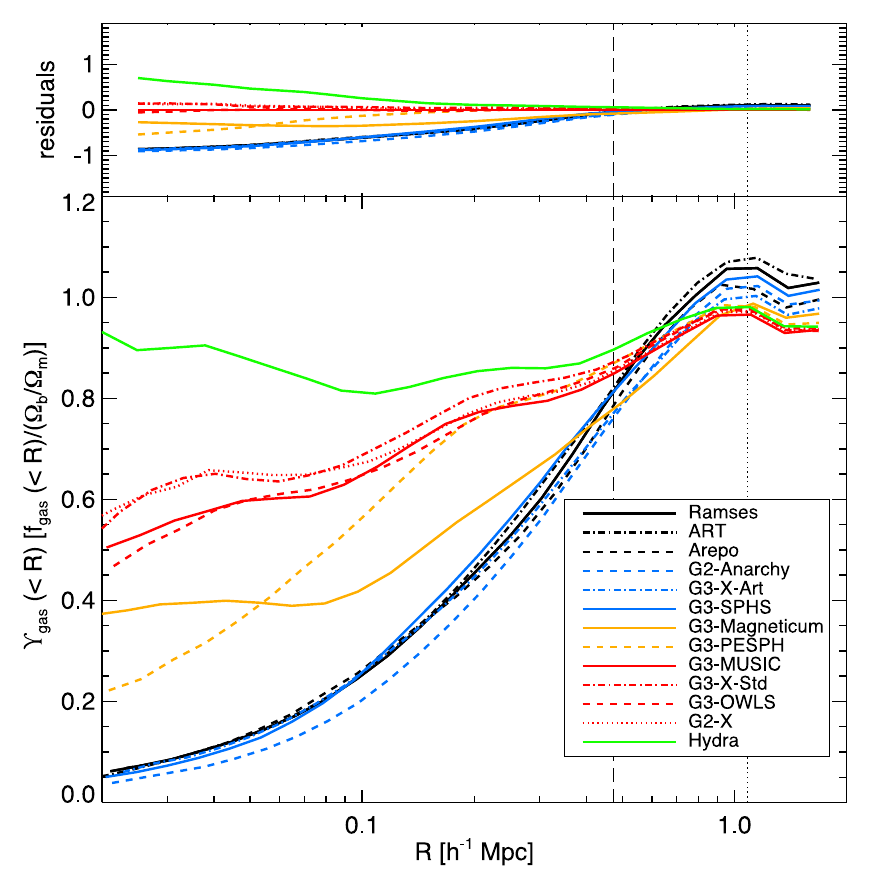}
\hfill
\includegraphics[width=0.51\textwidth]{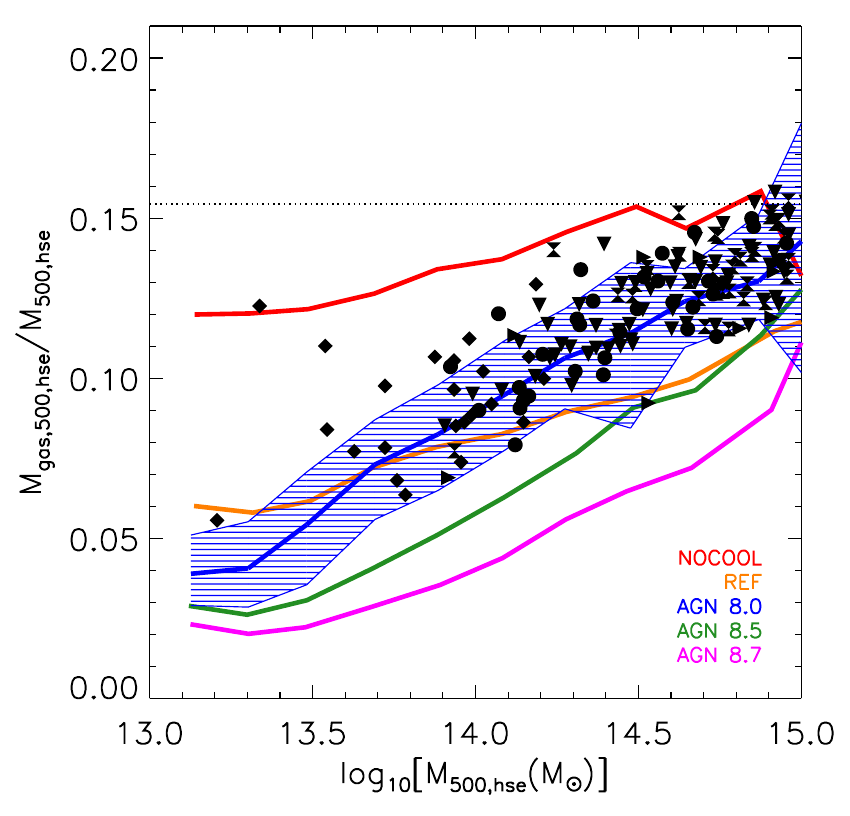}
\caption{\textit{Left:} Baryon fraction profiles for a massive galaxy cluster resimulated with 13 different codes from the same initial condition. The dotted vertical line indicates the value of $R_{500}$. Figure from \citet{nifty1}. \textit{Right:} ICM gas fraction profiles as a function of halo mass in simulations implementing various prescriptions for baryonic physics (non-radiative, red; cooling and star formation, yellow; AGN feedback, blue, green and magenta). The AGN feedback models range from gentle, self-regulated feedback (AGN 8.0, blue) to more bursty and energetic injection (AGN 8.7, magenta). The black points show a compilation of observed ICM gas fraction measurements. Figure from \citet{lebrun14}.}
\label{fig:depletion}
\end{figure*}

Because of their large mass and deep gravitational potential, the total baryon content in galaxy clusters is expected to reflect that of the Universe as a whole \citep{White93,Evrard97,Kravtsov05}. The total baryon fraction $f_b=(M_{\rm gas}+M_{\star})/M_{\rm tot}$ should thus match the cosmic baryon fraction estimated from primordial nucleosynthesis and the CMB power spectrum. Simulations using different hydrodynamical solvers and baryonic physics substantially agree in predicting that the depletion of baryons within $R_{200}$ during the hierarchical formation process should be small \citep[$\sim5\%$,][]{Planelles13,Sembolini13,lebrun14,wu15,nifty1,nifty2,hahn17,barnes17,barnes18,lovell18}. In particular, \citet{nifty1} resimulated the region surrounding a massive cluster ($M_{\rm vir}=1.1\times10^{15}M_\odot$) with 13 different hydrodynamical codes from the exact same initial conditions and compared the output. The comparison includes classical SPH ({\sc Gadget-2}), advanced SPH ({\sc Gadget-3}), AMR ({\sc Art, Ramses}), and moving-mesh ({\sc Arepo}) codes. In the left-hand panel of Fig. \ref{fig:depletion} the baryon fraction of the simulated cluster is shown as a function of radius. While in the central regions and out to $\sim300$ kpc the various codes do not converge, around $R_{500}$ and beyond they agree within a few percent. Thus, the baryon fraction within $R_{200}$ is very robustly predicted by numerical simulations, independent of the exact input physics or the numerical scheme. 

\subsubsection{Effect of non-gravitational processes}
\label{sec:fbar_ngrav}

Various works have also studied the impact of baryonic physics (cooling, star formation, supernova and AGN feedback) on the depletion of baryons within the virial radius. In the case where a large amount of non-gravitational energy is injected within the ICM, the gaseous atmosphere expands and a global depletion of baryons within the virial radius may occur. AGN feedback is the main source of non-gravitational energy in the ICM \citep[see][for a review]{mcnamara07}. The baryon depletion caused by AGN feedback is known to be important for haloes with masses below $\sim 10^{14}M_\odot$ \citep{Planelles13,lebrun14,wu15,lovell18}, for which the baryon budget falls short of the cosmic value by a factor $\sim2$. The right-hand panel of Fig. \ref{fig:depletion} from \citet{lebrun14} shows the hot gas fraction in several sets of numerical simulations implementing various prescriptions for baryonic physics (non-radiative, cooling and star formation, and three models for AGN feedback) and compares the results with published datasets. While the non-radiative run predicts little depletion for haloes in the range $10^{13}-10^{15}M_\odot$, the runs implementing additional physics largely differ for haloes of $M_{500}\leq10^{14}M_\odot$. Note that the run including cooling and star formation but no AGN feedback suffers from the overcooling problem, and predicts stellar fractions that are largely in excess of the measured values. At high mass ($M_{500}\sim10^{15}M_\odot$), all but the most extreme AGN feedback model converge to a very similar value for the hot gas fraction, indicating that baryonic effects are subdominant. 


\subsection{Mass estimates from mass proxies and scaling relations}
\label{sec:scaling}

The gravitational potential of galaxy clusters can be probed through observations of the ICM in X-rays and the SZ effect, or through the richness in optical/NIR wavelengths. One  expects  simple scaling relations between the mass and  global  ICM  properties such as the X--ray luminosity, $\LX$, or the SZ Compton parameter $\YSZ$, and galaxy content.   More specifically, the simplest models of structure formation, based on simple gravitational collapse, predict that galaxy clusters constitute a self-similar population. As discussed above (Eqn.~\ref{eq:delta}), the virialised part of a cluster  corresponds roughly to a fixed density contrast ($\Delta \sim 500$) as compared  to the critical density of the Universe, $\rhoc(z) $ at the redshift in question:
\begin{equation}
        \frac{\Mv}{  \frac{4\pi}{3}\,\Rv^{3} } =   \Delta\,\rhoc(z)
\label{eq:rho}
\end{equation}
\noindent with a strong similarity in the internal structure of virialised dark matter haloes within the corresponding radius, $\Rv$.  This reflects the fact that there is no characteristic scale in the gravitational collapse. The gas properties directly follow from the dark matter properties,  assuming that the gas evolution is purely driven by gravitation, i.e. by the evolution of the dark matter potential.  The internal gas structure is universal, as is the case for the dark matter. The gas mass fraction $\fgas$ reflects the Universal value, since the gas 'follows' the collapse of the dark matter. It is thus constant:
\begin{equation}
\frac{\Mgas}{\Mv}  = \fgas =  {\rm const.}
\end{equation}
\noindent  Furthermore, as the gas  is roughly in HE in the potential of the dark matter, the virial theorem  gives:
\begin{equation}
\TX = \beta  \frac{G \mu {\rm m_{\rm p}}\,\Mv }{\Rv}  
\label{eq:Tvir}
\end{equation}
\noindent where  $\mu$ is the mean molecular weight in amu for an ionised plasma, $m_{\rm p}$ is the proton mass, $\TX$ is the gas mean temperature, and $\beta$ is a normalization factor which depends on the cluster internal structure. Since this structure  is universal,  $\beta$ is a constant,  independent of redshift $z$ and cluster mass. 

Each cluster can therefore be defined by two parameters only:  its mass and its redshift.  From the basic equations, Eqn.~\ref{eq:rho}-\ref{eq:Tvir}, one can derive a scaling law for each physical property, $Q$, of the form $Q\propto A(z)\Mv^{\alpha}$, that relates it to the redshift and  mass. The evolution factor,  $A(z)$, in the scaling relations is due to the evolution of the mean dark matter (and thus gas) density, which varies with the critical density of the Universe, $
 \overline{\rhog} \propto \overline{\rhoDM}= \Delta \rhoc(z) \propto E^{2}(z)$. For instance, the gas  mass scales as $\Mgas \propto \Mv$, the temperature as $T_{\rm X}  \propto   E^{2/3}(z) \Mv^{2/3}$. The integrated SZ signal $\YSZ$, or its X-ray equivalent $\YX=\Mgas\,\TX$, introduced by \citet{kra06}, scales as $\YSZ \propto   E^{2/3}(z)\,\Mv^{5/3}$, while the (bolometric) X--ray luminosity scales as $\LX \propto E(z)^{7/3}\,\Mv^{4/3}$.
 
 \begin{figure*}
\hbox{
\includegraphics[width=0.48\textwidth, keepaspectratio]{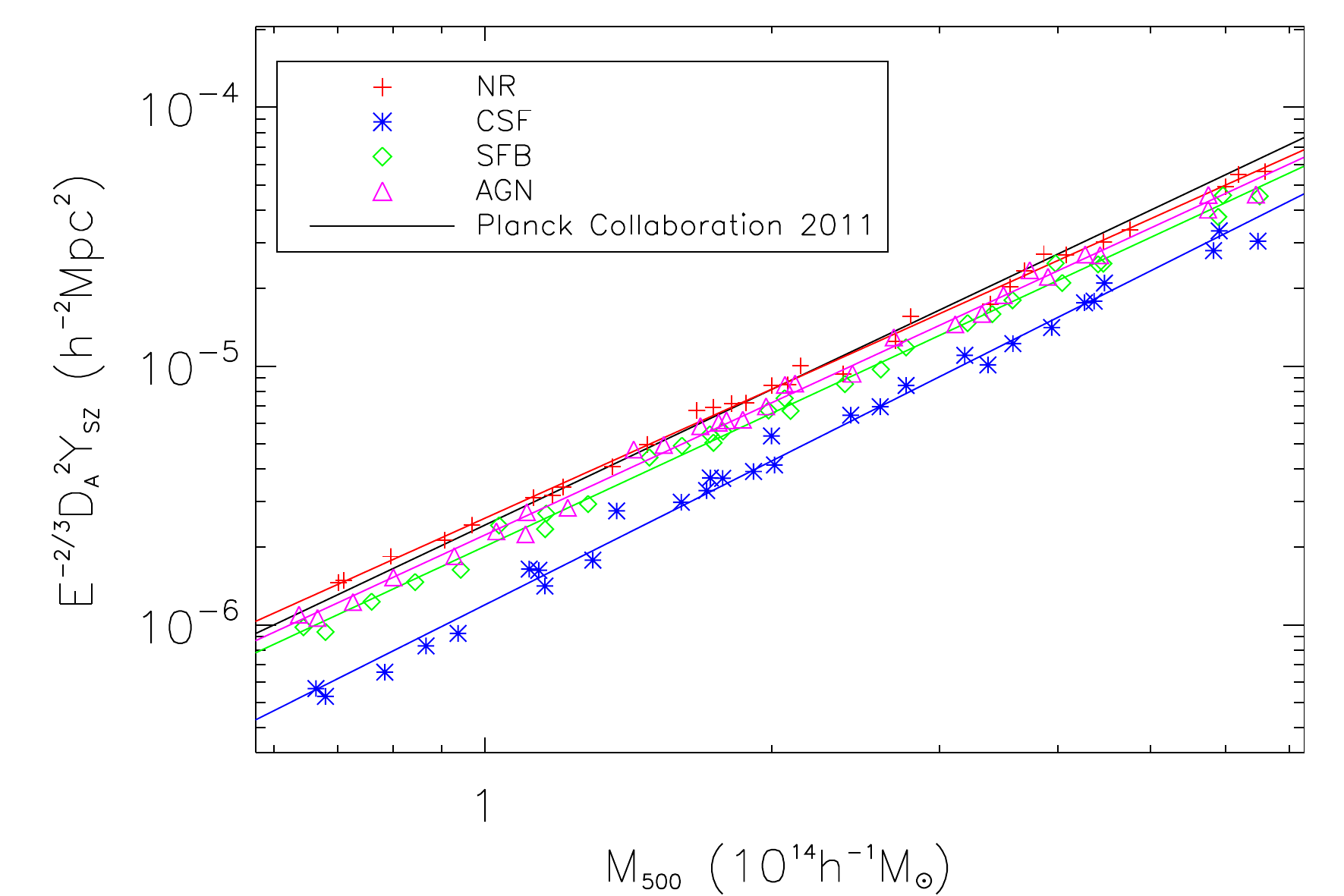}
\hfill
\includegraphics[width=0.48\textwidth, keepaspectratio]{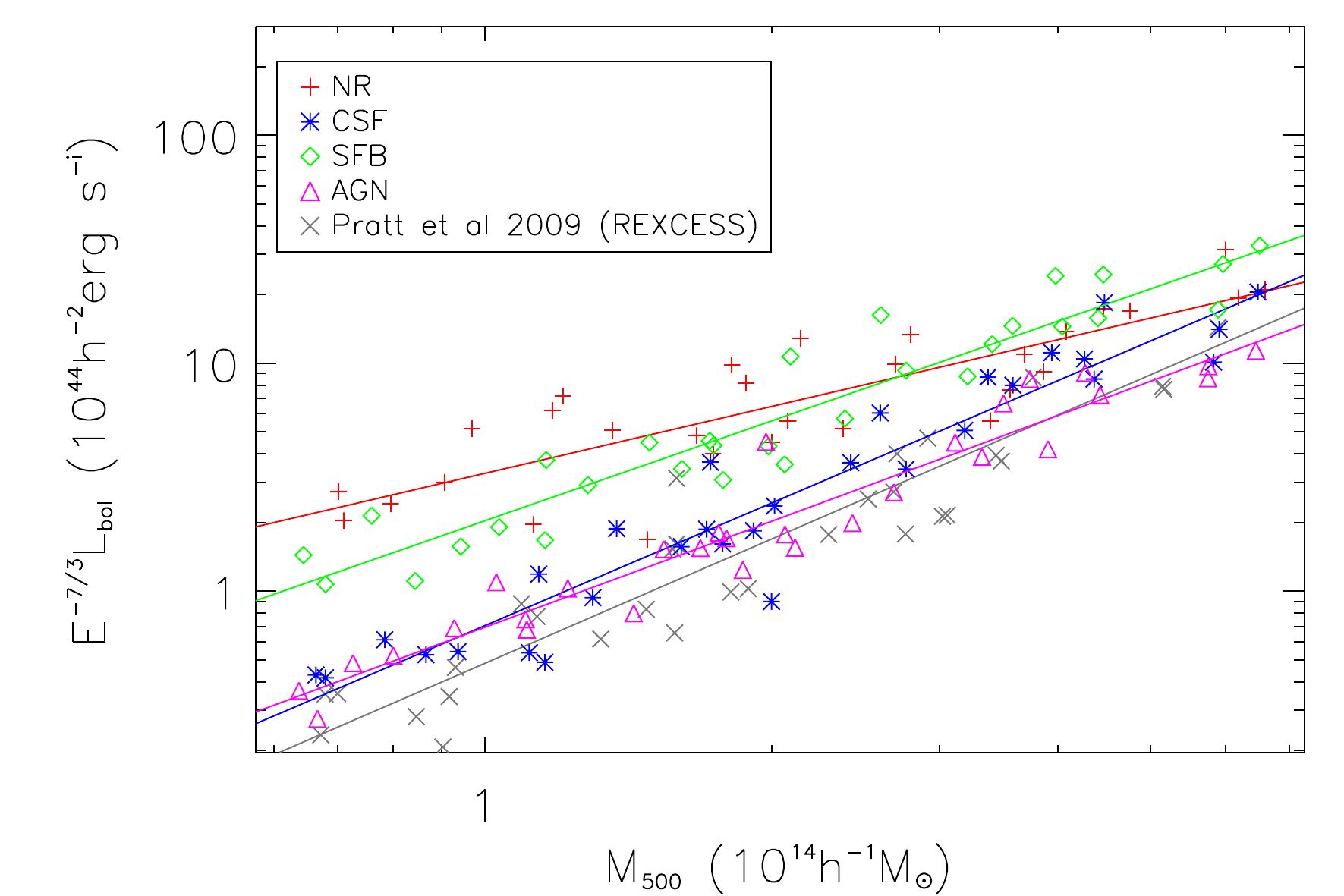}
}
\caption{Relation  between the SZ signal (left) or the  X--ray luminosity (right)  and the mass from the numerical simulations  of \citet{pike14}. Observational data points are from \citet{PEPXI} and \citet{pratt09}. The relations are plotted for  different implementation of the gas physics: the non-radiative  model (NR, red crosses), cooling and star formation model (CSF; blue stars), supernova feedback (SFB; green diamonds) and AGN models (magenta triangles).  These relations are compared to observations (black line and black crosses).  The  $\YSZ$  is proportional to the total thermal energy and the relation  between $\YSZ$ and the mass  depends weekly on cooling and feedback  processes. The scatter is  tight, reflecting the similarity in shape of the pressure profiles. In contrast, the X-ray luminosity depends on the square of the density and  is dominated by the core properties. It is  very sensitive to gas physics and presents a large scatter at a given mass, reflecting the large scatter in the scaled  density profile in the core of clusters. }
\label{fig:scalingtheor}
\end{figure*}
 
 These scaling relations then  allow estimation of the mass through so-called mass proxies,   i.e. global physical  properties, directly related to the mass, but easier to measure.   However, there are intrinsic limitations  to these 'cheap' mass estimates. Even in the simplest, purely gravitational model, the normalisation of the relations depends on the formation history, and must be derived from numerical simulations. Furthermore, each relation has an intrinsic scatter due to individual cluster formation histories \citep{poo07,yu15}. Even more importantly, non gravitational physics (cooling and galaxy/AGN feedback) affects the normalisation, slope, scatter, and  evolution of each relation. In particular, as $\LX$ and $\YSZ$  depend on  the gas content,  the  slope of the $\LX$--$M$ and $\YSZ$--$M$ relations is directly affected by the  mass dependence  of the baryon depletion (Sec.~\ref{sec:fbar_ngrav}). A large numerical simulation  effort  has been undertaken to understand   how these scaling relations depend on the gas physics \citep[e.g.][]{pike14, planelles14,lebrun14,tru18}, including their scatter and evolution \citep{lebrun17}.
 There is now a consensus that  AGN feedback is a key ingredient of  realistic models. A key recent advance is the development  of new cosmological hydrodynamical simulations, with calibrated sub-grid feedback models that are able to reproduce the observed gas and stellar  properties of  local clusters \citep{mccarthy17}.

 The most robust mass proxies correspond to the lowest-scatter relations that depend as little as possible on the gas physics. In this respect, the SZ signal, proportional to the integral of the pressure, or equivalently to the total thermal energy of the gas, is generally believed to be particularly well-behaved \citep[e.g.][]{das04,mot05}.  The SZ signal and the corresponding pressure profiles beyond the core are mostly governed by the characteristics of the underlying potential well,  with a weak dependence on dynamical state and on the poorly-understood non-gravitational  physics.  $\YSZ$ is thus considered to be a robust, low scatter mass proxy. In contrast,  the X--ray luminosity is more complex. The X--ray flux is sensitive to core properties, which presenting a large scatter and a strong dependence on thermodynamical state (Fig.\,\ref{fig:scalingtheor}). The gas mass may be a good mass proxy for the most massive systems, but not at low mass where it is strongly dependent on galaxy feedback.  A recent comparison of the properties of various mass proxies, seen from the point of view of numerical simulations, can be found in \citet{lebrun17}. Improved understanding of covariances among different observables is one of the important steps toward improving constraining power of upcoming multi-wavelength cluster surveys \citep[e.g.][]{sta10,shir16}.


\section{Observational mass estimation methods}
\label{sec:mem}

In this Section, we discuss the principal methods that are used to estimate individual cluster masses. Each method is briefly described, along with its underlying assumptions, and the various systematic uncertainties  and potential biases that can be  encountered in translating the observation into a mass measurement are discussed.

\subsection{Kinematics}

The first estimate of the mass of a cluster of galaxies \citep{Zwicky33,Zwicky37} was obtained by
applying the virial theorem to the distribution of cluster galaxies in projected phase-space (PPS), and it was based on the assumption that galaxies are unbiased tracers of the cluster mass distribution. If this assumption is relaxed, the virial mass estimate can vary by an order of magnitude or more \citep{Merritt87,Wolf+10}. It is therefore important not to make any assumption on the relative distribution of the cluster mass and the galaxies, even if several studies have shown that red, passive galaxies do indeed trace the total mass profile of clusters \citep[e.g.][]{vanderMarel+00,BG03}. Observational selection tends to make the bias in the spatial distribution stronger than the bias in the velocity distribution \citep{Biviano+06}, so it is more robust to estimate a cluster mass directly from the velocity distribution of cluster galaxies, using a scaling relation \citep[e.g.][]{Evrard+08,Munari+13,Ntampaka+15}, rather than using the virial theorem. The intrinsic scatter of the mass-velocity dispersion relation is $\leq 5$\%, but observational effects (see Sect.~\ref{sss:pbms}) increase the scatter to $\sim 40$\% \citep{WCS10,SMBD13}. 


\subsubsection{Methods} \label{sss:kinmet}
If $\gtrsim 100$ tracers of the cluster gravitational potential are available, cluster masses and mass profiles can be determined without any assumption about the spatial and/or velocity distribution of the tracers relative to the mass. One possibility is to relate the observed PPS distribution of galaxies to their intrinsic phase-space distribution via \citep[see, e.g.,][]{DM92}
\begin{eqnarray}
& & g(R,v_{\rm los}) = \nonumber \\
& & \qquad 2\int_R^\infty
\frac{r\,d r}{(r^2-R^2)^{1/2}}
\int_{-\infty}^{+\infty} 
\int_{-\infty}^{+\infty} 
f(E,L) \,d v_\theta \,d v_R \,,
\end{eqnarray} 
where $r$ is the radial distance from the cluster centre in 3D, $(v_R,v_{\theta})$ are Cartesian components of the velocity along the polar coordinates $(R,\theta)$ in the plane of the sky, and $v_{\rm los}$ is the line-of-sight velocity component (i.e. the one we observe via the redshift measurement). The intrinsic phase-space distribution $f$ is expressed in terms of the energy $E$ and angular momentum $L$. The gravitational potential is related to $f(E,L)$ through the Poisson equation. Since the shape of the $f(E,L)$ distribution function is not known from theory, it is generally estimated for haloes extracted from cosmological simulations \citep{Wojtak+08,Wojtak+09}. The $f(E,L)$ method has been used to estimate the mass profiles $M(r)$ of 41 nearby clusters by \citet{WL10} and a stack of sixteen $z=0.17-0.55$ clusters by \citet{vanderMarel+00}.

Another widely adopted method for the $M(r)$ determination is to search for a solution of the Jeans equation for a collisionless system of galaxies in dynamical equilibrium,
\begin{equation}\label{e:Jeans}
G \, M(r) = - r  \langle v_r^2 \rangle \left( \frac{d
\ln \nu}{d \ln r} + \frac{ d \ln \langle v_r^2 \rangle}{d \ln r} +2 \beta \right).
\end{equation}
where $\nu(r)$ is the cluster 3d galaxy number density profile, and $\langle v_r^2 \rangle$ is the mean squared radial velocity component, that reduces to the radial veocity dispersion $\sigma_r$ in the absence of bulk motions. $\beta(r)$ is the velocity anisotropy profile,
\begin{equation}
\beta(r) \equiv 1 - \frac{\langle v_{\theta}^2 \rangle}{\langle v_r^2 \rangle},
\end{equation}
where $\langle v_{\theta}^2 \rangle$ is the mean squared velocity component along one of the two tangential directions in spherical coordinates, that reduces to the tangential velocity dispersion $\sigma_{\theta}$ in the absence of bulk motions. Since most clusters of galaxies do not rotate \citep{HL07}, it is usually assumed that the two tangential components of the velocity are identical. The Abel integral equation relates $\nu(r)$ to the observable (projected) galaxy number density profile $N(R)$ \citep{BT87}, under the assumption of spherical symmetry. On the other hand, one cannot directly determine $\sigma_r(r)$ from the observable $\sigma_{\rm{los}}(R)$, since knowledge of $\beta(r)$ is required. This is the so-called \textit{mass-anisotropy degeneracy} \citep[MAD hereafter,][]{BM82} and it is the critical point of this method. To solve the MAD, one can use the mean $\beta(r)$ of cluster-size haloes extracted from cosmological simulations \citep{MBM10,lau10}; $M(r)$ then follows directly from the observables in a non-parametric approach \citep{MB10,Wolf+10}. Other possibilities to solve the MAD problem is to use the fourth-order (kurtosis) Jeans equation \citep{Lokas02,LM03,RF13}, or to separately solve the Jeans equation for different tracers, e.g. early-type and late-type galaxies, since they may have different $\beta(r)$ for the same $M(r)$ \citep{Battaglia+08,BP09}. The Jeans equation has been used to determine $M(r)$ of many individual clusters or stacks of several clusters \citep[e.g.][]{CYE97,BG03,LM03,KBM04,BP09,Lemze+09}.

\texttt{MAMPOSSt} \citep[\textit{Modelling of Anisotropy and Mass Profiles of Observed Spherical Systems,}][]{MBB13} is a hybrid method that solves the Jeans equation Eqn.~(\ref{e:Jeans}) to compute the probability of observing a galaxy in a given $(R,v_{\rm{los}})$ position in PPS, by assuming models for $M(r)$ and $\beta(r)$ and a shape (e.g. a Gaussian) for the 3D velocity distribution (and not, as is usually done, for the line-of-sight  velocity). The probability of observing a galaxy with velocity $v_{\rm{los}}$ at the projected radius $R$ is:
\begin{eqnarray}
& & p(v_{\rm{los}}|R) =  (2\pi)^{-1/2}\, \nonumber \\ 
& & \times \int_0^\infty (\nu/ \sigma_{\rm{los}})\,\exp [-v_{\rm{los}}^2/(2\,\sigma_{\rm{los}}^2)] \, dz \, / \,
\int_0^\infty \nu\, dz 
 \ ,
\label{e:mamp}
\end{eqnarray}
where $z$ is the direction of the line-of-sight. The 3D number density profile $\nu(r)$ comes from the observed (projected) number density profile $N(R)$ via the Abel integral. The line-of-sight velocity dispersion comes from $\sigma_r$ and $\beta$ via:
\begin{equation}
\sigma_{\rm{los}}^2(R,r) = [1-\beta(r)\,(R/r)^2]\;\sigma_r^2(r) \ ,
\label{e:slos}
\end{equation}
and $\sigma_r$ is obtained from $\beta(r)$ and $M(r)$ as a solution of the Jeans Eqn.~(\ref{e:Jeans}) \citep{vanderMarel94},
\begin{equation}
\nu \, \sigma_r^2 = - G \int_{r}^{\infty} \nu(\xi)
M(\xi)/\xi^2 \, \exp [ 2 \int_{r}^{\xi} \beta \,
d \eta/\eta] \, d \xi\, .
\label{e:sr}
\end{equation}
The best-fit parameters of the input models for $M(r)$ and $\beta(r)$ are obtained by maximising the product of the $p(v_{\rm{los}}|R)$ probabilities. \texttt{MAMPOSSt} has been used to determine several individual or stack cluster mass profiles \citep{Biviano+13,MBM14,Durret+15,Balestra+16,Biviano+16,Verdugo+16,Biviano+17a,Biviano+17b}. In combination with gravitational lensing (see Sect.~\ref{ss:lensing}) \texttt{MAMPOSSt} has also been used to constrain the nature of gravity \citep{Pizzuti+16,Pizzuti+17} and the equation of state of dark matter \citep{Sartoris+14}.

The \texttt{Caustic} method \citep{DG97,Diaferio99,Serra+11} has been developed to estimate cluster masses beyond the virial region, i.e. outside the domain of validity of the methods described above, that all rely on the asumption of dynamical equilibrium. This method defines the caustic in PPS by identifying steep density gradients in PPS along the velocity axis. N-body simulations show that the caustic amplitude ${\cal A}(R)$ can be used to estimate $M(r)$ via
\begin{equation}
G \, [M(r)-M(r_0)]=\int_{r_0}^r {\cal A}^2(x) \, {\cal F}_{\beta}(x) \, dx \, .
\label{eq:cau}
\end{equation}
${\cal F}_{\beta}(x)$ is a radial varying function of both $\beta(r)$ and the gravitational potential itself. Eqn.~\ref{eq:cau} can be solved by assuming a constant value for ${\cal F}_{\beta}$. This assumption is violated within the virial region, leading to a mass over-estimate \citep{Serra+11}, but it is a valid one outside the virial region, where numerical simulations indicate ${\cal F}_{\beta} \approx 0.5-0.7$ \citep[][and references therein]{GMK13}. Since Eqn.~(\ref{eq:cau}) is differential in $M(r)$, one can obtain the mass profile out to a given radius $r_0$ by another technique, e.g. \texttt{MAMPOSSt}, and then use the \texttt{Caustic} method to determine $M(r>r_0)$ \citep{BG03,Biviano+13}. The \texttt{Caustic} method has been extensively used to determine cluster mass profiles \citep[e.g.][]{BG03,Biviano+13,Geller+14,Guennou+14}.


\subsubsection{Sources of systematic uncertainty} \label{sss:pbms}

In Sect~\ref{sss:kinmet} we have already mentioned the systematic uncertainties that are specific to each individual method. Here we discuss how these and other issues propagate into systematic effects in the resulting mass estimate. The typical level of systematic uncertainty  in cluster mass estimates inherent in current methods, assuming typical data-sets of $\sim 100$ cluster members, is summarised below in percentages (a value of 0 means the bias can be fully corrected):\\

\noindent\begin{minipage}{\columnwidth}
\begin{itemize}
\item mass-anisotropy degeneracy \dotfill $8\%$ 
\item uncertainty in ${\cal F}_{\beta}$ \dotfill $15 \%$ (Caustic-method specific)
\item dynamical equilibrium \dotfill $30 \%$ (irrelevant for Caustic method)
\item interlopers \dotfill $10 \%$
\item spatial incompleteness \dotfill $0\%$
\item triaxiality \dotfill $30\%$
\end{itemize}
\end{minipage}
\\

\paragraph{\bf Dynamical equilibrium:}
Deviation from dynamical equilibrium can result from ongoing major mergers between a cluster and a massive accreting subcluster. \citet{TNM10} find that a cluster-subcluster collision may lead to a factor $\sim 2$ mass over-estimate from kinematics, for $\sim 1$ Gyr after each core passage of the subcluster, but only if the collision axis is aligned with the line-of-sight. Most of the effects of the collision are erased after a dynamical timescale. Observationally, deviation from dynamical equilibrium can be identified by the analysis of the shape of the velocity distribution of cluster galaxies \citep{Biviano+06,RLT11,RPHL18}.

\paragraph{\bf Interlopers:}
Interlopers can be defined in two ways: (1) galaxies that are located in projection within a given radius from the cluster centre, but are outside the sphere of same radius, or (2) galaxies that are unbound to the cluster. While interloper-removal techniques have become increasingly sophisticated with time \citep[e.g.][]{YV77,Fadda+96,Wojtak+07,MBB13}, it is impossible to reduce contamination by interlopers to zero and, at the same time, retain all the real members in the sample \citep{MBM10}. Comparison to numerical simulations indicate that contamination by interlopers and incompleteness of real cluster members tend to overestimate cluster masses at the low-mass end and underestimate cluster masses at the high-mass end \citep{Wojtak+18}.

\paragraph{\bf Spatial incompleteness:}
A particular observational set-up (e.g. caused by fiber collision of slit positioning) can cause a spatially-dependent incompleteness of the spectroscopic sample. If not properly corrected, this incompleteness induces an error in the determination of $\nu(r)$. On the other hand, the velocity distribution of cluster galaxies is mildly, if at all, affected by spatially-dependent incompleteness. If the incompleteness cannot be corrected it is more robust to base the cluster mass estimate on the velocity distribution only \citep{Biviano+06}, or to use the complete photometric sample with background subtraction, to estimate $\nu(r)$ \citep[see, e.g.][]{Biviano+13}.

\paragraph{\bf Triaxiality:}
All clusters are triaxial, and the velocity dispersion tensor is elongated along the same direction as the galaxy spatial distribution. If the cluster major axis is aligned along (perpendicular to) the line-of-sight, the observed velocity dispersion will be higher (lower) than the average of the three components of the velocity dispersion tensor \citep{Wojtak13}. The cluster projection angle relative to the cluster major axis is generally very difficult to determine \citep{Sereno07}, so triaxiality becomes an important source of (systematic) error in the cluster mass estimate, up to $\sim 60$\%, but generally much lower than this \citep{Wojtak13}. Stacking several clusters together is a simple and effective way of getting rid of the triaxiality problem \citep[e.g.][]{vanderMarel+00}.


\begin{figure*}
\includegraphics[width=0.33\textwidth, keepaspectratio]{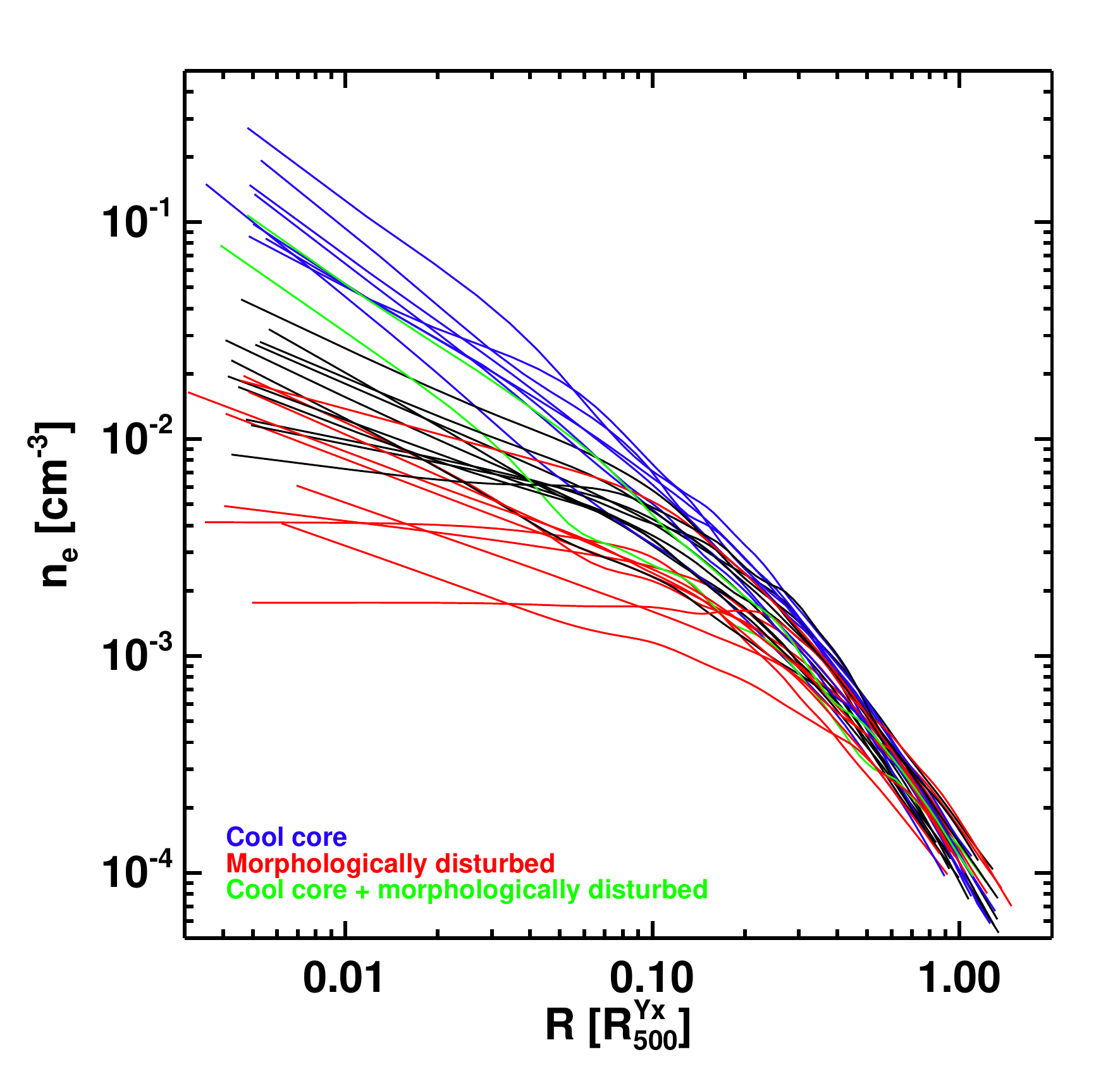}
\includegraphics[width=0.33\textwidth, keepaspectratio]{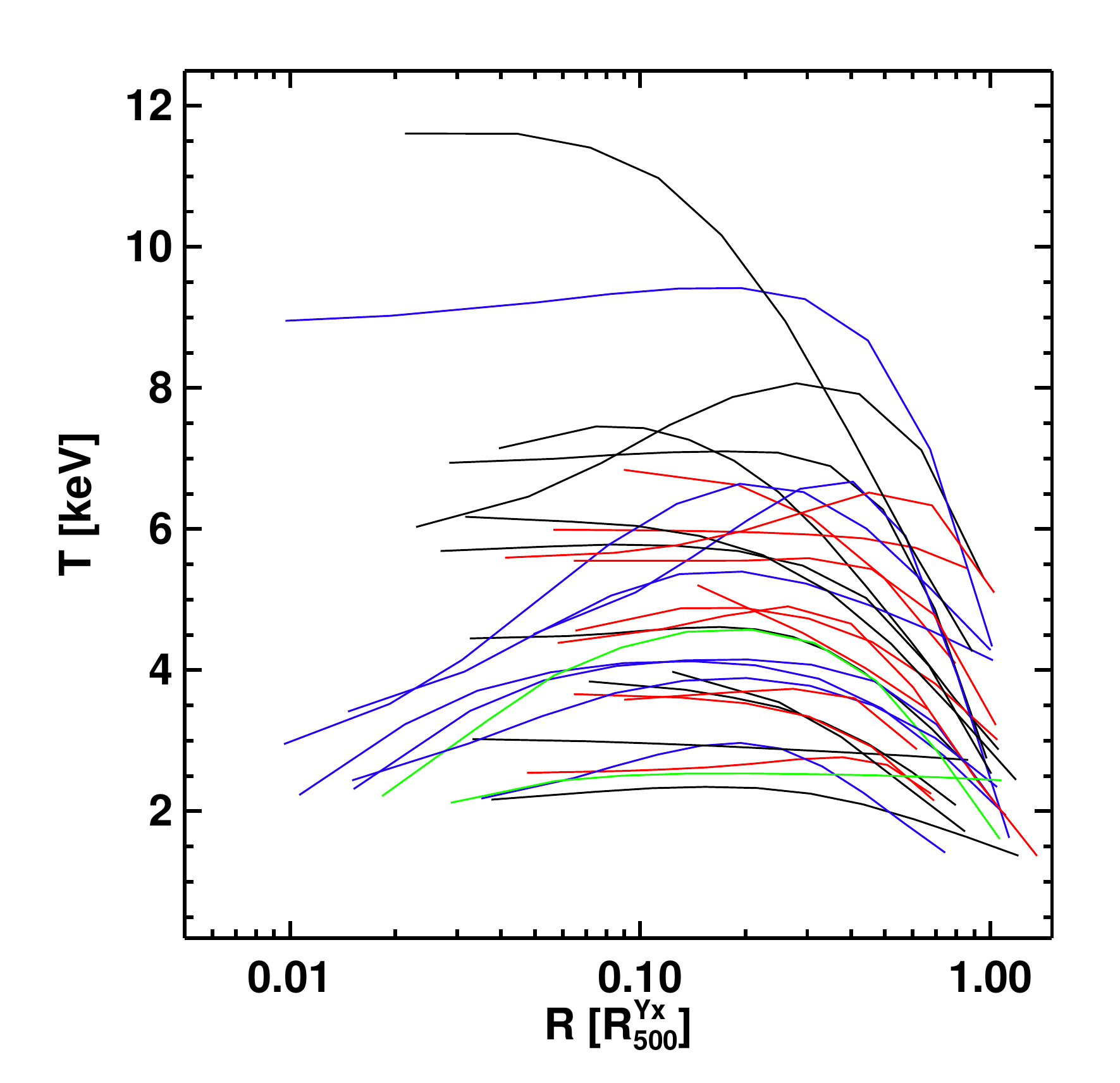}
\includegraphics[width=0.33\textwidth, keepaspectratio]{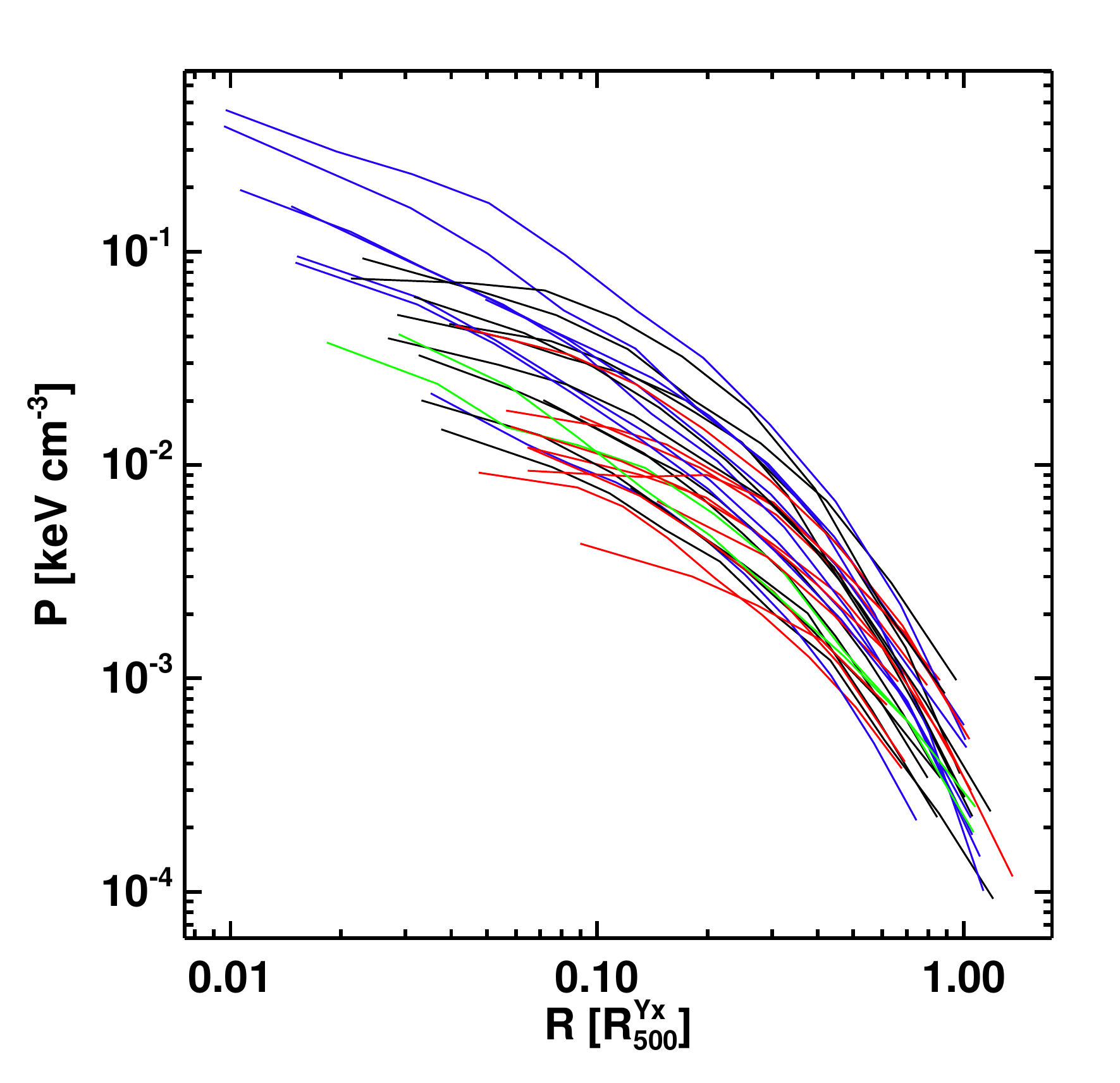} 
\caption{Left to right: Reconstructed gas density, temperature, and pressure profiles from the REXCESS  sample \citep{Pratt10,arn10}, plotted as a function of $R_{500}$. The profiles are colour-coded as a function of dynamical state as defined in \citet{pratt09}: cool core (blue), morphologically disturbed (red), cool core and morphologically disturbed (green) and intermediate (black).
} 
\label{fig:physprof}
\end{figure*}

\subsection{X-ray and hybrid SZ} 
\label{sec:X-ray}

Upon reaching equilibrium, the thermodynamical properties of the ICM satisfy the HE relation between the ICM pressure $P_{\rm gas}$, the ICM density $\rho_{\rm gas} = \mu m_p n_{\rm gas}$ and the potential (see Eqn. \ref{eq:hee}). We discuss in Sect.~\ref{sec:icm_energy} the insights gained from numerical simulations for when the condition of HE is not satisfied, and what this implies for the mass reconstruction. Measurement of cluster masses in X-rays, through the hydrostatic assumption, gained significant traction after the launch of ROSAT in 1990, owing to the easy availability of spatially resolved density profile observations. 


\subsubsection{Method}
\label{sec:xraymethod}

Assuming a spherically symmetric distribution, one can write the HE equation:
\begin{equation}
M_{\rm tot}(<r) = - \frac{r \, P_{\rm gas}}{\mu m_{\rm u} G \, n_{\rm gas}} \frac{d \log P_{\rm gas}}{d \log r}, 
\label{eq:mhe}
\end{equation}
where $G$ is the gravitational constant, $m_{\rm u} = 1.66 \times 10^{-24}$ g is the atomic mass unit, 
and $\mu= \rho_{\rm gas} / (m_{\rm u} n_{\rm gas}) \approx (2 X +0.75 Y +0.56 Z)^{-1} \approx 0.6$ 
is the mean molecular weight in atomic mass unit for an ionized plasma;  $X$, $Y$ and $Z$ being the mass fraction for hydrogen, helium and other elements, respectively. For a typical metallicity of 0.3 times Solar abundance, and assuming the abundance table of \citet{ag89},  $X+Y+Z=1$, with $X \approx 0.716$ and $Y \approx 0.278$.

Assuming the ICM follows the equation of state for a perfect gas ($P_{\rm gas} = k T_{\rm gas} n_{\rm gas}$, where $k$ is the Boltzmann constant), the directly-observable physical quantities in X-rays are the radial density $n_{\rm gas}$ and temperature $T_{\rm gas}$ of the plasma (e.g. Fig.~\ref{fig:physprof}).
The gas density can be obtained from the geometrical deprojection of the X-ray surface brightness, extracted in thin annuli. Corrections for contaminating gas clumps can be obtained by masking substructures (which are spatially resolved with \xmm\ and \cxo), and by measuring the azimuthal median, instead of the azimuthal mean \citep{zhu13,mor13,eckert+15}. The radial gas temperature distribution is built from spectra extracted in annuli that are wider than those used for the surface brightness, and can be modelled with an absorbed thermal component. 

A relatively recent development is the availability of spatially-resolved SZ electron pressure profiles,  $P_{\rm gas}$, which  can be obtained from geometric deprojection of the azimuthally-averaged integrated Comptonization parameter $y$ \citep[e.g.][]{mro09,planck13,sayers+16,romero+17,rup18}.
Joint deprojection of SZ and X-ray images can be applied to avoid the use of X-ray spectroscopic data \citep[e.g.][]{lar06,ameglio07,adam16,rup17,shitanishi+18,ghi18b}, and also for total mass estimates \citep[e.g.][]{ameglio09,adam16,rup17,rup18}.

\begin{figure*}
\includegraphics[width=\textwidth, keepaspectratio]{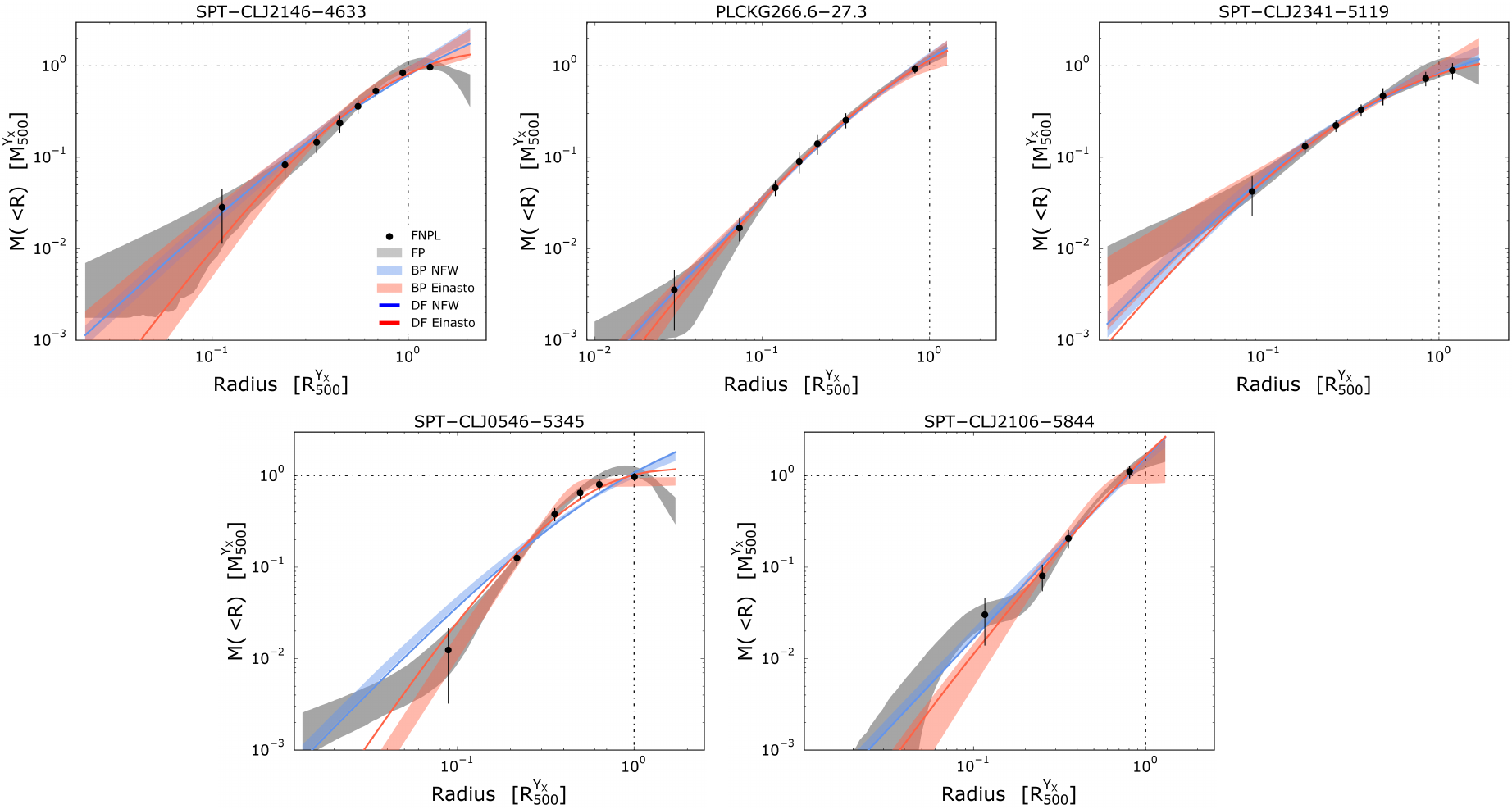}

\caption{3D total mass profiles of the clusters in the sample analysed by \citet{bartalucci18} at $z \sim 1$, with  several mass distribution reconstruction methods overplotted.} 
\label{fig:bmprof}
\end{figure*}

The two main approaches adopted to solve Eqn.~\ref{eq:mhe} are known as the {\it backward} and {\it forward} methods. In the {\it backward} method, a parametric mass model is assumed and combined with the gas density profile to predict a gas temperature profile $T$, which  is then compared to the measured $T_{\rm m}$ in the spectral analysis.
In the {\it forward} method, functional forms are fitted to the deprojected gas density or surface brightness  profiles, and to the temperature 
\citep[e.g.][]{pra02,vik06,pointecouteau05} or pressure profiles \citep[e.g.][]{pra16,ghi18a,ettori18}, 
with no assumptions on the form of the gravitational potential.  In all cases, the procedure takes  into account projection and PSF effects (the latter can be neglected for \chandra\ data). The HE equation (Eqn.~\ref{eq:mhe}) is then 
applied to evaluate the radial distribution of the mass.  More details on the traditional methods applied to X-ray data to estimate the mass profile can be found in \citet{ettori+13}. 

The forward method has several variants, as described by \citet{bartalucci18}. The fully parametric method \citep[e.g.][]{vik06} directly uses the best-fitting analytic density and temperature or pressure models.  Another approach,  the non-parametric  like method \citep[e.g.][]{pra16},  uses parametric models only to correct the observed temperature profiles for projection and PSF effects, and to smooth the density gradients.   

The various methods have been compared by \citet{bartalucci17} and \citet{bartalucci18}. They showed that the density profile is exceptionally robust to both the method used for its reconstruction and to instrumental systematic effects. They found that mass profile estimates are also insensitive to the reconstruction method  in the radial range of the measured temperature profile. On the other hand, the mass uncertainty does depend on the method, with fully parametric methods yielding the smallest uncertainties. The mass estimate also depends on the method when extrapolation is required, especially in the case of irregular profiles  (see Fig.~\ref{fig:bmprof}). 

If SZ data are available, the likelihood can also include a comparison with $T = P_{\rm SZ} /n_{\rm gas}$. This method takes advantage of the larger extension of the SZ signal in constraining the mass profile model \citep[e.g.][]{planck13,ghi18a, ettori18}. As combination with SZ data does not need spectroscopic temperature measurements, this method also allows for hydrostatic mass profile estimates out to higher redshift \citep{rup18}.


\subsubsection{Sources of systematic uncertainty}

The hydrostatic mass estimate depends on the direct measurement of the gas density profile from X-ray data, combined with the radial profile of either the X-ray spectroscopic temperature, or the SZ-derived pressure of the ICM.  Any bias on these measurements propagates into systematic effects on the resulting mass estimate, which can be roughly summarised in percentages as follows:\\

\noindent\begin{minipage}{\columnwidth}
\begin{itemize}
\item assumption of spherical symmetry \dotfill few \%
\item hydrostatic mass bias \dotfill $< 10$ - 30\%
\item gas temperature inhomogeneities \dotfill few - 10-15\%
\item gas clumping \dotfill few \%
\item absolute X-ray temperature calibration \dotfill 15-20\%
\end{itemize}
\end{minipage}
\\

\paragraph{\bf Spherical assumption:} The biases induced by the assumption of spherical symmetry were investigated by \citet{buo12}, who found that while the mean bias is small ($\lesssim 1\%$), substantial variations can occur on a cluster to cluster basis, depending on the exact viewing orientation.

\paragraph{\bf Hydrostatic mass bias:}
The fundamental assumption of the X-ray and SZ analyses described above is that the gas is in HE in the dark matter potential. As discussed  in Sect.~\ref{sec:icm_energy}, numerical simulations are unanimous in predicting that such an assumption is likely to lead to an underestimate of the mass due to neglect of bulk motions and turbulence in the ICM. The effect will naturally be most important in dynamically disturbed systems (up to $\sim 30\%$), and least important in relaxed objects ($\lesssim 10\%$). The actual magnitude of this so-called `hydrostatic bias' is difficult to ascertain both numerically (see Sect.~\ref{sec:icm_inhom}) and observationally, although great progress has recently been made in this area and is discussed below in Sect.~\ref{sec:stateoftheart}.

\paragraph{\bf Gas temperature inhomogeneities:}
An issue that can potentially affect the X-ray analysis is the presence of temperature inhomogeneities in the gas (Sect.~\ref{sec:icm_inhom}). If a significant amount of cool gas is present, then a single temperature fit will be biased towards lower temperatures, which will in turn have an impact on the mass estimate.  Usually, X-ray spectra are accumulated within annular regions and their spectral shape is fitted assuming that the gas temperature within the considered shell is uniform. Essentially all X-ray instruments used thus far for estimating ICM temperatures possess an effective area that peaks around 1~keV and declines steeply above $\sim5$ keV. This renders current X-ray telescopes intrinsically more sensitive to the gas in the temperature range $\approx1-3$ keV \citep{Mazzotta04,vikhlinin06b}. If the gas distribution within a given shell is multiphase, the X-ray spectra fitted assuming that the plasma is single-phase should in principle underestimate the mean gas-mass-weighted temperature. This effect can be enhanced in cases where the cooler gas phase coincides with infalling substructures, which feature an increased gas density with respect to their environment and thus contribute strongly to the total X-ray flux. The percentages listed above come from numerical simulations \citep{rasia14}, but estimates of the effect vary widely owing to differences in numerical schemes and input physics. For example, simulations with heat conduction always predict more homogeneous temperature distributions, minimising any bias, while simulations lacking non-gravitational input from supernovae and AGN predict long-lasting, dense cool cores that will strongly contribute to any bias. While most observational studies attempt to make allowances for temperature inhomogeneities by masking detected structures and taking the line-of-sight gradient into account with a spectral-like temperature weighting while deprojecting, there may remain an additional source of temperature inhomogeneities which current studies cannot pinpoint. 

\paragraph{\bf Gas clumping:}
A further issue is gas clumping, i.e. deviations from an isotropic distributions induced by substructures, which can potentially bias measurements of the gas density, and thus the mass, when the X-ray signal is azimuthally averaged over concentric annuli. Current limits from X-ray observations \citep{eckert+15,morandi13,urban14,tchernin16} agree with the predictions from numerical simulations \citep[e.g.][]{roncarelli13}. Observational constraints on gas clumping are described in detail in the chapter of this series related to galaxy cluster outskirts \citep{wal19}.

\begin{figure}
\includegraphics[width=1.025\columnwidth]{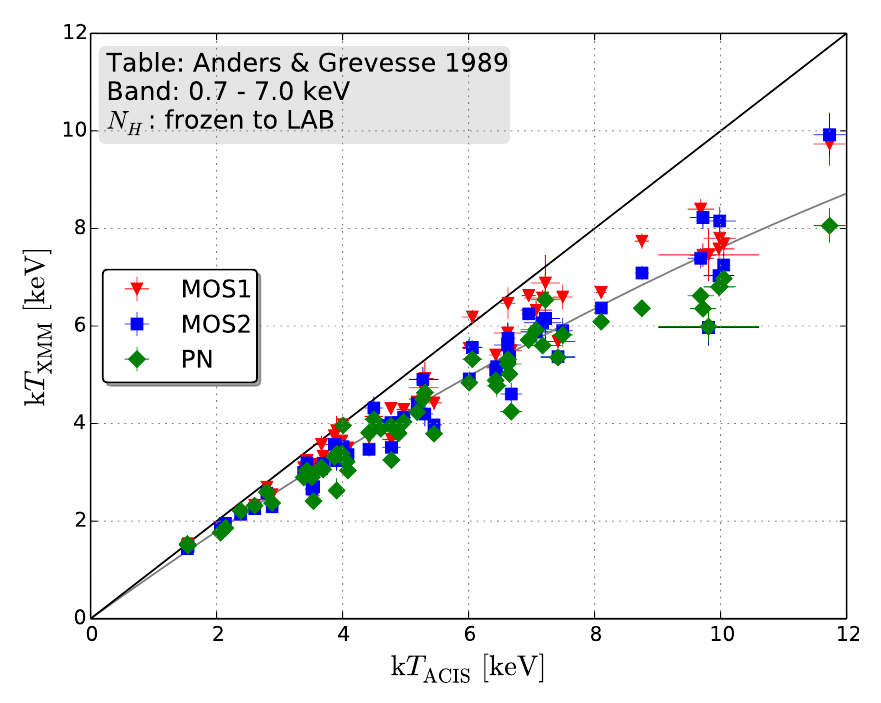}
\hfill
 \caption{ICM gas temperatures measured with the three \xmm\ instruments (MOS1, MOS2, and PN) plotted against temperatures measured with \emph{Chandra}/ACIS for the same regions. The black line is the one-to-one relation. Reproduced from \citet{schellenberger15}.} \label{fig:hse_systematics}
\end{figure}

\paragraph{\bf Absolute X-ray calibration:}
A final issue is the absolute calibration of X-ray instrumentation. In recent years, it has become apparent that ICM temperatures estimated with different X-ray instruments (in particular \xmm\ and \emph{Chandra}) show a systematic offset \citep{nevalainen10,Mahdavi13,Martino14,donahue14,schellenberger15}. The observed differences result from the calibration of the effective area of the two telescopes, which is inconsistent at the 5-10\% level. In Fig. \ref{fig:hse_systematics}, taken from \citet{schellenberger15}, we show a comparison between spectroscopic temperatures measured with \xmm\ and \emph{Chandra} for the same regions. While for temperatures below 5 keV the offset between the two is small, at high temperatures the measured temperatures differ by 15-20\%. Hydrostatic masses estimated with X-ray data only are principally proportional to the gas temperature. The corresponding uncertainty propagates linearly to the hydrostatic mass in the first approximation  (for a mass at a fixed overdensity, the scaling is roughly $\propto M\sim T^{1.5}$). However the hydrostatic mass also depends on the temperature gradient. In this context, we note that \citet{Martino14} actually found a very good agreement (within 2\%) between hydrostatic masses estimated from \chandra\ and \xmm. 

One possible way of investigating the issue is to compare in a systematic way the spectroscopic X-ray temperatures with the temperatures estimated by combining the gas density with the pressure measured through the SZ effect, $\eta_T=T_{\rm X}n_{\rm X}/P_{\rm SZ}$. \citet{bourdin17} measured $\eta_T=1.02_{-0.03}^{+0.02}$ with a low-redshift \xmm\ sample. A very similar value, $\eta_T=1.04\pm0.08$, was reported for the X-COP sample \citep{ghi18b}. \citet{adam17} performed a detailed comparison of temperatures in the hot cluster MACS\,J0717.5+3745 between \xmm, \emph{Chandra} and \emph{NIKA}, and found that the joint X/SZ temperatures lie somewhat in between, $\eta_T\sim0.9$ and $\eta_T\sim1.1$ for \xmm\ and \emph{Chandra}, respectively. The statistical quality of such comparisons is expected to increase substantially in the near future given the growing number of systems with available deep SZ data.


\subsection{Weak lensing analysis} 
\label{ss:lensing}

The deep potential well of a galaxy cluster weakly and coherently
distorts the shapes of background galaxies through the differential deflection of
light rays. A statistical treatment of the coherent distortion pattern 
allows us to measure cluster masses without assumptions about their
physical nature or dynamical state. This is the well-known weak gravitational lensing (WL hereafter) effect, which has recently become extremely competitive as a means to estimate cluster masses. In general, wide-field cameras installed on large ground-based telescopes are the best instruments for weak-lensing analysis;
e.g. Suprime-cam and Hyper Suprime-Cam (HSC) on the Subaru telescope, and
MegaCam of the Canada-France-Hawaii Telescope, and Dark Energy Camera
(DECam) of the Victor M. Blanco 4-meter Telescope. 
Specifically, large mirrors can observe galaxies up to $z\sim1$ in short
observing times, and the wide field-of-view cameras cover out to the virial radius with superb image
quality. The advent of
the prime-focus cameras installed on large mirror telescope has made a tremendous progress of
weak-lensing analysis for the last decade.


\subsubsection{Method}

Images of background source galaxies are distorted by the tidal
gravitational field. Image distortion of background source galaxies is expressed by the complex shear, $\gamma\equiv \gamma_1 + i
\gamma_2$.
The complex shear is related to the convergence $\kappa$, through
\begin{eqnarray}
 \gamma (\btheta) = \frac{1}{\pi}\int d^2 \btheta'
  D(\btheta-\btheta')\kappa(\btheta')
\end{eqnarray}
with
\begin{eqnarray}
 D(\theta)=\frac{\theta_2^2-\theta_1^2-2i\theta_1\theta_2}{|\theta|^4},
\end{eqnarray}
where \mbox{\boldmath $\theta$} is an apparent angular position. 
Here, the convergence $\kappa$ is the dimensionless projected mass
density, given as 
\begin{eqnarray}
 \kappa(\btheta)=\frac{\Sigma(\btheta)}{\Sigma_{\rm cr}(z_{\rm l},z_{\rm
  s})}, \label{eq:kappa}
\end{eqnarray}
with the dimensional projected mass density $\Sigma$ and the critical surface mass density 
\begin{eqnarray}
\Sigma_{\rm cr}(z_l,z_s)=\frac{c^2}{4\pi G}\frac{D_{\rm s}}{D_{\rm
 l}D_{\rm ls}}. \label{eq:sigma_cr}
\end{eqnarray}
Here $D_{\rm l}$ is the angular diameter distance to the lens,
and $D_{\rm s}$ and $D_{\rm ls}$ are the angular diameter distances from the
observer to the sources and from the lens to the sources, respectively.

The complex ellipticity of individual galaxies is defined as \citep{Bartelmann01},
\begin{eqnarray}
 \varepsilon=\frac{Q_{11}-Q_{22}+2iQ_{12}}{Q_{11}+Q_{22}+2(Q_{11}Q_{22}-Q_{12}^2)^{1/2}}
\end{eqnarray}
\citep[e.g.][]{KSB,Okabe08,Kitching08,Oguri12,Heymans12,Miller13,Applegate14,Umetsu14,
Hoekstra15,Okabe16b,DESWL16,Okura18} or 
\begin{eqnarray}
 \chi=\frac{Q_{11}-Q_{22}+2iQ_{12}}{Q_{11}+Q_{22}}
\end{eqnarray}
\citep[e.g.][]{Bernstein02,Hirata03,HSCWL1styr}, where $Q_{ij}$ is the
quadruple moment of the surface brightness. 
The observed ellipticites, $\varepsilon,\chi$ are distorted by the
gravitational lensing effect, and expressed in the weak-limit ($\kappa\ll 1$ and
$\gamma\ll 1$), as follows, $\varepsilon \rightarrow
\varepsilon_s+g$ and $\chi\rightarrow \chi_s+2g$,
where the subscript $s$
denotes the intrinsic (unlensed) ellipticity and $g$ is the reduced shear
\begin{eqnarray}
g=\frac{\gamma}{1-\kappa}.
\end{eqnarray}

Since the gravitational lensing signals in the central regions of
galaxy clusters are somewhat strong, one in general uses the reduced shear, $g$, rather than
the shear $\gamma$ for cluster mass measurements.
Assuming that orientations of intrinsic ellipticity are random ($\langle
\varepsilon_s \rangle=0$ and $\langle \chi_s \rangle=0$ ), the
gravitational lensing signal can be measured by an ensemble average of
background galaxies, $\langle g \rangle \simeq \langle \gamma \rangle
\simeq \langle \varepsilon \rangle \simeq \langle \chi \rangle /2$.
The statistical uncertainty of the shear component, $\sigma_g \simeq
(\langle \varepsilon_s^2 \rangle/N)^{1/2}$, decreases with increasing the
number of background galaxies, $N$. 
Therefore, weak-lensing analysis requires a large number of background galaxies. 

Weak-lensing observables do not provide direct estimates of three-dimensional masses of clusters 
because the lensing signal probes the two-dimensional projected mass
distribution.
One therefore estimates $M_\Delta$ by fitting a three-dimensional model to
the data. For this purpose, a tangential distortion profile as
a function of cluster-centric radius is widely used in weak-lensing mass measurements.
This quantity is computed in a given annulus by azimuthally averaging the measured galaxy
ellipticities. In recent studies, the expression of a dimensional
component, $\Delta \Sigma_+$, is being widely used
rather than a dimensionless expression, $g_+$, thanks to recent updates
of photometric redshifts. The tangential components of reduced shear in
the $i$-th radial bin are estimated as 
\begin{eqnarray}
    \langle \Delta \Sigma_+ \rangle (R_i) =\frac{\sum_n \Sigma_{{\rm cr},n} \varepsilon_{+,n}\,w_n}{\sum_n w_n},
    \label{eq:DSimga_+}
\end{eqnarray}
where the subscript $n$ denotes the $n$-th galaxy located in the annulus 
spanning $R_1<R_i<R_2$ and $w_{n} \propto \Sigma_{{\rm cr},n}^{-2}$ is
the weighting function considering the intrinsic ellipticity and shape
measurement errors. We here use the notations of projected radii $R$ and three-dimensional radii $r$.
When $\Sigma_{\rm cr}$ is computed by integrating
the full probability function, $P(z)$, $\Sigma_{\rm cr} \equiv \langle
\Sigma_{{\rm cr}}^{-1}\rangle^{-1}$ where the bracket denotes the
average over redshifts. 
The projected distance from a given cluster centre, $R_i$, is defined by the weighted
harmonic mean radius of sparsely distributed galaxies
\begin{eqnarray}
    R_i=\frac{\sum_n w_n}{\sum_n w_n R_n^{-1}}, 
\end{eqnarray}
\citep{Okabe16b}.
When one corrects the measured values using the multiplicative
calibration bias $m$ for individual galaxies \citep{Heymans06,Massey07},
the measured ensemble shear becomes $\Delta \Sigma_{+,i} \rightarrow \Delta
\Sigma_{+,i}/(1+K_i)$, where the correction factor, $K$, is described by 
\begin{eqnarray}
 K_i=\frac{\sum_n m_n w_n}{\sum_n w_n}.\label{eq:Kcor}
\end{eqnarray}
When one computes the tangential shear profile in comoving coordinates,
all the equations are computed with the conversion factors of 
$\Sigma_{\rm cr}^{\rm c}\equiv \Sigma_{\rm cr}(1+z_l)^{-2}$ and $R^{\rm c}\equiv (1+z_l) R$.

Given the tangential shear profile, the log-likelihood is expressed by 
 \begin{eqnarray}
-2\ln {\mathcal L}&=&\ln(\det(C_{ij})) + \\
 &&\sum_{i,j}(\Delta \Sigma_{+,i} - f_{{\rm model}}(R_i))C_{ij}^{-1} (\Delta
 \Sigma_{+,j} - f_{{\rm model}}(R_j)), \nonumber
 \label{eq:likelihood} \end{eqnarray}
where the subscripts $i$ and $j$ are the $i-$ and $j-$th radial bins.
Here, $f_{\rm model}$ is the reduced shear prediction for a specific
mass model.
The covariance matrix, $C$, in Eqn.~\ref{eq:likelihood} is given by:
\begin{eqnarray}
 C&=&C_g+C_s+C_{{\rm LSS}}+C_{\rm int},\label{eq:C_WL}
\end{eqnarray}
\citep[e.g.][]{Gruen15,Umetsu16,Miyatake18}.
Here $C_g$, $C_s$, and $C_{\rm LSS}$ are the shape noise, the photometric redshift error, and the covariance matrix of 
uncorrelated large-scale structure (LSS) along the line-of-sight \citep{Schneider98}, respectively. 
The covariance matrix of
uncorrelated large-scale structure (LSS), $C_{\rm LSS}$, along the
line-of-sight \citep{Schneider98} is given by
\begin{eqnarray}
 C_{{\rm LSS},ij}=\Sigma_{\rm cr}^2(z_l,\langle z_s \rangle) \int
  \frac{ldl}{2\pi} C^{\kappa\kappa}(l) J_2(l R_i/D_l) J_2(l R_j/D_l), \label{eq:CovLSS}
\end{eqnarray}
where $J_2(l\theta_i)$ is the first kind and second order Bessel
function \citep{Hoekstra03} and $C^{\kappa\kappa}(l)$ is the weak-lensing power spectrum,
obtained by
\begin{eqnarray}
 C^{\kappa\kappa}= \frac{9H_0^2\Omega_m^2}{4c^4} \int^{\chi_s}_0 d\chi
  \left(\frac{\chi_s-\chi}{\chi_s a(\chi)}\right)^2 P_{\rm nl}(l/\chi;z),
\end{eqnarray}
\citep[e.g.][]{Schneider98,Hoekstra03}.
Here, $\chi_s$ is the comoving distance for the source at the average source redshift, $\langle z_s \rangle$. The latter is calculated
by $\mathcal L$ (Eqn. \ref{eq:Lz}) averaged over the radial range of the tangential shear profile. 
$P_{\rm nl}$ is the non-linear matter power spectrum \citep[e.g.][]{Smith03,Eisenstein98}.
$C_{\rm int}$ accounts for the intrinsic variations of projected
cluster mass profiles such as halo triaxiality and the presence of
correlated haloes \citep{Gruen15}.
The intrinsic covariance becomes a significant component of the
uncertainty budget of WL mass measurements as the mass increases and the data quality improves.
Since this component strongly depends on the prior and realisations, one
should carefully consider applications and limitations to the data. 
In general, each paper clearly specifies which components in the
covariance matrix are considered, and thus it is important to undertake a careful
reading to understand each analysis.

The model for the dimensional reduced tangential shear, $f_{\rm model}$,
is expressed by
\begin{eqnarray}
 f_{\rm model}(R_i)=\frac{\bar{\Sigma}_{\rm model}(<R_i)-\Sigma_{\rm model}(R_i) }{1-{\mathcal L}_i
  \Sigma_{\rm model}(R_i)}, \label{eq:fmodel}
\end{eqnarray}
with
\begin{eqnarray}
\mathcal{L}=\frac{\sum_n \Sigma_{{\rm cr},n}^{-1} w_n}{\sum_n w_n}.\label{eq:Lz}
\end{eqnarray}
Here, $\bar{\Sigma}$ and $\Sigma$ are the averaged
surface mass density within the radius and the local surface mass
density at the radius, respectively. 
The average source redshift, $\langle z_s \rangle$, is calculated by $\mathcal L$.
The denominator describes the non-linear correction in terms of the reduced
tangential shear, and can be also rewritten by  $(1-{\mathcal L}
\Sigma_{\rm model} )^{-1}\simeq 1 + {\mathcal L}  \Sigma_{\rm model}$.


\subsubsection{Sources of systematic uncertainty}

A weak-lensing analysis is generally composed of four steps: shape
measurement, estimation of photometric redshifts, selection of 
background galaxies, and mass modelling.
Systematic errors inherent in the steps can be roughly summarised in
percentage terms as follows:
\\

\noindent\begin{minipage}{\columnwidth}
\begin{itemize}
\item accuracy of shape measurements \dotfill $\sim$ a few - $10$ \%
\item accuracy of photometric redshifts \dotfill $\lesssim$ sub - a few  \%
\item background galaxies in shape catalogues \dotfill $\lesssim40$ \%
\item mass modelling   \dotfill  $\sim 10$ \%.
\end{itemize}
\end{minipage}
\\

The first three systematic errors depend on the 
technical details adopted in each paper and/or the data quality.
The last is related to both assumed mass models and intrinsic
cluster physics,
such as the distribution of internal structures, halo triaxiality, outer slope of
dark matter halo and surrounding environments.
A key effort of recent studies of cluster weak-lensing analysis is
how to control these systematic issues. 

\paragraph{\bf Shape measurements:}

Shape measurement methods can be categorised into several types:
moment measurements in real space or Fourier space, 
model fitting through maximum likelihood method or Bayesian approach, and machine learning 
\citep[e.g.][]{KSB,Mandelbaum15}. 
The anisotropic PSF ellipticities can be decomposed into three components:  optical aberration, atmospheric
turbulence and chip-misalignment \citep{Hamana13}, of which the optical aberration is the major contributor.
The PSF anisotropy is corrected via the stellar ellipticity
$\varepsilon^*$, where  an asterisk denotes stellar objects.
A good star and galaxy separation is essential in this procedure. 
Since both galaxies and stars are sparsely distributed over images, a model function of the distortion
patterns, such as bi-polynomial functions, is computed by fitting stellar ellipticites. 
 However, the isotropic PSF correction cannot be tested by the imaging data
itself, thus mock analysis of simulated images is essential to
assess the reliability of shape measurement technique of faint small galaxies
\citep[see the details][]{Heymans06,Massey07,Bridle10,Kitching12,Kitching13,Mandelbaum15,Mandelbaum17}.
The STEP programme \citep{Heymans06,Massey07} introduces a formula to
describe the accuracy of measurement pipelines, defined by
\begin{eqnarray}
 g-g_{\rm input}=mg_{\rm input}+c
\end{eqnarray}
where $g$ and $g_{\rm input}$ are the measured and input
shear, $m$ is a multiplicative bias and $c$ is a residual additive term.
Based on their pipeline tests, a multiplicative bias can correct the measured shear (Eqn. \ref{eq:Kcor}) if necessary.
Potential systematic uncertainties can be
examined using cross-correlations between maps derived from different
quantities \citep[e.g.][]{Vikram15,Oguri18}: E-mode and B-mode maps with
galaxy ellipticites, raw stellar ellipticities, residual stellar ellipticites,
and foreground galaxies.

\paragraph{\bf Photometric redshifts:}

It is important to estimate the redshifts
of background galaxies because the lensing efficiency is proportional to
$\beta\equiv D_{\rm ls}/{D_{\rm s}}$ (Eqns. \ref{eq:kappa} and \ref{eq:sigma_cr}) 
at a fixed cluster redshift ($z_l$). This quantity 
is significantly affected by changing source redshifts for objects at $z\simgt 0.4$. 
As it is not realistic to obtain spectroscopic redshifts for all 
background galaxies, photometric redshifts (photo-$z$) are always used. 

Typically, weak-lensing analysis of individual clusters uses two- or three-band imaging due to
limitations in observing times. Such a minimum combination of bands cannot in
principle be used to estimate photometric redshifts, one instead determines them
by matching magnitudes of source galaxies with those in photometric redshift
catalogues in the literature
\citep[e.g.][]{Okabe10b,Oguri12,Hoekstra12,Applegate14,Hoekstra15,Okabe16b,Umetsu16}.
To be more precise, the value of $\beta$
of the $i$-th background galaxy ($\beta_i$) or the average value
($\langle \beta \rangle$) in the target fields
is computed by taking into account adequate weights assigned to the
background galaxies. The most widely-used external photometric catalogue 
is from the COSMOS survey \citep{Ilbert13}, for which the limiting magnitude
$i'\simeq 27.5$ is sufficiently deep for galaxies used in weak-lensing analysis.
The COSMOS photometric redshifts based on the thirty bands are
well-calibrated by comparing with spectroscopic
values. Some papers also compute $\beta_i$ or $\langle \beta \rangle$
using the full probability function $P(z)$ from the COSMOS photometric catalog. 
The photometric redshift distribution can also be computed from 
pointed observations with five- or four-band imaging \citep[e.g.][]{Applegate14,Ziparo16}, and wide area galaxy surveys, e.g. the Canada France Hawaii Telescope Legacy Survey \citep[CFHTLS;][]{CHFTPhotoz06}, the Hyper SuprimeCam Survey \citep[HSC-SSP;][]{Tanaka18}, the Dark Energy Survey 
\citep[DES;][]{DESPhotoz14,DESPhotoz18}, and the Kilo-Degree Survey \citep[KiDS;][]{KIDSPhozoz17}.
The advantage of this method is that the photo-$z$ are estimated using the same data as those used for the shape measurements.

An accurate characterisation of the true underlying redshift distribution of galaxies is one of the major challenges. The photo-$z$ estimations are roughly categorised into two types: template-fitting methods \citep[e.g.][]{Arnouts99,Ilbert13}, and machine-learning methods \citep[e.g.][]{ANNz,MLZ14}.
Both template-fitting and machine learning methods are complementary and
necessary to each other \citep[e.g.][]{Salvato18}.
To test the performance of photo-$z$ estimations, one can apply
a few standard quantities such as bias (a systematic offset between photo and spectro-$z$), scatter, and outlier fraction.
 Given five broad-band filters, a sub-percent level bias between photo- and spectro-$z$ is typically achieved 
\citep[e.g.][]{CHFTPhotoz06,Tanaka18, DESPhotoz14,DESPhotoz18,
KIDSPhozoz17}, with a scatter of a few percent 
after $N$ $\sigma$ clipping.
Since weak-lensing analysis of galaxy clusters uses a large
number of galaxies at $z\sim1$, the statistical uncertainty of average
photometric redshift would be reduced by $\sim N^{-1/2}$ where
$N\simgt \mathcal O(10^4)$ is the number of the background galaxies.
Therefore, the statistical uncertainty of cluster masses caused by
photo-z estimations is typically in the order of sub percent.
Such a photo-$z$ uncertainty effect on cluster mass measurements 
can be considered in the error covariance matrix (Eqn. \ref{eq:C_WL}).

\paragraph{\bf Background Galaxy Selection and dilution effects:}

Contamination of  background catalogues by unlensed galaxies leads to an 
underestimate of the weak-lensing signal, and thus it is of critical 
importance to obtain a secure background catalogue.
The main source of contaminated unlensed galaxies are faint 
galaxies belonging to the cluster, rather than foreground objects
\citep{Broadhurst05}.
The number density of cluster galaxies increases with decreasing projected cluster-centric radius.
The ratio of cluster galaxies to background galaxies, $f_{\rm mem}$,
increases with decreasing projected
cluster-centric radius, and thereby dilutes the shear signal more at
smaller radii than at larger radii, resulting in a significant underestimate in the concentration parameter of the universal mass density profile.
This is often referred to as a dilution effect
\citep[e.g.][]{Broadhurst05,Umetsu10,Umetsu14,Umetsu15,Okabe10b,Okabe13,Okabe16b,Medezinski10,Medezinski17}.
The number of cluster members increases as cluster richness
increases, while the ratio of cluster member galaxies to background galaxies decreases as cluster redshift decreases \citep{Okabe16}.
 Therefore, the dilution effect is a redshift-, richness-, and radially-dependent phenomenon. 
\citet{Okabe16b} have shown that dilution of lensing signals for massive clusters at
$z\sim0.2$ can reach up to $\sim40\%$ at small radii, which is
significantly larger than the systematic errors of shape measurements
and photometric redshifts.  
Therefore, background selections are
the dominant source of systematic bias in weak-lensing measurements of
galaxy clusters. 

Corrections for dilution can take the form of the co-called `boost factor correction', which  attempts to correct lensing signals for a number density excess,
$(1+n_{\rm non-bkg}/n_{\rm bkg})$,
under the assumption of a radially uniform distribution of background
galaxies \citep[e.g.][]{Applegate14,Hoekstra15,Melchior17}. However, the assumption of a flat observed number density profile of
background galaxies ignores magnification bias
\citep[e.g.][]{Broadhurst95,Umetsu14} -- i.e.\ the depletion
of the number density of background galaxies at small radii due to lensing magnification.
In addition, as the boost factor correction and the concentration parameter
are highly degenerate at small radii, this approach cannot constrain
the overall mass profiles of galaxy clusters.

Another approach is to obtain a pure background source catalogue using colour
information \citep[e.g.][]{Okabe08,Umetsu14,Umetsu15,Okabe10b,Okabe13,Okabe16b,Medezinski10,Medezinski18}
or photometric redshifts \citep{Applegate14,Medezinski18}. The basic idea is to select a colour space region in which the contribution from cluster member galaxies
is  negligible by monitoring the consistency of the information from
colour, lensing signal, and the external
photometric redshift catalogue. The advantages of this technique are the consistency assessment between different datasets of galaxy
 colour, lensing information and photometric redshifts; the
 quantitative control of the purity of background galaxies;
 and no assumption of specific cluster mass models or a radial distribution for the background galaxies.
\citet{Okabe16}  have shown that  lensing signals corrected by the boost factor, with
    the assumption of the uniform background distribution, are 
    significantly underestimated compared to those derived from the pure
    background catalogue.
A final method is to use photometric redshifts directly computed by
the same dataset, simply selecting with the criteria $z_s>z_{\rm min}$.
Here $z_{\rm min}$ is the minimum redshift defined by authors. 
With the full probability function, $P(z)$, background galaxies can be
also defined as 
\begin{eqnarray}
 p_0 < \int_{z_{\rm min}}^\infty P(z) dz, 
\end{eqnarray}
where $<$ means that the
probability beyond $z_{\rm min}$ is higher than the requirement for
background selection, $p_0$ \citep[e.g.][]{Heymans12,Applegate14,Medezinski18}.

\paragraph{\bf Mass modelling of the tangential shear profile:}

Given mass models, such as NFW (Eqn. \ref{eq:rho_nfw}) and Einasto (Eqn. \ref{eq:Einasto})
one can analytically or numerically compute the local $\Sigma(R)$ and averaged surface mass density within a radius  $\bar{\Sigma}(<R)$ by integration of the three-dimensional mass density along the line-of-sight. 
The NFW form has an analytic expression for $\Sigma$ and $\bar{\Sigma}$ \citep{Bartelmann96}, while the other models mentioned above require numerical integrations. In addition to the cluster halo model, the projected mass density of the outer density profiles (i.e. a two-halo term)
can also also considered when tangential shear profiles extend far into the outskirts of galaxy clusters \citep{Oguri11a,Oguri11b}. Such a two-halo term is sometimes shown in stacked lensing profiles.

The haloes of real clusters are not perfectly spherical, but have many 
subhaloes and a triaxial structure. 
Lensing-projection bias caused by such intrinsic properties induces  bias
and scatter into weak-lensing mass measurements. For instance, if the major
axis of a triaxial halo is aligned along the line-of-sight, this leads to an overestimate of the 
halo concentration \citep{Oguri04b}.  The presence of massive subhaloes enhances the
local surface mass density and consequently underestimates the
tangential shear \citep{Okabe14a}. 
Since both the angular resolution and the signal-to-noise ratio of a tangential
shear profile of an individual cluster are relatively low,
it is very difficult to uncover all the internal properties
through lensing information alone. 

\citet{Becker11} have estimated weak-lensing masses using the tangential
shear profile of simulated haloes considering the shape noise only, and found
that the bias and scatter in $M_{500}$ for massive clusters are $\sim
-5\%$ and $\sim 30\%$, respectively. 
\citet{Oguri11b} have shown using numerical simulations
that weak-lensing masses, $M_{\rm vir}$ are underestimated by up to $5-10\%$ and a
choice of the outer boundary for fitting affects mass estimates. 
\citet{Meneghetti10} have compared weak-lensing masses at three
overdensities of $\Delta=2500,500$ and $200$ with input true mass from
numerical simulations, and found that the mean masses agree with the
input value but there is $~16\%$ scatter in realisations.
\citet{Okabe16b} have shown, based on a method to adaptively choose
the radial ranges for fitting, that the geometric mean of WL masses agrees with
the input masses for massive clusters with $\sim5\%$ scatter.
Since the assumed set-up parameters for observing conditions, such as
cluster mass ranges and redshifts, the number density of background
galaxies, and the fitting method, are all different in the literature, 
it is difficult to quantitatively compare results.


\begin{figure*}
\begin{center}
 \includegraphics[width=0.7\textwidth]{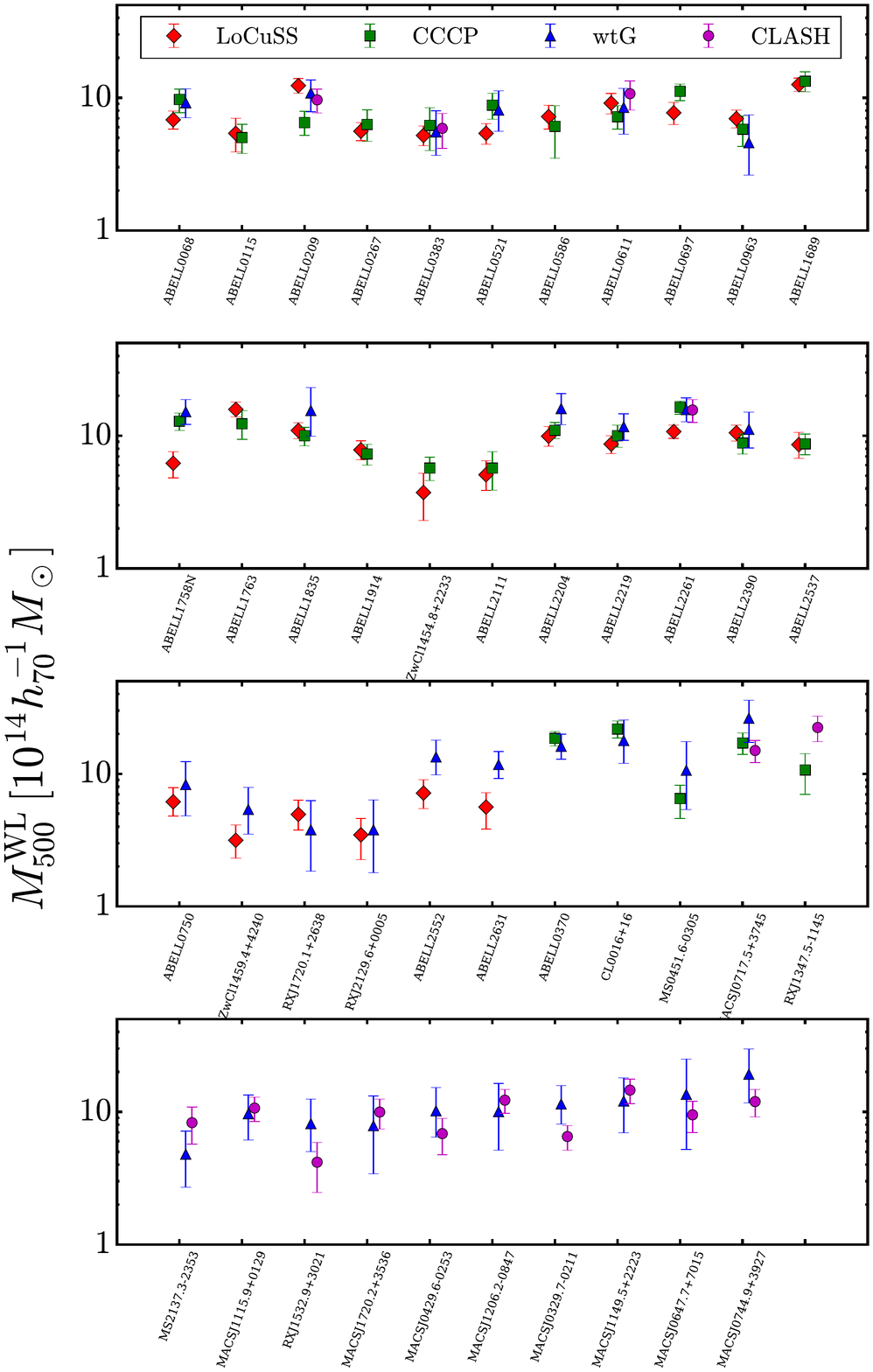}
\caption{Weak lensing total mass comparisons at $\Delta=500$ for LoCuSS \citep{Okabe16b}, CCCP \citep{Hoekstra15}, WtG  \citep{Applegate14}, and CLASH \citep{Umetsu16}.}
\label{fig:Mwl_com}
\end{center}       
\end{figure*}


\section{Recent advances}
\label{sec:stateoftheart}

In this Section, we discuss recent advances in lensing and X-ray methods in addressing the various outstanding questions and problems outlined above. We also discuss recent mass measurement comparisons between methods.

\subsection{Lensing}

\subsubsection{Results from new samples}

In lensing, possibly the most significant recent advance is the ready availability of mass measurements and profile shape parameters for moderately-large samples (many 10s) of objects. In this context, weak-lensing mass measurements for individual clusters have been carried
out by several projects, e.g. the Local Cluster Substructure Survey
\citep[LoCuSS;][]{Okabe10b,Okabe13,Okabe16b},  the Canadian Cluster Comparison Project \citep[CCCP;][]{Hoekstra12,Hoekstra15}, the Cluster Lensing And Supernova survey with Hubble
\citep[CLASH;][]{Merten14,Umetsu14,Umetsu16}, and Weighing the Giants
\citep[WtG;]{vonderLinden14a,Kelly14,Applegate14}.
The LoCuSS project presented Subaru weak-lensing mass measurements of 50
clusters, selected in X-ray luminosity from the RASS, in the redshift range of
$0.15-0.3$. The CCCP project complied CFHT data of 50 clusters at redshifts $0.15 < z
< 0.55$; 30 out of 50 clusters were selected to have {\it ASCA} X-ray temperatures
of $kT > 5$ keV.  CLASH presented results for a sample of 16
X-ray-regular and 4 high magnification galaxy clusters at $0.19<z<0.69$,
combined with Subaru and {\it HST} data.  WtG carried out Subaru
weak-lensing analysis for 51 of the most X-ray luminous galaxy clusters at $0.15<z<0.7$.

The weak-lensing analysis philosophies for the four projects demonstrate 
some strong differences, as summarised in Table \ref{tab:comp}.
The LoCuSS  project \citep{Okabe16b} made a pure background catalogue by checking
the consistency between colour, lensing strength and photo-$z$ in the 
colour-magnitude plane. They treated mass and concentration as free
parameters and carried out tangential shear fitting with
various combinations of radial ranges and number of bins, to choose
a set close to the average mass and concentration, 
because sparse distributions of background galaxies and intrinsic
cluster properties such as substructures might affect mass estimations.
The CCCP project \citep{Hoekstra15} selected background galaxies in
colour-magnitude planes, adopted a boost factor correction, and restricted
the fit to $0.5-2h_{70}^{-1}$ Mpc to minimise lensing bias
\citep{Becker11}. They assumed the mass-concentration relation of
\citet{Dutton14} because of the radial range of the fit.
The uncertainty in the determination of photometric redshifts
is the largest source of systematic error for their mass estimates.
The CLASH project \citep{Umetsu16} selected background galaxies in the colour-colour
plane and combined information on the tangential shear profile, the magnification bias, and
the projected mass estimated by {\it HST} strong lensing  for
the mass measurements. They did not
employ a boost factor to compensate for contamination of their
background galaxy catalogues. The halo concentration for the NFW
model was treated as a free parameter. Their covariance error matrix is composed of
the shape noise, photo-$z$ error,  uncorrelated LSS lensing, and the
intrinsic scatter.
The WtG \citep{Applegate14} selected background galaxies in the 
colour-magnitude plane and corrected tangential shear profiles ($0.75-3h_{70}^{-1}\,{\rm Mpc}$) with a boost
factor profile using priors from the  X-ray gas distribution of
\citet{Mantz10}. They assumed a concentration parameter of
$c_{200}=4$. The uncertainty of mean source redshifts is
negligible in their analysis. They also selected background galaxies
using the full probability function of the photometric redshifts for a
subsample of clusters with five-band imaging.

\begin{table*}
 \caption{Summary of weak-lensing analysis methods of LoCuSS \citep{Okabe16b}, CCCP
 \citep{Hoekstra15},      CLASH \citep{Umetsu16} and  WtG \citep{Applegate14}.
  Column {\it Method} denotes either tangential
    shear fitting ($g_+$), or  joint fitting using tangential shear
    profiles, strong-lens and the magnification bias (SL, $g_+$ \& $\mu$).
    Column {\it Calibration factor} is the shear-calibration factor, with
    `Yes' indicating that such a factor was applied to the shear
    signal before fitting mass models, and `No' indicating
    otherwise.  Column {\it Boost factor} is the correction factor by the
    number density caused by imperfect background selection -- Yes/No
    indicates whether or not this factor was calculated and applied to
    the data. Column {\it Radial bins} gives the choice of radial binning
      scheme for the fitting of the shear profile. $c_\Delta$ states whether the concentration parameter
    was a free parameter in the fit, or fixed, or scaling with the
    mass. Noise denotes treatments of covariance matrix in the fitting (Eqn. \ref{eq:CovLSS}). } \label{tab:comp}
\begin{center}
{\small 
    \begin{tabular}{ccccccc}
      \hline
      \hline
 Name & {\small Method}
      & {\small Calibration}
      & {\small Boost}
      & {\small Radial}
      & $c_\Delta$
      & Noise\\
      &
      & {\small factor}
      & {\small factor}
      & {\small bins}
		     &
			  & \\
      \hline
     LoCuSS 
      & $g_+$
      & No/Yes
      & No
      & Adaptive
		     & Free
			& $C_g+C_s+C_{\rm LSS}$\\
     CCCP 
      & $g_+$
      & Yes
      & Yes
      & Fixed
		     & Scaling
			 & $C_g+C_s$\\
     CLASH 
      & SL, $g_+$ \& $\mu$ 
      & Yes 
      & No
      & Fixed
		     & Free
			 & $C_g+C_s+C_{\rm LSS}+C_{\rm int}$\\
     WtG
      & $g_+$
      & Yes
      & Yes
      & Fixed 
		     & Fixed 
			  & $C_g$\\
      \hline
    \end{tabular}
}
  \end{center}
\end{table*}

Comparison of the cluster mass measurements
between these different surveys is of paramount importance for cluster
cosmology experiments.
Mass comparisons can be expressed in terms of the geometric mean $\exp\left(\langle
\ln(Y/X)\rangle\right)$, or fitting with the lognomal quantities ($\ln
X$ and $\ln Y$), because the two quantities are interchangable. 
\citet{Hoekstra15}, \citet{Umetsu16}, and \citet{Okabe16b} found that
the latest weak-lensing
masses of CCCP, CLASH and LoCuSS are in excellent agreement, within $\sim5\%$, and that 
the WtG masses are somewhat larger than the others ($\sim10-15\%$).
A comparison of $M_{500}$ is shown in Figure \ref{fig:Mwl_com}. 
All the projects use only two- or three-band imaging, nevertheless
weak-lensing masses estimated from different methods show good agreement, which is
promising for further weak-lensing surveys for galaxy clusters.
\citet{Okabe16b} pointed out that the mass discrepancy between
the WtG and the others is caused by a shallow number density profile
for the boost factor. They found a number density excess in the
boost-factor profile even outside
$R_{200}$, which may incorrectly enhance lensing signals and
overestimate cluster masses.  

\begin{figure*}
 \includegraphics[width=0.425\textwidth]{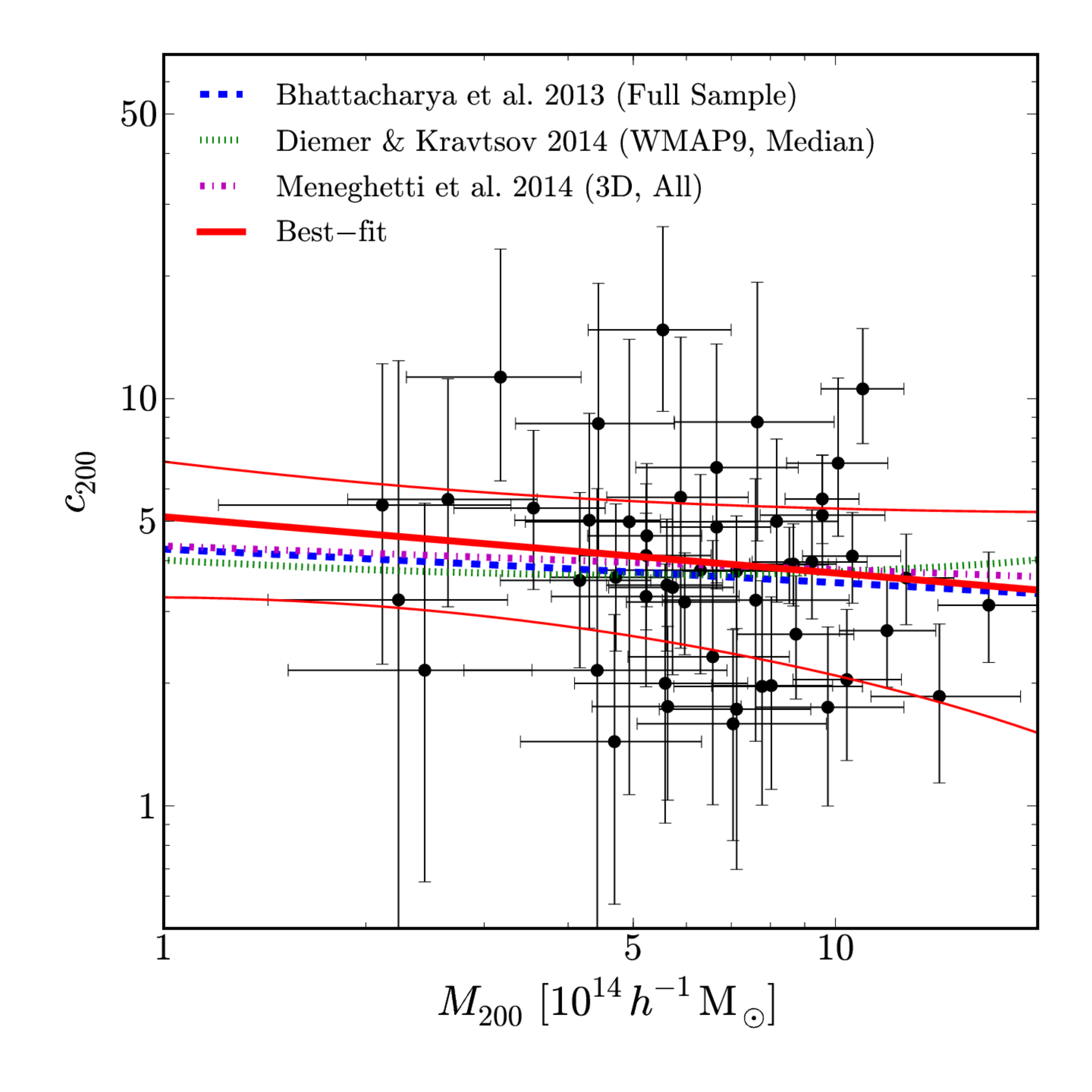}
 \hfill
 \includegraphics[width=0.5375\textwidth]{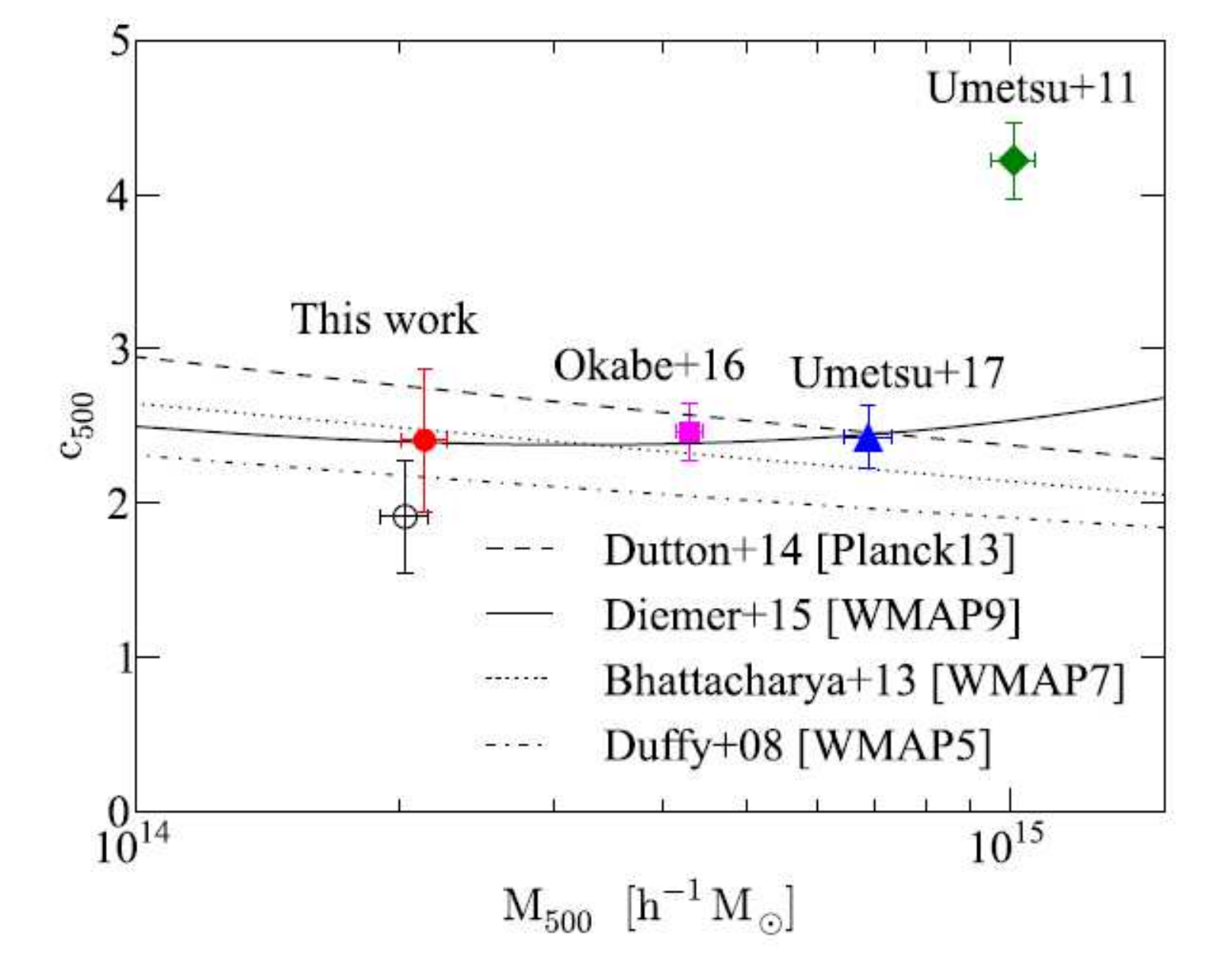}
\caption{({\it Left}): The observed distribution of the concentration parameters
    $c_{\rm 200}$ as a function of the cluster masses $M_{\rm 200}$ for
    50 clusters \citep{Okabe16b}. The errors denote 68\% confidence intervals. The thick
    and thin lines (red) are the best-fit function and the errors,
    respectively.  The dashed blue, dotted green and dotted-dashed
    magenta lines are the mean mass-concentration relation from recent
    numerical simulations of \citet{Bhattacharya13}, \citet{Diemer14}
 and \citet{Meneghetti14} at $z_l=0.23$, respectively.
 ({\it Right}): constraints on the massâconcentration relation for
 shear-selected clusters \citep{Miyazaki18}. The open and filled circles
 denote the halo concentration computed with and without the dilution
 effect, respectively.  The filled triangle shows the results for 16
X-ray-selected clusters at an average redshift of $0.34$ obtained from a
 strong and WL analysis of \citet{Umetsu17}.
The filled square is from \citet{Okabe16b}, estimated from 50 X-ray luminous
(LoCuSS) clusters at redshifts between 0.15 and 0.3,
 The filled diamond shows the results for a sample of four strong-lensing selected
superlens clusters at an average redshift of $0.32$ from a strong
and WL analysis of \citet{Umetsu11}. }
\label{fig:MC}       
\end{figure*}


\subsubsection{Mass and concentration}

\paragraph{\bf NFW models}
A weak-lensing study is a powerful direct way to constrain the
mass and concentration relation, because tangential shear profiles 
computed from the wide-field data easily cover the entire radial extent of galaxy clusters, in contrast to X-ray observations which typically
cover out to $\sim R_{500}$. The purity of background galaxies in shape
catalogues is the most important issue for studies of mass-concentration relation. 
\citet{Okabe13} have shown that the concentration parameter is
significantly underestimated by the inclusion of unlensed cluster member
galaxies in a shape catalogue. 
The contamination from member galaxies should be at the $1\%$
level, otherwise the concentration parameter is underestimated
\citep{Okabe10b}. 
The CLASH project \citep{Umetsu14,Merten15} have shown through a joint
shear and magnification study and a strong- and weak-lensing study that
the concentration for 20 X-ray clusters at $z\sim0.35$ is in a good agreement with a
recent prediction \citep{Meneghetti14}. 
\citet{Okabe16b} have found that the mass and concentration relation for 50 X-ray selected
clusters at $z\sim0.23$ is in good agreement with
those of three independent numerical simulations (the left panel Figure
\ref{fig:MC}). 
A fitting formula of the mass and concentration relation
should take account of the correlation between the errors on
concentration and mass by calculating the error covariance matrix.
The intrinsic scatter of halo concentration could be considered if necessary.

\citet{Cibirka17} have carried out a stacked lensing analysis for 27
richness selected galaxy clusters at $z\sim0.5$ and found a good
agreement with expectations for shape and evolution. 
\citet{Miyazaki18} have discovered 67 galaxy clusters through peak-finding
in weak-lensing mass maps reconstructed from the high number density of
background galaxies ($n_g\sim25$\,[arcmin$^{-2}$]) of the HSC-SSP survey
\citep{HSC1styr}. 
The clusters in the resulting catalogue are referred to as `shear-selected clusters', and represent one of the first applications to the HSC of this potentially powerful selection method which is complementary to X-ray, SZ and optical selection.
They have carried out a stacked lensing analysis and found that the halo
concentration for shear-selected clusters agrees well with those for
X-ray selected clusters. This indicates that shear-selected clusters
are less biased by halo orientation, 
in contrast to high concentration parameter for strong-lensing clusters
\citep[e.g.][]{Broadhurst05,Umetsu11}. 

Current observational studies probe the relation between mass and
concentration only in narrow redshift ranges. This is purely caused by 
dataset limitations. On-going and future optical surveys such as the DES \citep[][]{DES16}, HSC-SSP  \citep[][]{HSC1styr} and the Large Synoptic Survey Telescope \citep[LSST;][]{Ivezic08} will detect large samples of galaxy clusters in wide mass and redshift ranges, and will enable us to constrain the redshift evolution
of the mass and concentration. Moreover, sample selection biases will be investigated in detail.


\begin{figure*}
  \includegraphics[width=0.525\hsize]{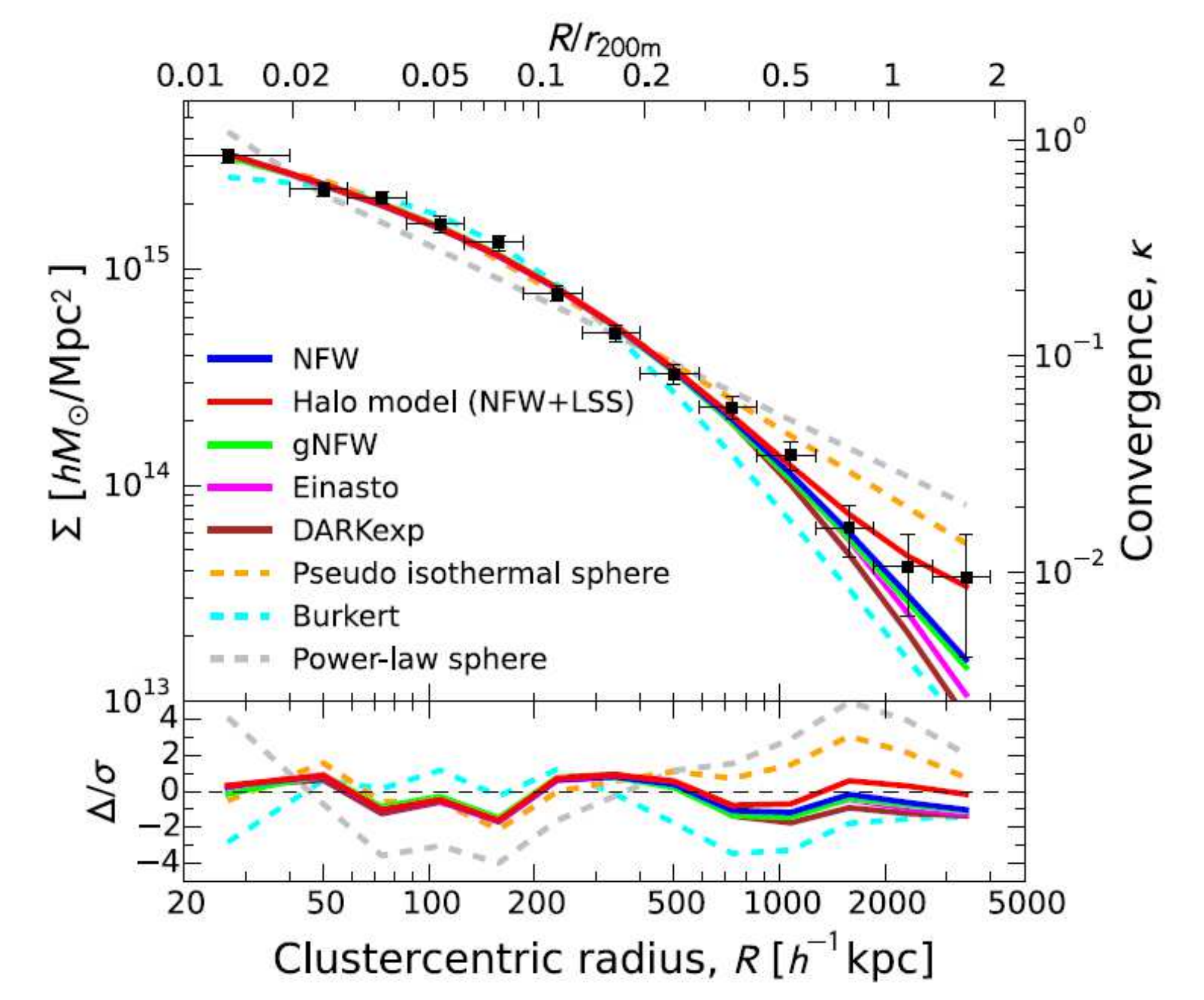}
\hfill
  \includegraphics[width=0.42\hsize]{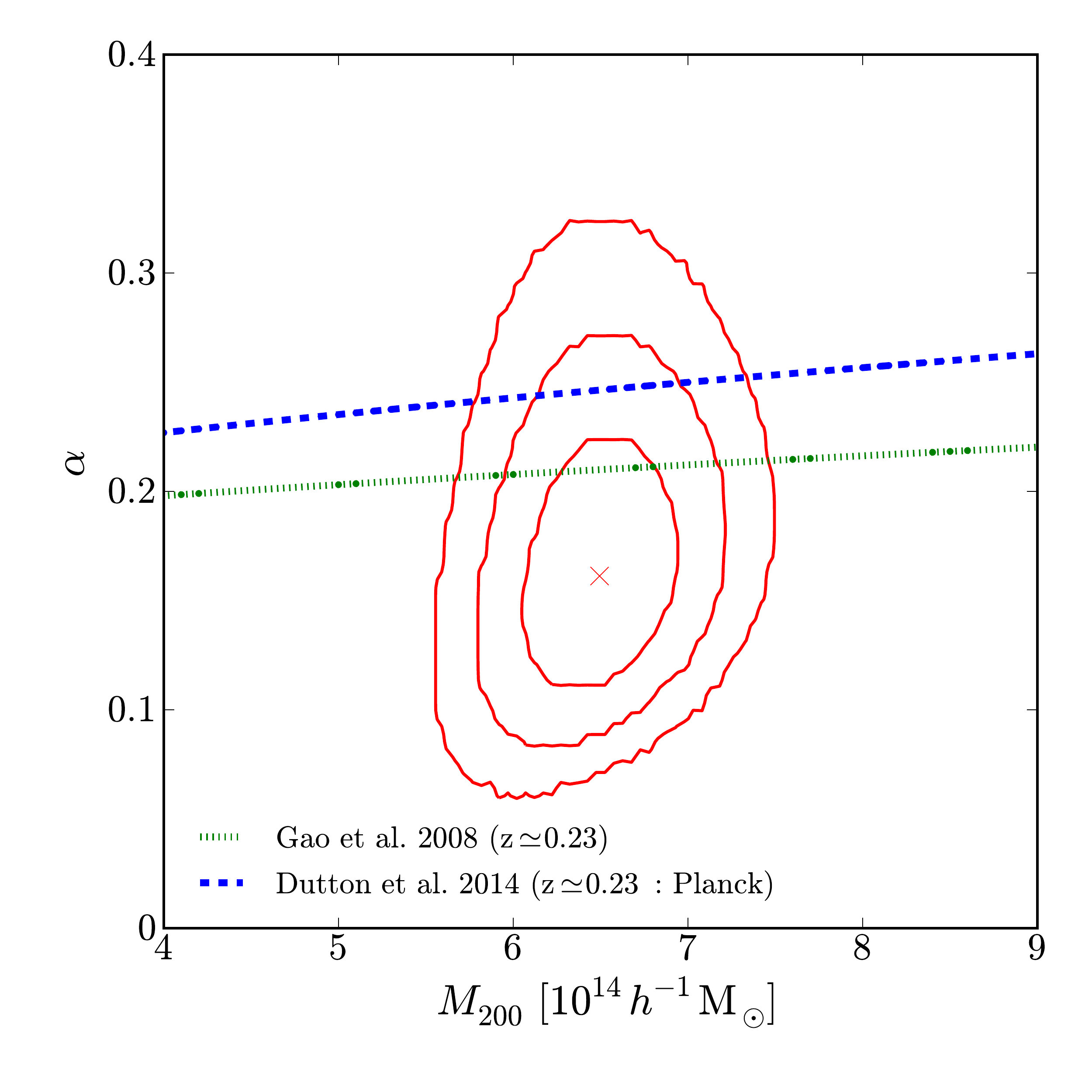}
 \caption{{\it Left} Comparison of models to the ensemble-averaged
 surface mass density $\Sigma(R)$ \citep{Umetsu16} (black squares)
 obtained for 16 X-ray-selected clusters.  Models with a probability, $p$, corresponding to $\chi^2$ is higher than 0.05
  are shown with solid lines, while those with $p<0.05$ are shown with
 dashed lines. The
 blue, red and magenta solid curves show the best-fit for NFW, NFW+two
 halo term and Einasto profiles, respectively.  
 The lower panel shows the deviations in units of $\sigma$ of the best-fit
 profiles with respect to the observed $\Sigma$ profile.
 {\it Right}:$\alpha$-mass relation \citep{Okabe16b}.
    The cross denotes the best-fit parameters and the contours show the 68.3\%, 95.4\%, and 99.7\% confidence
    levels. Blue dashed and green dotted lines are from \citet{Dutton14} and \citet{Gao08},
    respectively.  }
  \label{fig:Einasto}
\end{figure*}

\paragraph{\bf Einasto models}

As a next step for mass modelling, one aims to measure the shape parameter, $\alpha$, of the
\citet{Einasto65} profile (Eqn. \ref{eq:Einasto}) which describes the spherically averaged mass density profile for
simulated haloes better than the NFW profile \citep{Navarro04}.
Since it is very difficult to distinguish between the NFW and Einasto
profiles with the tangential profiles of individual clusters, one in general
adopts the NFW model for individual mass measurements. 
On the other hand, a stacked lensing profile
\citep[e.g.][]{Okabe10b,Okabe13,Umetsu14,Umetsu16,Umetsu17,Okabe16b,Cibirka17,Miyazaki18}
is a powerful route to constrain the average mass density profile. 
First, the average distortion or projected mass profiles are less
sensitive to internal substructures and the asphericity of the individual
cluster mass distributions and also to uncorrelated large-scale
structure along the same line-of-sight. 
This is because these structures are averaged out via the
stacking, under the assumption that the universe is statistically
homogeneous and isotropic. 
Second, stacking procedures improve the signal-to-noise ratio of lensing profiles.
Since the lensing signals at larger radii could be detected, one usually
adopts a main halo and the two halo term.

\citet{Umetsu16} have computed the stacked $\Sigma$ profile for 16 X-ray
selected clusters and constrain $\alpha=0.232_{-0.038}^{+0.042}$ (left
panel of Figure \ref{fig:Einasto}).
\citet{Okabe16b} have compared the relation between the shape parameter
and mass for 50 X-ray selected clusters with predictions of numerical simulations
\citep{Dutton14,Gao08}, and found that they are in agreement with each
other (right panel of Figure \ref{fig:Einasto}).  More precise
observational constraints on the density profile shape of clusters,
including on the mass dependence of the Einasto profile parameters, awaits
larger cluster samples from on-going or future surveys.


\subsection{X-rays and hybrid SZ}

\begin{figure*}
\includegraphics[bb=260 0 530 207, clip,scale=1.,width=0.55\textwidth, keepaspectratio]{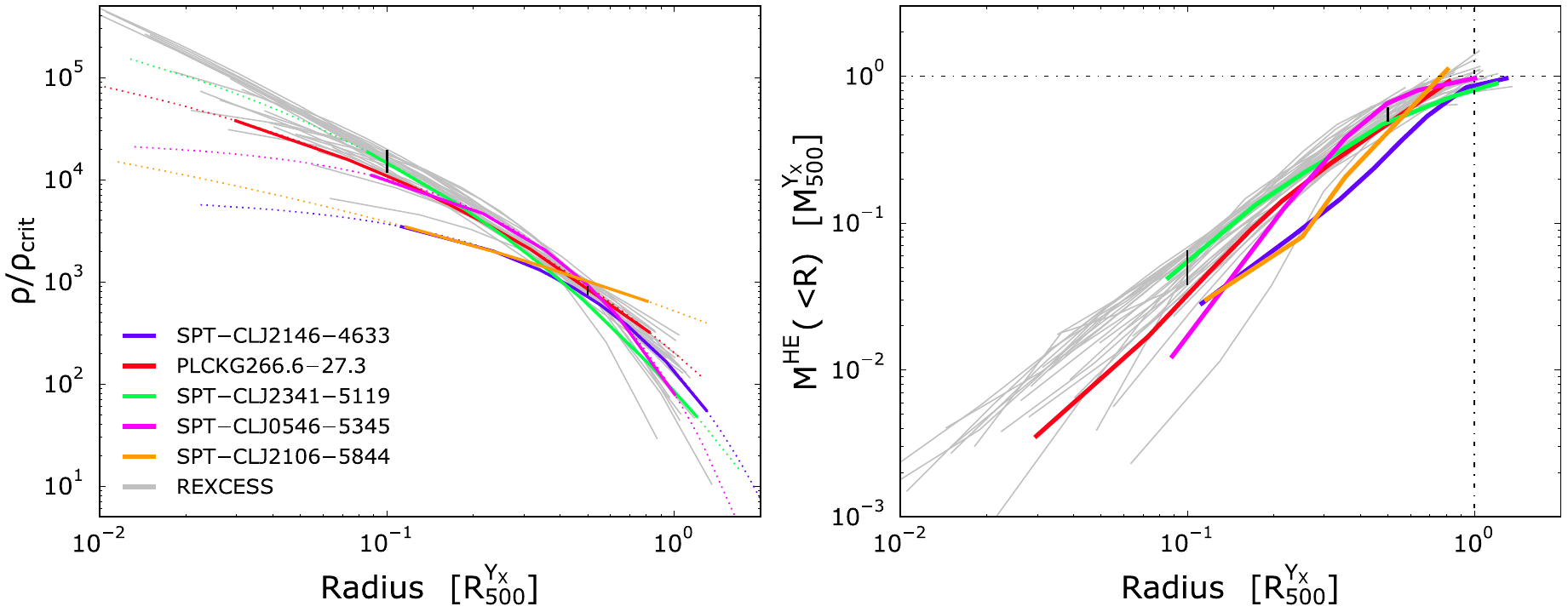}
\hfill
\includegraphics[width=0.445\textwidth, keepaspectratio]{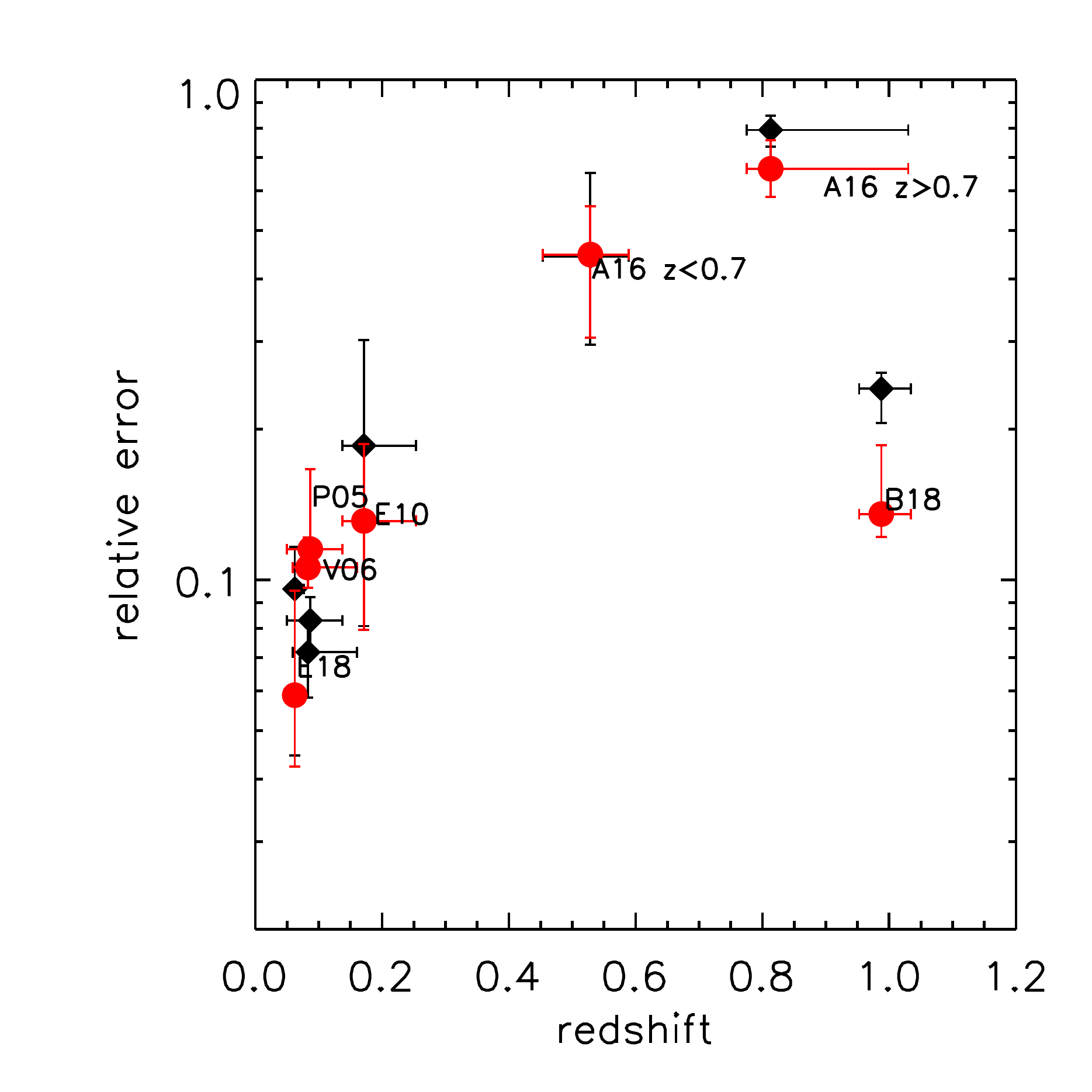}
\caption{{\it Left:} Mass profiles for a sample of five clusters at $z\sim1$ derived by \citet{bartalucci18} using \xmm\ and \chandra\ data (coloured lines), compared to local REXCESS mass profiles (grey lines).  {\it Right:} Relative errors (median, and 1st and 3rd quartiles) on the mass (red dots) and NFW concentration (black diamonds) estimated in the following studies: P05 \citep{pointecouteau05}, V06 \citep{vik06}, E10 \citep{ettori10}, A16 \citep{amodeo16}, E18 \citep{ettori18}, B18 \citep{bartalucci18}. 
} \label{fig:Mz}
\end{figure*}

\subsubsection{Hydrostatic mass  and mass profiles }

The launch of \xmm\ and {\it Chandra} opened the way to precise spatially resolved X-ray spectroscopy, enabling measurement of both the gas density and the temperature profiles, and thus the total mass profile using the HE equation. \citet{pointecouteau05} and \citet{vik06}, using  respectively \xmm\ and {\it Chandra}, measured high precision mass profiles for small samples ($\sim 10$) of local ($z<0.15$) relaxed clusters ($\Mv>10^{14} \Msol$). \citet{buo07} extended this work into the group regime \citep[see also][]{gas07},  while \citet{ettori10} studied a larger sample, albeit with lower precision, using the \xmm\ archive (44 clusters at $z<0.3$). 

The consistent picture that emerges from these observations is that the dark matter profile is indeed cuspy. Fits with parametric models usually reject profiles with a finite core or are inconclusive \citep[see also][]{buo04,voi06}. Generally, self-similarity of shape is also evident from all techniques, although there is no quantitative assessment of the intrinsic scatter. All quantitative tests of $\Lambda$CDM predictions are based on parametric profile fitting with the NFW profile. X-ray determinations of the $c-M$ relation are consistent with theoretical predictions, and have even been used to provide independent constraints on $\Omega_{m}$ and $\sigma_8$ \citep{buo07,ettori10}. When constraints can be put on more general profiles, such as the generalised NFW or Einasto profiles \citep[e.g.][]{voi06,man16}, the central logarithmic slope has been found to be consistent with unity, i.e. with an NFW profile. 

More recent studies have pushed the measurements for relaxed clusters to higher redshifts, e.g. the studies of \citet[][34 relaxed clusters at  $0.06<z<0.7$]{sch07}, \citet[][40 relaxed objects at $0.1 < z < 1.1$]{man16}, or \citet[][an archival sample at $0.4<z<1.2$]{amodeo16}, but the individual mass profiles in these studies generally have large uncertainties,  particularly at the highest redshifts.
The evolution factor of the corresponding $c-M$ relations, expressed as $(1+z)^\alpha$, is consistent with theoretical expectations, but with large uncertainties ($\alpha = 0.71\pm0.52$, and $\alpha= 0.12 \pm 0.61$, respectively). 
 \citet[][see Fig.~\ref{fig:Mz}]{bartalucci18} have recently reconstructed the hydrostatic mass profiles of the five most massive ($M_{500} > 5 \times 10^{14} M_{\odot}$) SZ-selected clusters at high redshift ($z\sim1$), combining deep observations  from \xmm\ and {\it Chandra}.  Using both forward and backward methods, they investigated halo shape parameters such as sparsity and concentration, measured to high accuracy. 
 Comparing to local clusters, they found  hints for evolution in the central regions (or for selection effects). The total baryonic content is in agreement with the cosmic value at $R_{500}$. Comparison with numerical simulations shows that the mass distribution and concentration are in line with expectations. \citet{bartalucci18} also investigated the sparsity of their sample, finding good agreement with expectations \citep[see also][]{corasaniti18}. Typical uncertainties on the NFW concentration as a function of redshift are illustrated in the right hand panel of  Fig.~\ref{fig:Mz}.


\subsubsection{New results from combination with SZ}

As detailed in Sect.~\ref{sec:xraymethod}, a recent observational development is the ready availability of spatially-resolved SZ electron pressure profiles, which  can be obtained from geometrical deprojection of the azimuthally-averaged integrated Comptonization parameter. The power of the SZ effect is that it directly measures the line-of-sight pressure. However, measurement of other key thermodynamic quantities such as temperature and entropy requires access to the gas density. This is trivial to obtain from X-ray imaging. Previous studies \citep[e.g.][]{bas10,planck13} have been limited to a few massive local systems due to the intrinsic faintness of the SZ signal and the 1-2 orders of magnitude difference in angular resolution between X-ray and SZ observations. 

\begin{figure*}
\includegraphics[width=0.435\textwidth, keepaspectratio]{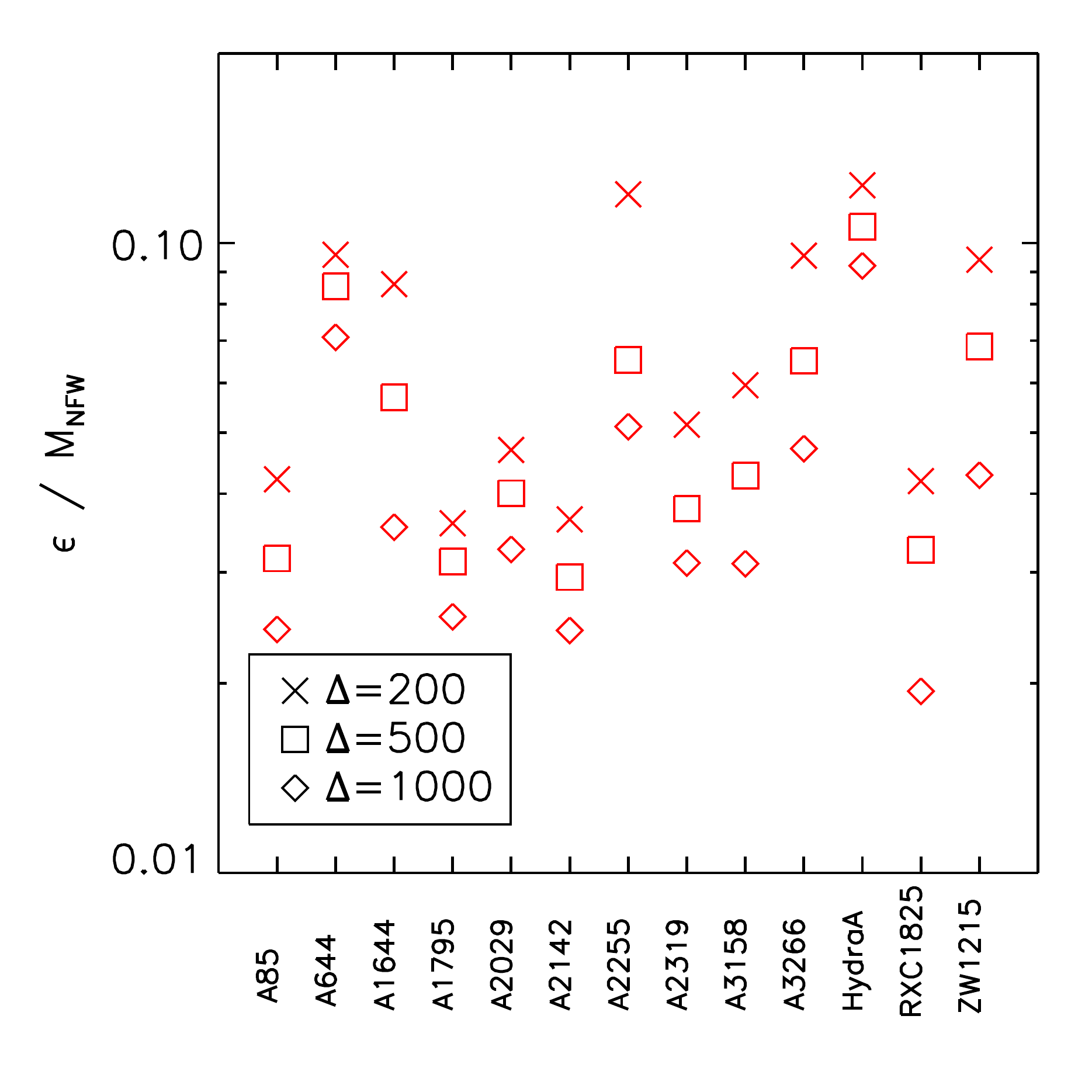}
\hfill
\includegraphics[bb=25 0 525 395 ,clip,scale=1.,width=0.53\textwidth, keepaspectratio]{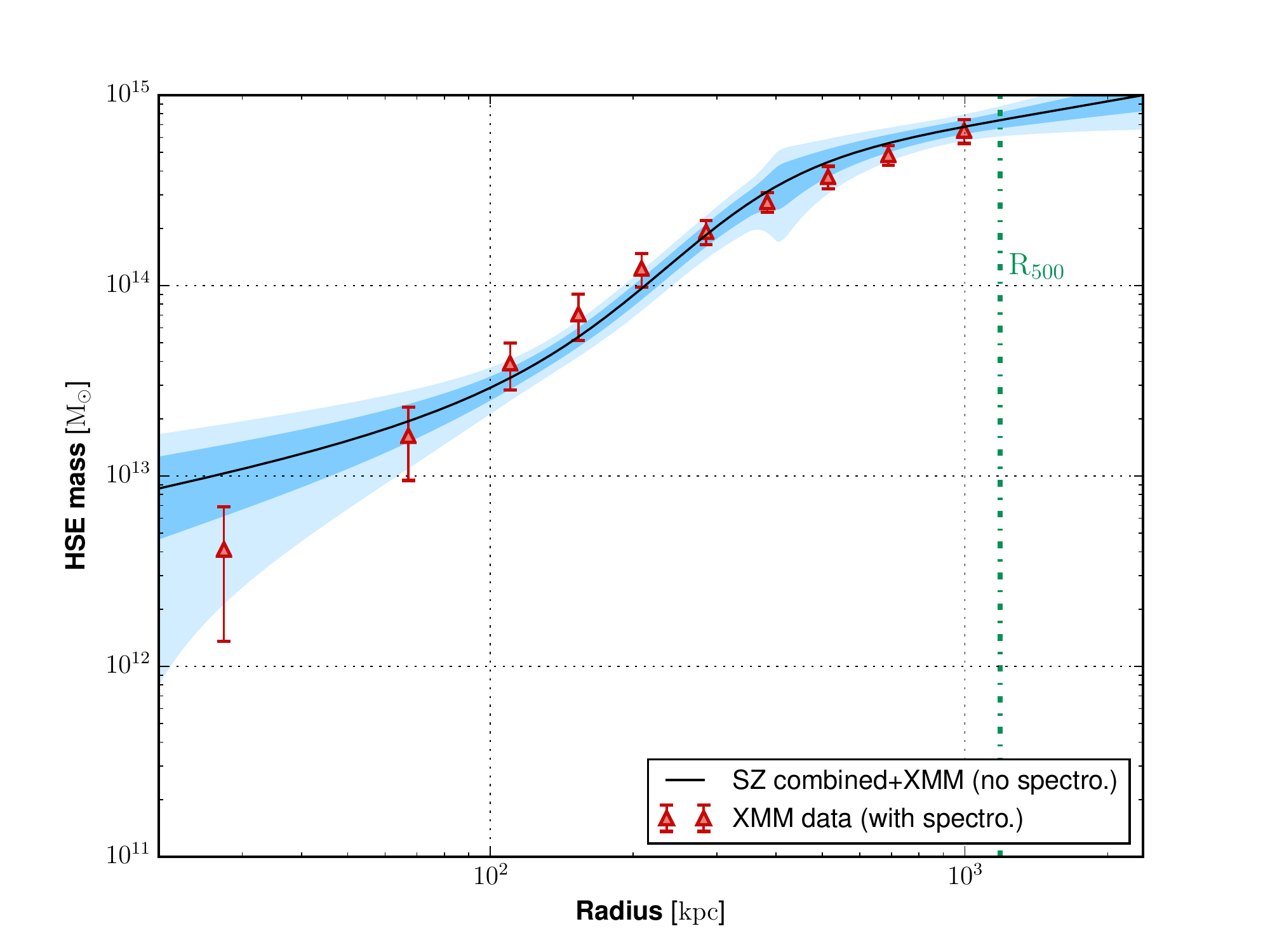}
\caption{{\it Left:} Relative statistical errors on the hydrostatic masses measured at $R_{200}$ in the X-COP sample from \citet{ettori18}. {\it Right:} HE mass profile of PSZ2\,G$144.83+25.11$ at redshift $z = 0.58$, derived from \xmm\ density and temperature profiles (triangle), compared to the mass profile derived from  the \xmm\ density combined with the NIKA2 pressure profile (dark blue region). Figure from \citet{rup18}.
} \label{fig:MSZ}
\end{figure*}

For local systems ($z<0.2$),  in the most recent works published on the hydrostatic mass, 
improved data quality and refined analysis techniques have allowed to reach statistical uncertainties on the reconstructed mass 
of about 10\% at $R_{200}$ (see Fig.~\ref{fig:MSZ}).
In nearby ($z<0.1$) massive systems, the X-COP collaboration \citep[e.g.][]{ghi18b, ettori18, eckert18}
has been able to reconstruct hydrostatic mass profiles out to $2 R_{500}$ by combining X-ray and SZ data. They find that  (i) the NFW mass model provides, on average, the best-fitting mass model in reproducing the observed radial profiles
of relaxed massive nearby systems, with relative errors at $R_{200}$ lower than 10\% (see Fig.~\ref{fig:MSZ}), 
(ii) alternative models of gravity that do not require any dark matter contribution
(such as MOND or Emergent Gravity) show significant tensions when compared with the prediction from the HE equation,
(iii) estimates of the dark matter distribution obtained for the same objects with different techniques (as e.g. lensing, galaxy dynamics, scaling laws) are consistent with the hydrostatic mass with differences in the order of 15\%.

At higher redshifts ($z>0.5$), the new sensitive, high resolution SZ instruments such as NIKA2 and Mustang/Mustang2, are potentially game-changers. For example, the angular resolution of NIKA2 is comparable to that of \xmm, over a 6.5' diameter field of view, finally opening the way to effective exploitation of the X-ray/SZ synergy. As an example, \citet{rup18} recently published a novel non-parametric X-ray/SZ analysis of the cluster PSZ2 G144.83 +25.11 at $z = 0.59$. The 150 GHz image at $< 18^{\prime \prime}$ resolution showed a clear extension to the SW that may be a merging subclump. Excluding this region, the radial profiles resulting from the combination of the density from \xmm\ and the SZ pressure from NIKA2 and \planck\ were in excellent agreement with those obtained from the X-ray data alone  (see Fig.~\ref{fig:MSZ}). The resulting hydrostatic mass profile provides competitive constraints to the X-ray only analysis. 

\begin{figure*}
\includegraphics[width=0.49\textwidth]{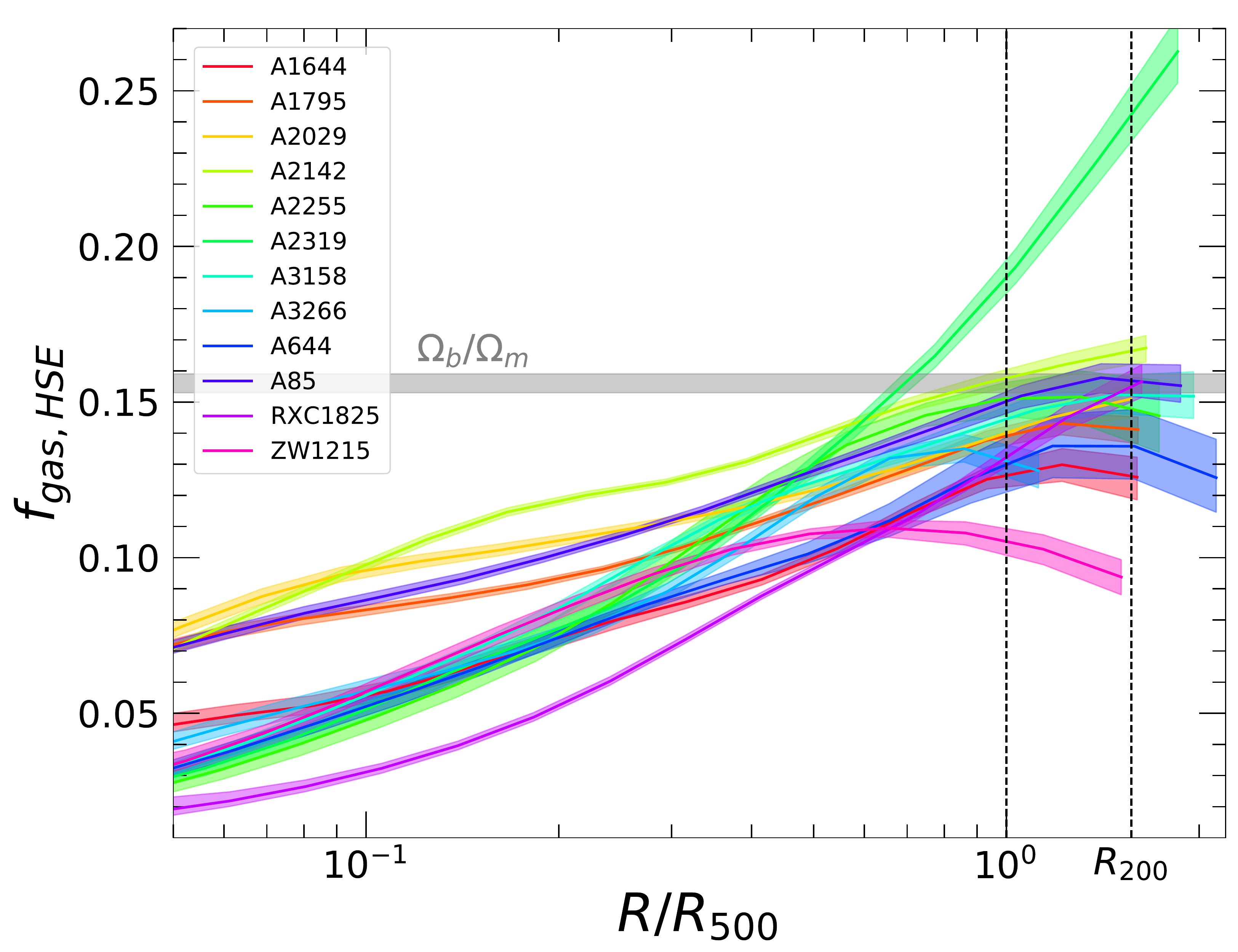}
\hfill
\includegraphics[width=0.49\textwidth]{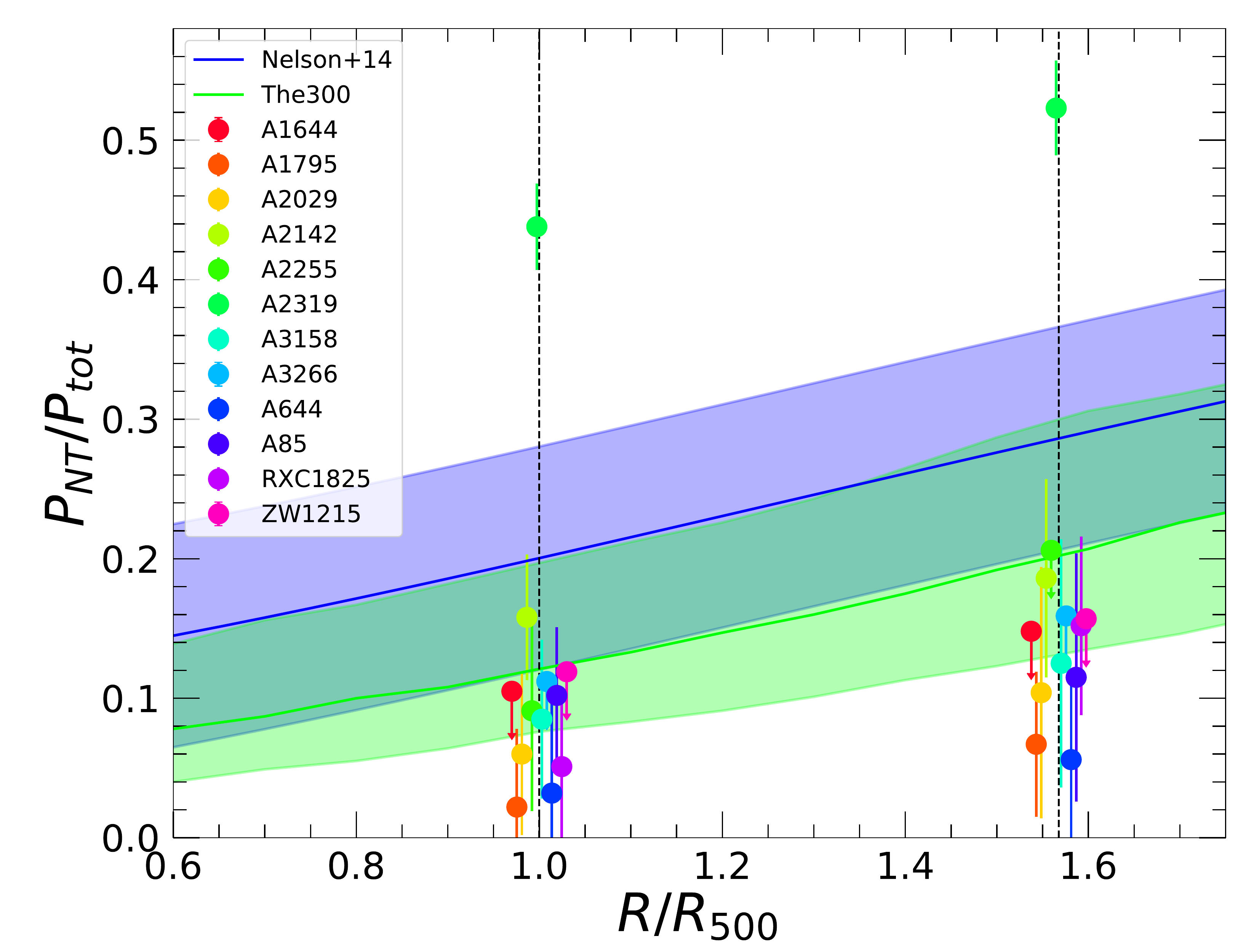}
\caption{\emph{Left:} Hydrostatic gas fraction profiles $f_{\rm gas,HE}(R)=M_{\rm gas}(<R)/M_{\rm HE}(<rR$ for 12 clusters in the X-COP sample \citep[reproduced from][]{eckert18}. The dashed and dash-dotted vertical lines represent the position of $R_{500}$ and $R_{200}$, respectively. The horizontal shaded area show the cosmic baryon fraction from \emph{Planck} CMB \citep{P15XIII}. \emph{Right:} Non-thermal pressure fraction at $R_{500}$ and $R_{200}$ inferred by comparing the measured hydrostatic gas fraction of X-COP clusters with the expectations, taking into account baryon depletion and stellar fraction. The blue and green lines and shaded areas are the predictions for the random-to-thermal pressure fraction from two sets of numerical simulations \citep{Nelson14,Cui18}.}
\label{fig:xcop_fgas}
\end{figure*} 


\subsubsection{Baryon budget and gas fraction}

As described in Sect. \ref{sec:depletion}, galaxy clusters are expected to be fair archives of the baryon budget in the Universe. \emph{Planck} data constrain the cosmic baryon fraction with a statistical precision of just 2\% \citep[$f_{b}=0.156\pm0.003$,][]{P15XIII}. Thus, the baryon fraction of massive clusters within their virial radius is in principle known with a high level of accuracy, and measurements are highly sensitive to the accuracy of the estimated mass. The ICM contains the vast majority of the baryons, with stars within galaxies and intracluster light typically contributing $1-1.5\%$ of the total mass within $R_{200}$ \citep[e.g.][]{gonzalez07,gonzalez13,Leauthaud12,lagana13,coupon15,eckert16,chiu17}. 

Measurements of the ICM gas fraction using hydrostatic mass estimates typically infer gas fractions of 10-15\% within $R_{500}$ \citep{Vikhlinin06,Allen08,ettori09,Pratt10,mantz14}, in good agreement with the expected cosmic baryon budget. However, recent observations extending measurements out to $R_{200}$ and beyond have reported excesses in the gas fraction over the cosmic value when using hydrostatic masses \citep{Simionescu11,Kawaharada10,Ichikawa13,ghi18a}. Such differences disappear when computing gas fractions with weak-lensing masses \citep{Okabe14a}. On the other hand, several studies have also reported hydrostatic gas fractions consistent with expectations all the way out to the virial radius \citep{tchernin16,Walker12a,Walker13}. 

Recently, \citet{eckert18} reported ICM gas fractions estimated using the HE assumption for the X-COP sample, a sample of 12 clusters with deep \xmm\ X-ray and \emph{Planck} SZ data out to $R_{200}$ and beyond. In the left-hand panel of Fig. \ref{fig:xcop_fgas} we show the gas fraction profiles estimated through a joint fit to \xmm\ and \emph{Planck} SZ data. With the exception of one system, A2319, for which substantial non-thermal pressure support was detected  \citep[][]{ghi18a}, all measurements converge towards a gas fraction at the virial radius that is very close to the expected baryon budget.


\subsection{Non-thermal pressure, feedback, and the validity of the hydrostatic assumption}

\subsubsection{Constraints from X-ray and SZ observations}
\label{sec:fgasxcop}

The gas fraction can be used to put constraints on the non-thermal pressure support if it is assumed that the deviations from the expected true (i.e. Universal) value originate from random isotropic gas motions (see Eqn. \ref{eq:mhse_tot}). The ratio of hydrostatic to true gas fraction is related to the non-thermal pressure fraction $\alpha=P_{\rm rand}/P_{\rm tot}$ as \citep[see Sect. 3.2 of][]{eckert18},
\begin{equation}
\frac{f_{\rm gas,HE}}{f_{\rm gas,true}}=\left(1-\frac{P_{\rm th}\,R^2}{(1-\alpha)\, \rho_{\rm gas} GM_{\rm HE}}\frac{d\alpha}{dR}\right)(1-\alpha)^{-1}.
\end{equation}
Taking into account the expected depletion of baryons induced by hydrodynamical processes and the stellar fraction, constraints on the amount of pressure in the form of random gas motions can be obtained. 

In the right-hand panel of Fig. \ref{fig:xcop_fgas} we show the inferred non-thermal pressure fraction for the 12 X-COP clusters, which is then compared with the predictions from two different sets of numerical simulations ($\Omega_{500}$, \citealt{Nelson14}; The300, \citealt{Cui18}). The median non-thermal pressure fraction is 6\% at $R_{500}$ and 10\% at $R_{200}$, which can be translated into a typical Mach number $\mathcal{M}_{3D}=\sigma_{v}/c_s\approx0.33$ at $R_{500}$. 

While the ICM gas fraction is very sensitive to the hydrostatic mass bias, one may argue whether the assumption that the true baryon fraction should match the cosmic baryon fraction with small ($\sim5\%$) corrections can be violated. This can occur if a large amount of non-gravitational energy is injected within the ICM, in particular by AGN feedback (see Sect. \ref{sec:depletion}). Given the measured gas fractions for the X-COP sample (see Fig. \ref{fig:xcop_fgas}), a large hydrostatic bias ($>20\%$) would imply that a substantial amount of baryons have been driven outside of the virial radius even for the most massive local clusters. These systems contain a total thermal energy of several $10^{63}$ ergs, implying that feedback energies in excess of $10^{62}$ ergs are required to substantially deplete the overall baryon fraction. Such an energy input corresponds to an overall AGN luminosity of $\sim10^{45}$ ergs/s injected continuously over 10 Gyr, assuming 100\% coupling with the ICM and neglecting cooling losses. 

The recent high spectral resolution results from {\it Hitomi} have provided an unprecedented view of gas motions in the Perseus cluster \citep{hitomi16,hitomi18}.  Although the purpose of these observations was to obtain constraints on the interaction between the central AGN and the surrounding ICM, these unique data have given insight into the level of turbulence close to the core of Perseus. They have shown that even in the presence of the AGN the turbulent line broadening is rather modest ($164 \pm 10$ km s$^{-1}$). Better constraints will be obtained from XRISM and Athena (Sect.~\ref{forward:TNG}).

The above illustrates that the constraints on departures from HE and the gas depletion due to feedback are linked on a fundamental level, and can be used more as a consistency check than as an absolute constraint.  For example, an extreme HE bias of $\sim 60\%$, as would be suggested from the tension between {\it Planck} SZ cluster counts and CMB, would imply a level of gas depletion that is completely at odds with reasonable feedback prescriptions in cosmological simulations. The two issues should thus be addressed self-consistently in both observations and simulations.


\subsubsection{Constraints from X-ray and optical observations}
\label{sec:xrayWL}

A comparison of HE and WL masses for a large number of clusters is a
useful route to test the validity of the HE assumption.
The mass bias, $b_{\rm WL}$, relative to the WL mass, can be estimated by the geometric mean for targeted clusters, 
\begin{eqnarray}
 1-b_{\rm WL}= \exp \left[ \left(\sum_i \ln \left(\frac{M_{\rm HE}}{M_{\rm WL}}\right)_i
           w_i\right)\left(\sum w_i\right)^{-1} \right], 
\end{eqnarray}
where $w_i$ is a weight function ($w_i=1$ for a uniform weight), or
equivalently by fitting the lognormal quantities. 

\citet{Mahdavi13} compared X-ray masses with weak-lensing masses for 50
CCCP clusters and found that the average mass ratio of X-ray to WL
masses is $1-b_{\rm WL}=0.88\pm0.05$ at $R_{500}$. \citet{Hoekstra15} subsequently updated the CCCP WL masses and reported masses on average $19\%$ higher. Thus,
 applying a factor 1.19 correction to the denominator of the CCCP
 implies an average mass ratio $1-b_{\rm WL}\sim 0.74$.

 \citet{Smith16} have complied 50 LoCuSS clusters at $0.15 < z < 0.3$ and
found the mean ratio of X-ray to lensing mass $1-b_{\rm WL}=0.95\pm0.05$, where X-ray
masses \citep{Martino14} used spectroscopic-like temperature
\citep{Mazzotta04} and WL masses are from Subaru/Suprime-Cam \citep{Okabe16b}.
We note that the \citet{Martino14} X-ray masses are on average $\sim14\%$
higher than those of \citet{Mahdavi13}.

\citet{Applegate16} have investigated a lensing to X-ray mass ratio for
12 relaxed clusters from the WtG project, using WL masses
\citep{Applegate14} and {\it Chandra} masses.
They reported $b_{\rm WL}-1=0.967^{+0.063}_{-0.092}$ and
$1.059^{+0.092}_{-0.096}$ at $R_{2500}$ and $R_{500}$, respectively.

\citet{Siegel18} carried out a joint analysis of \chandra\ X-ray
observations, Bolocam thermal SZ observations, HST strong-lensing data,
and Subaru/Suprime-Cam weak-lensing data for 6 CLASH clusters, and
constrained the nonthermal pressure fraction at $R_{500}$ to be
$<0.11$ at 95\% confidence. 

A recent analysis of a sample of  16 massive clusters by \citet{mau16}  suggested that the mass profiles obtained independently from X-ray hydrostatic and caustic  (Sect.~\ref{sss:kinmet}) methods agree to better than 20\% on average across the radial range probed by the observations. At $R_{500}$, they measure a mass ratio $M_{\rm X} / M_{\rm C} \gtrsim 0.9$, implying a low or zero value of the hydrostatic bias if the caustic masses are assumed to be equivalent to the true mass. Interestingly, \citet{mau16} found no dependence of the $M_{\rm X} / M_{\rm C}$ scatter on dynamical state.

\begin{figure*}
\begin{center}
 \includegraphics[angle=90,width=0.75\textwidth]{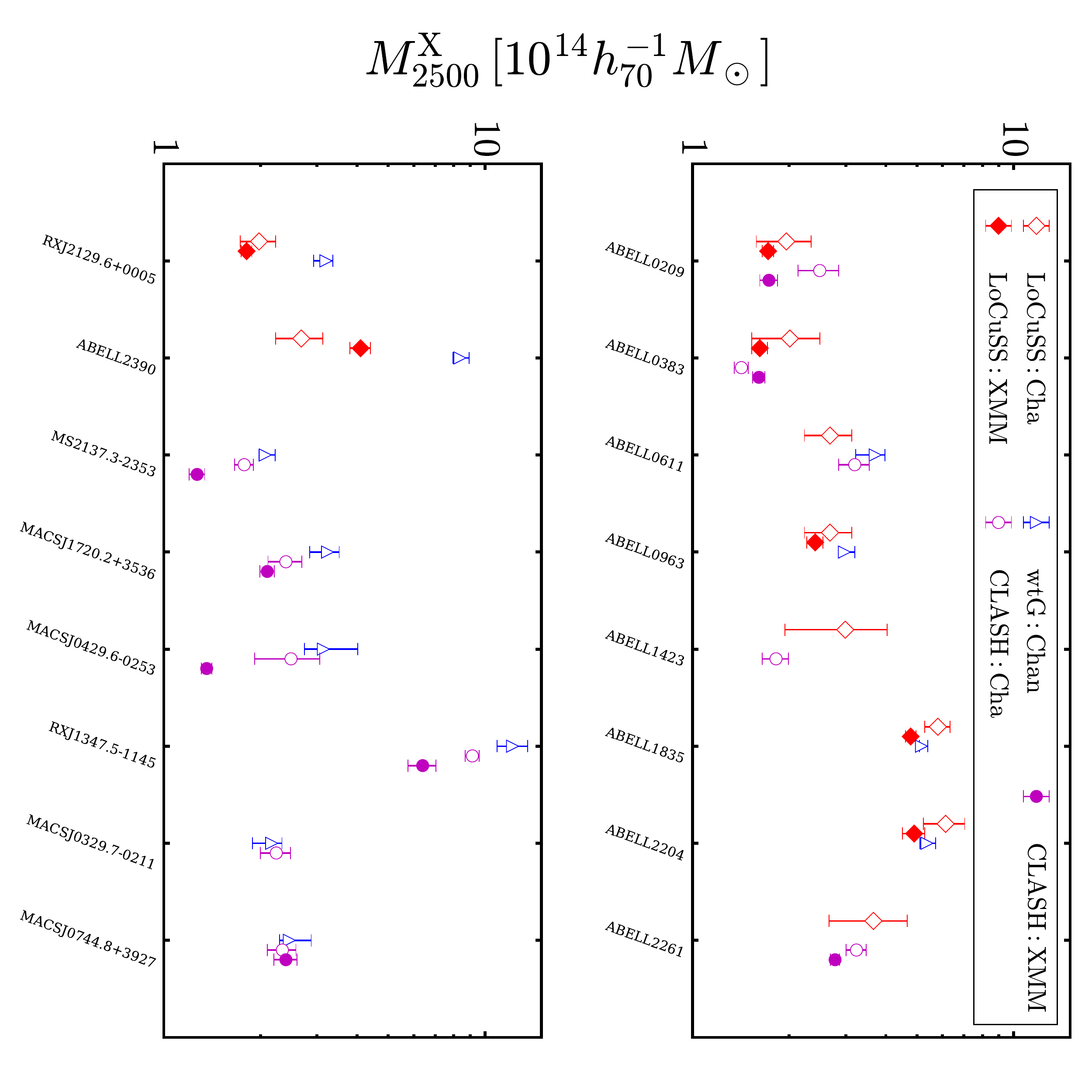}
\caption{X-ray mass comparisons of LoCuSS (diamonds), WtG (triangles) and CLASH (circles) at $\Delta=2500$. Open and solid symbols are from {\it Chandra} and \xmm\ observations.}
\label{fig:Mx_com}       
\end{center}
\end{figure*}

To further illustrate the above, we compare in Fig.~\ref{fig:Mx_com} the X-ray mass measurements from LoCuSS \citep{Martino14}, WtG \citep{Applegate16} and CLASH \citep{Siegel18} at $\Delta=2500$. 
Since the CCCP X-ray masses \citep{Mahdavi13} are measured within radii determined by the WL mass measurement, we do not include them in the comparison. The {\it Chandra} and \xmm\ results are denoted by open and solid symbols, respectively. Scatter between each X-ray measurement is significantly larger than that of WL mass measurements (Fig. \ref{fig:Mwl_com}). \citet{Sereno2015b}, compiling WL and X-ray masses from the literature, have found that the intrinsic scatter of HE masses ($\sim20-30$ per cent) is larger than that of WL masses ($\sim 10-15$ per cent). The {\it Chandra} masses are generally systematically higher than those from \xmm\ due to the absolute temperature calibration issues described above. The WtG masses are also systematically higher than those of other projects. 

\subsection{Halo triaxiality}

Halo asphericity and orientation induces significant scatter in
 the projected lensing signals. 
Simultaneous modelling of the mass, concentration, shape, and
 orientation using lensing data and/or independent data has been proposed by
 various papers
 \citep[e.g.][]{Oguri05,DeFilippis05,Sereno07,Corless09,Sereno11,Sereno13b,Umetsu15}.
 The previous review by \citet{Limousin13} gives a good summary of the technique. 
 Lensing information probes the structure and morphology of
 the matter distribution in projection. X-ray
observations provide us with the characteristic size and orientation of
the ICM in the sky plane. The elongation
of the ICM along the line-of-sight can be constrained from
the combination of X-ray and thermal SZ observations, because of a
difference of emissivity. Therefore, the triaxial model can be
constrained by combining these complementary data.  
Recently, \citet{Sereno13b} have developed a parametric triaxial
framework to combine and couple independent morphological constraints from lensing
and X-ray/SZ data, using minimal geometric assumptions.
\citet{Umetsu15} applied the technique to A1689 and found that the mass
distribution is elongated with an axis ratio of $\sim 0.7$ in projection
and the thermal gas pressure contributes to $\sim 60\%$ of the total
pressure balanced with the mass.
\citet{Chiu18} have carried out a three-dimensional triaxial analysis
\citep[see also][]{Umetsu18} for 20 CLASH clusters and obtained a joint ensemble constraint on the
minor-to-major axis ratio $q=0.652_{-0.078}^{+0.162}$.
Assuming priors on the axis ratios derived from numerical simulations,
they found that the degree of triaxiality for the full samples prefers
a prolate geometry for cluster haloes.
\citet{Sereno18} have measured based on a full 3D analysis of lensing, X-ray and SZ measurements 
shapes of X-ray selected CLASH clusters and found that it is in a good agreement with numerical simulations for dark matter only.

\begin{figure*}
 \includegraphics[width=0.48\textwidth]{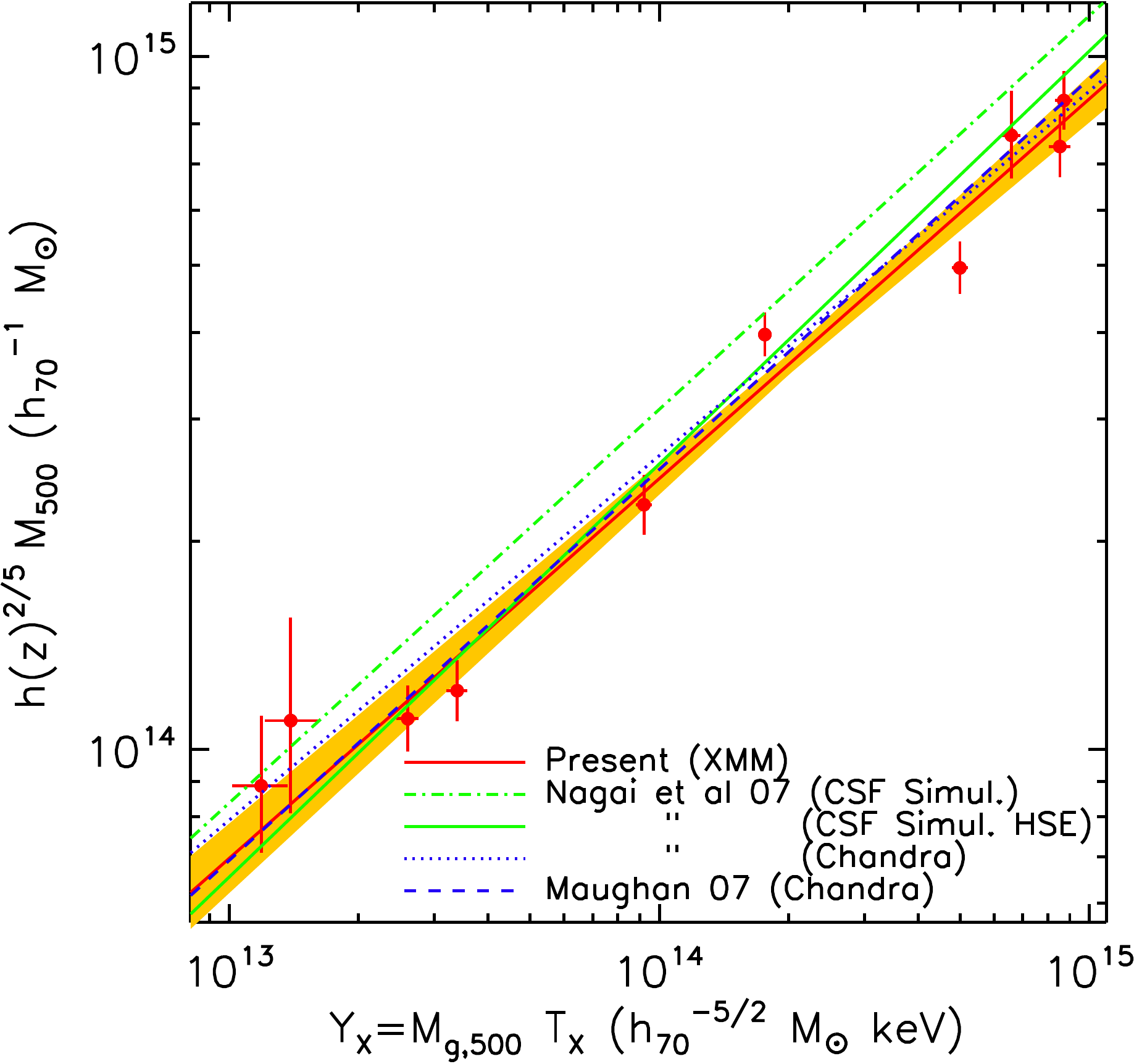}
 \hfill
 \includegraphics[width=0.48\textwidth]{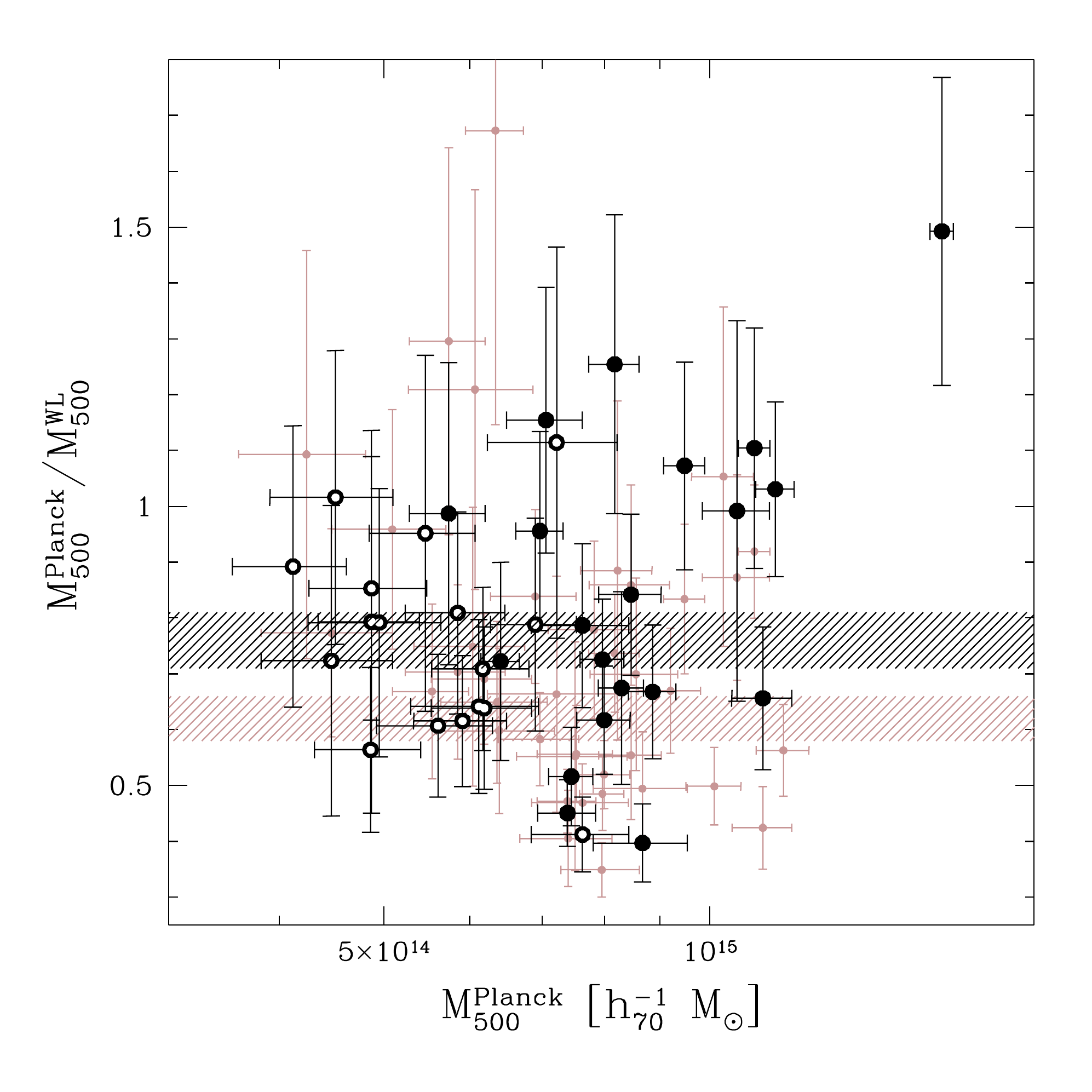}
\caption{({\it Left}): The observed $M_{500}$--$\YX$ relation from \xmm\ observations of ten relaxed clusters \citep[][red points with $1\sigma$ uncertainty envelope]{arn07} compared to the predicted relation from numerical simulations including cooling and galaxy feedback \citep[green dot-dash line, true mass; solid green line, HE mass, from][]{Nagai07}. The observed relations from \chandra\ are also shown \citep{Nagai07,mau07}.
 ({\it Right}): Ratio of the hydrostatic and the weak lensing mass estimates as a function of mass, from \citet{Hoekstra15}. The CCCP sample yields an average value of $0.76 \pm 0.05$ (dark hatched region), while the average for the WtG measurements is $0.62 \pm 0.04$ (pink region). }
\label{fig:proxycomp}       
\end{figure*}

\subsection{Mass estimates from mass proxies}

The mass estimate is  essential both when using clusters to constrain cosmology, and for studying structure formation physics. Mass proxies play an important role in this context. They can provide statistically more precise mass measurements, especially at high redshift and low mass, as compared to WL or hydrostatic masses, and in some cases the masses from mass proxies may even be less biased (e.g. for highly disturbed systems far from HE). 

We recall also that in cluster surveys, objects are never detected directly by the mass, but through their observable baryon signature (i.e. through a mass proxy).  The calibration of the corresponding mass proxy scaling relations is always needed to understand the selection function (i.e. the probability that a cluster of a given mass is detected via its given baryon signal). This step  is necessary even if individual masses were available for all objects in subsequent follow up, in order to relate the theoretical mass function to the observed number counts. 

There is a vast amount of literature on the subject of scaling relations,  a discussion of which is beyond the scope of the current paper. A recent observational review can be found in \citet[][]{gio13}. We summarise here some important recent advances.

\begin{itemize}

\item The precise measurements available from \xmm\ and {\it Chandra} have enabled excellent convergence in X-ray scaling relations, calibrated using hydrostatic masses of well-chosen samples of relaxed clusters, to minimize the HE bias. This is illustrated in the left hand panel of Fig.~\ref{fig:proxycomp}, which shows the $M_{500}$--$\YX$ relations from \citet{vik06} and \citet{arn07} are consistent at the $1\sigma$ level, with a normalisation that differs by less than $5\%$. These measurements have also allowed exploration of the scatter about the scaling relations (for relaxed objects), confirming that the X-ray luminosity is a high-scatter mass proxy except when the core is excised \citep[][]{mau07,pratt09,man18}, and that $\YX$ is a low-scatter proxy \citep{arn07}. 

\item These scaling relations were then exploited for the cosmological analysis of the X-ray selected sample of \citet{vik09} and the new SZ samples from {\it Planck}. \citet{PCXX2014} 
combined the \MY\ relation obtained on a sample of {\it relaxed} clusters with masses derived from  HE  equation \citep{arn10}, and the $\YX$--$\YSZ$ relation calibrated  on the sub-sample of 71 {\it Planck} clusters in the cosmological sample  with archival \xmm\ data. They introduced a mass bias parameter, $b$, allowing for any difference between the X--ray determined masses and true cluster halo mass: $M_{\rm HE,X} = (1-b)\,\Mv$. This factor encompasses {\it all} systematics in our knowledge of the exact relationship between the SZ signal and the mass \citep[see the extensive discussion in the Appendix of ][]{PCXX2014}. Such a difference can arise from cluster physics, such as a violation of HE or temperature structure in the gas, or from observational effects, essentially instrumental calibration. Even with a fiducial $(1-b)=0.8$ motivated by numerical simulations, this calibration yielded the well-known tension between cosmology from cluster number counts and the {\it Planck} CMB cosmology.
 
\item This generated a large effort to recalibrate the relation between {\it Planck} $\YSZ$ and mass using next-generation WL data from CCCP, WtG, LoCuSS, CLASH, etc, as described above in Sect.~\ref{sec:xrayWL}. The resulting $1-b$ from these lensing data is summarised in Table~2 of \citet{planckSZ15}, and ranges from $0.688\pm0.072$ to $0.780\pm0.092$, with systematic differences between studies on the $\sim 10\%$ level \citep[][see right-hand panel of Fig.~\ref{fig:proxycomp}]{Hoekstra15}. 

\item In parallel, a large effort has been undertaken on the calibration of optical mass proxies based on the richness, exploiting  large-area optical surveys such as SDSS. These studies have developed new, robust mass proxies based on richness \citep[e.g.][]{ryk12,ryk14}, and calibrated them using WL stacking techniques  \citep[e.g.][]{roz09}. 

\item Another key effort has been the critical comparison of various mass estimates, obtained from proxies and/or from direct lensing and/or X-ray analysis, and published in the literature \citep[e.g.][]{roz14a,ser15a,ser15b,gro16}. These studies  use the various samples to better understand the relation between different proxies and their associated biases, and have proposed new calibrations based on an approach that aims for consistency between different analyses \citep{roz14b,ser17}. These studies have underlined the necessity to  properly take into account the covariance between various quantities, and the need for treatment of the Eddington/Malmquist bias effects due to the scatter about the relations.

\end{itemize}

 \begin{figure*}
 \includegraphics[width=0.45\textwidth]{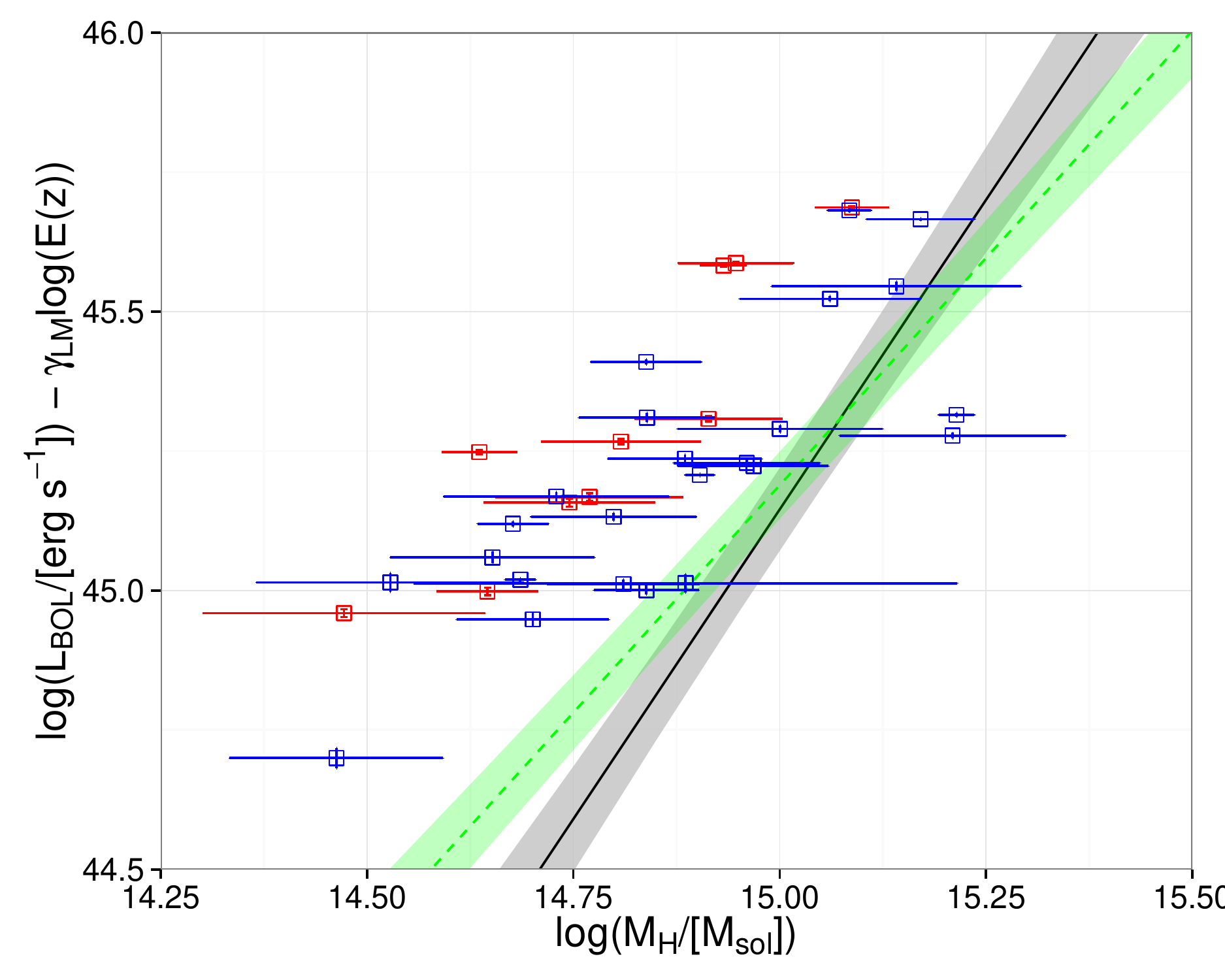}
 \hfill
 \includegraphics[width=0.535\textwidth]{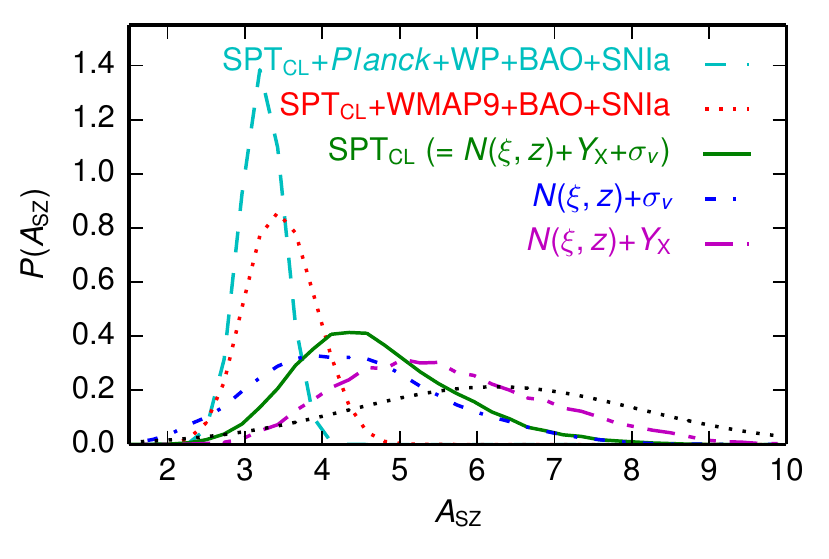}
\caption{({\it Left}): Illustration of the importance of selection biases in scaling relations, and the need to take them into account in cosmological analyses using the cluster population. \citet{gil17} showed that neglect of selection effects in their sample would lead to a 40 per cent underestimate in the mass for a given luminosity. The solid line in this plot is the estimated relation when selection effects are accounted for. Red points in this plot represent relaxed systems; blue points represent disturbed objects. The green line and shaded area represent the best-fitting relation of \citet{man10b} and the corresponding $1\sigma$ uncertainty. 
 ({\it Right}): The normalisation of the $M_{500}$--$Y_{\rm SZ}$ relation depends on the underlying cosmological prior \citep{boc15}.}
\label{fig:biases}       
\end{figure*}

The last point is especially important when calibrating the relation between the mass and the survey observable using external mass estimates of a sub sample. This is illustrated in the left hand panel of Fig.~\ref{fig:biases}, which shows the importance of the Malmquist bias in the survey observable. Here \citet{gil17}, analysing a complete sample of 34 luminous X-ray clusters,  found that not correcting for selection biases would result in a 40 per cent underestimation of the mass of a cluster at a given luminosity.
As this bias depends on both the cluster mass function and the survey selection function, it is becoming more common to perform a fully-consistent joint analysis of the scaling laws and cosmological constraints \citep[e.g.][]{man14,deh16}. 

However, there is a certain degeneracy between the normalisation of the scaling laws and the cosmology, as discussed further below.  In these analyses, it is critically important to understand where the constraints are really coming from when imposing consistency on a multi-dimensional data set. An example is shown in the  right hand panel of Fig.~\ref{fig:biases}, taken from \citet{boc15},  which illustrates how the normalisation of the scaling relation depends on the prior on the underlying cosmology.


\section{Impact on cosmology with clusters}
\label{sec:impact}

\subsection{Clusters as cosmological probes and the sensitivity to the mass scale}

The methods used to estimate cosmological parameters with clusters include use of the baryon fraction (and its evolution), cluster number counts (and their evolution), and the internal cluster shape. All of these approaches  require a robust and precise mass estimate.  The first two methods have so far provided the most competitive constraints, so in this Section we summarise the nature of these studies, and discuss their sensitivity to the mass scale. 

\subsubsection{The baryon fraction}
In galaxy clusters, the relative amount of baryons and dark matter should be close to the cosmic baryon fraction $\Omega_{\rm b}/\Omega_{\rm m}$, provided that the measurement has been performed over a sufficiently large volume inside which the effects of baryonic physics can be neglected. 
Using X-ray observations, the dominant component of the baryon budget can be well constrained and combined with total mass measurements (from e.g. lensing signal or assuming the HE of the X-ray emitting plasma with the gravitational potential) to recover the gas mass fraction $f_{\rm gas} = M_{\rm gas}/M_{\rm tot}$. 
When combined with the total stellar content to give the total baryon fraction, $f_{\rm b}$, some fundamental cosmological parameters can be constrained. This is because, as first proposed in \citet{White93}, the cosmic matter density $\Omega_{\rm m}$ is equal to $Y_{\rm b} \Omega_{\rm b} / f_{\rm b}$, where $Y_{\rm b}$ is the depletion parameter indicating the fraction of cosmic baryons that fall into the cluster halo as estimated from hydrodynamical cosmological simulations \citep[see e.g.][]{Planelles13}.
In addition, if $f_{\rm gas}$ is adopted as a standard ruler and assumed to be constant as function of cosmic time in the `correct' cosmology, constraints can be obtained on the dark energy component $\Omega_{\Lambda}$ \citep[see][]{sasaki96}. Here the cosmological constraints come from the dependence of the observed $f_{\rm gas}$ value on the angular distance, $\fgas \propto D_{\rm A}(z))^{3/2}$ (for X-rays; for SZ, $\fgas \propto (D_{\rm A}(z))$) .
 
These methods all need a robust and precise calibration of the total gravitating mass. The error on $\Omega_{\rm m}$ and  $D_{\rm A}(z)^{3/2}$  depends linearly on the mass uncertainty  with $f_{\rm gas}$, which can be translated into a corresponding accuracy on $\Omega_{\Lambda}$ via the functional dependence of $D_{\rm A}(\Omega_{\rm m},\Omega_{\Lambda})$ in the $z$ range under consideration.   Application of the methods have typically relied  on  X-ray masses derived from the HE equation (of better statistical precision than lensing mass) and are thus directly sensitive to corresponding systematic effects in the HE mass estimates, in particular the HE bais.  Furthermore, a good understanding of the depletion factor and its evolution  is required.  This quantity is expected to be more robust  for massive systems, where gravity dominates the energy budget over other physical phenomena such as AGN and SN feedback and gas cooling.  To minimise these systematics, the methods have  been essentially applied using the most massive relaxed systems \citep[e.g.][]{lar06,Allen08,ettori09,mantz14}. 

Conversely, the methods can provide a consistency check of the mass estimates in galaxy clusters, if the cosmological parameters are adopted from independent techniques (such as modelling of the temperature anisotropies in the CMB, or SN). Indeed, a knowledge of $\Omega_{\rm b}$ and $\Omega_{\rm m}$, combined
with the measurements of $M_{\rm gas}$, which is one of the better constrained quantities from X-ray observations, limits the level of systematics on the measurement of $M_{\rm tot}$ (see Sect.~\ref{sec:fgasxcop}).

\begin{figure*}
\begin{center}
  \includegraphics[scale=0.7]{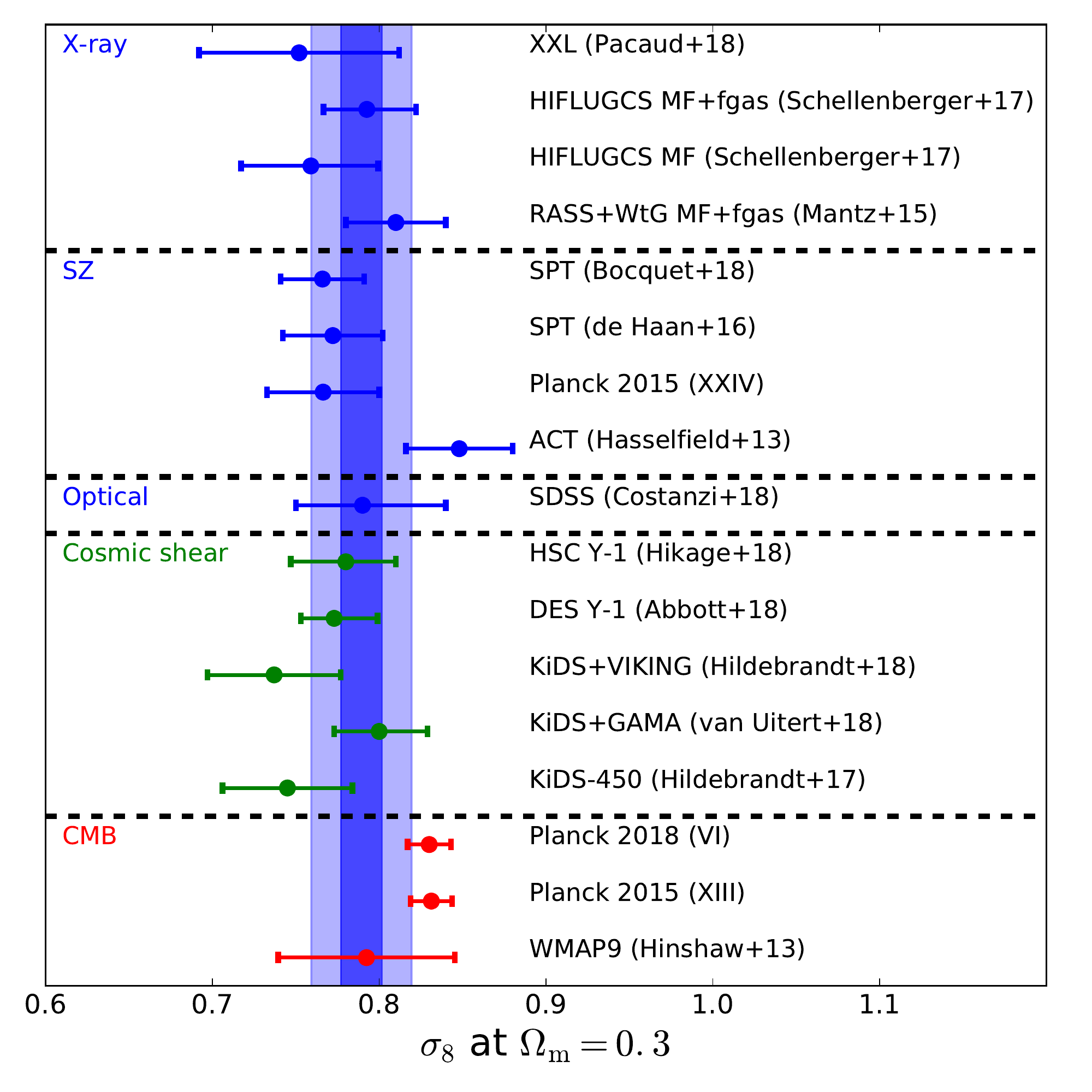}
\caption{Constraints on $\sigma_8$ at $\Omega_{\rm M}=0.3$ from the cluster mass function (sometimes combined with $f_{\rm gas}$ constraints) are shown with blue symbols. Standard deviation ($=0.033$) and error ($=0.012$) around the unweighted mean ($=0.789$) of all seven independent cluster analyses are shown as light and dark blue shaded bands, respectively. Also shown are constraints from WL/cosmic shear/galaxy clustering (green symbols) and from CMB (red symbols). Details on all the constraints are provided in Sections~\ref{impact:results} and \ref{impact:summary}. Note that analysis details differ for the various works. Adapted from \citet{sr17b}.}
\label{fig:impact:S8}
\end{center}       
\end{figure*}

\subsubsection{The mass function}

The mass function, defined as the number of haloes of a given mass per unit volume, can be written as:
\begin{equation}
\label{eq:mf}
\frac{dN}{dM}(M,z)=f(\sigma_{\rm M})\,\frac{\rho_{\mathrm
  m}(z=0)}{ M}\, \frac{   d\ln\sigma_{\rm M}^{-1} }{dM} \,,
\end{equation}
where  ${\rho}_\mathrm{m}(z=0)$ is the mean matter density at $z=0$, and  $\sigma_{\rm M}$ is the power spectrum of density perturbations \citep{jen01,tin08}.   Its dependence on mass and redshift can be written as:
\begin{eqnarray}
 & & \sigma_{\rm M} (M,z) =  \sigma_{\rm M} (M,z=0) D_{\rm grow}(z) \nonumber \\
 & & \qquad {\rm with}  ~~ \sigma_{\rm M} (M,z=0) \sim  \sigma_{\rm 8}  \left(\frac{M}{M_8}\right)^\alpha
\end{eqnarray}
where $D_{\rm grow}(z)$ is the growth factor, and considering that the present day power spectrum $ \sigma_{\rm M} (M,z=0) $ is close to a power law at cluster scales, with $\alpha  \sim -1/3$.  The logarithmic derivative term in Eqn.~\ref{eq:mf}  is approximately constant, and the mass function depends on mass and $\sigma_{8}$ essentially  as:
\begin{equation}
\frac{dN}{dM} \propto f(\sigma_{8}\, M^\alpha)/M 
\end{equation}
The mass function is thus very sensitive to $\sigma_8$, via  the exponential behaviour of the function 
\begin{equation}
f(\sigma)\propto\left[1+(\sigma/b)^{-a}\right] \exp(-c/\sigma^2). 
\end{equation} 
 The determination of $\sigma_{8}$  will thus essentially be degenerate with any mass bias, expressed as $M_{\rm obs}=(1-b)\,M_{\rm true}$, along the degeneracy line $\sigma_{8} (1-b)^{-\alpha }= {\rm cst}$, i.e. $\sigma_8 \propto (1-b)^{1/3}$  or $(1-b) \propto 1/\sigma_{8}^3$. 


\subsection{Recent results on cosmological constraints from galaxy clusters and their dependence on the mass determination}
\label{impact:results}

In the last $\sim$5 years there has been significant progress in cluster cosmology, including new samples selected in different wavebands and improved treatments of cluster masses. In the following we call a mass calibration procedure `internal' if mass measurements are available for (a subset of) the clusters in the sample used for the cosmological tests; we call it  `external' if the mass calibration is based on other clusters, e.g. a scaling relation from the literature. As discussed above, the cosmological parameters $\sigma_8$ and $\Omega_{\rm M}$ are particularly strongly affected by systematic mass uncertainties 
therefore we list those constraints as well.

Figure~\ref{fig:impact:S8} shows galaxy cluster constraints on $\sigma_8$ from the last $\sim$5 years, plus selected other constraints (cosmic shear/galaxy clustering and CMB). The Figure shows results from the following studies.

\begin{itemize}
\item {\bf X-ray samples:}
\label{impact:results:X}

\paragraph{\citet{pcg16,ppm18}:} These mass function constraints are based on the XXL sample of 178 X-ray selected galaxy clusters detected out to $z=1$. A mostly internal WL mass calibration of a subset of clusters is performed. For $\Omega_M=0.3$ the authors approximately find $\sigma_8=0.752 \pm 0.060$. These uncertainties are large because they use a large prior on $h=0.7\pm 0.1$,
use only the redshift distribution (which peaks around $0.3<z<0.5$), and allow for a large prior on the evolution of cluster luminosities in the scaling laws ($L(T,z)/L(T,z=0) = E(z)^{0.47\pm 0.68}$).

\paragraph{\citet{sr17b}:} The sample of the X-ray brightest clusters in the sky is used (HIFLUGCS, 64 clusters with $\bar{z}=0.05$). The focus of this work is not on using large numbers of clusters but on taking advantage of very high-quality data for all objects in the sample (i.e., 100\% internal mass `calibration'). This includes on average about 100 ks of \chandra\ data and 200 cluster galaxy velocities per cluster. Hydrostatic masses are taken from \citet{sr17a} and dynamical masses from \citet{zrs16}. Furthermore, a comparison to \planck\ `SZ masses' is undertaken.
They find $\sigma_8(\Omega_{\rm M}/0.3)^{0.5} = 0.793_{-0.026}^{+0.029}$ for their default results combining mass function, and $f_{\rm gas}$, and $\sigma_8(\Omega_{\rm M}/0.3)^{0.5} = 0.759^{+0.040}_{-0.042}$ when using the mass function alone.

\paragraph{\citet{mva15}:} The mass function is determined using 224 clusters from RASS-selected samples, extending to $z=0.5$. \chandra\ and ROSAT X-ray data for 94 clusters are used for gas mass determinations. Weak gravitational lensing data from the WtG project are used directly for the 27 internal clusters and for 23 further clusters through the (therefore partially external) gas mass--total mass relation. Furthermore, the $f_{\rm gas}$ constraints described in the paragraph below are incorporated.
Shear profiles are compared to an NFW model with $c=4$ for the lensing mass determination, the cosmology-dependence of the predicted NFW model is accounted for.
They find $\sigma_8(\Omega_{\rm M}/0.3)^{0.17} = 0.81 \pm 0.03$.

\paragraph{\citet{mantz14}:} 
The gas mass fraction, $f_{\rm gas}$, for a sample of 40 massive, relaxed clusters is determined at a range of redshifts. The nearby clusters are used to constrain $\Omega_{\rm M}$ and the apparent evolution of the gas mass fraction to constrain dark energy parameters. Gas masses are obtained from \chandra\ X-ray observations; total masses from X-ray hydrostatic analysis and, for a subset of 12 clusters, weak gravitational lensing observations from the WtG project. The radial range 0.8--1.2\,$R_{2500}$ is exploited for these measurements.
The cosmology-dependence of the $f_{\rm gas}$ measurements is taken into account self-consistently, including modelling the radial dependence of $f_{\rm gas}$ as an {\it average} power law in the relevant range. The mass calibration is internal. They find $\Omega_{\rm M} = 0.27 \pm 0.04$.

\begin{figure*}
  \begin{minipage}[c]{0.62\textwidth}
    \vspace{0pt}
    \includegraphics[width=\textwidth]{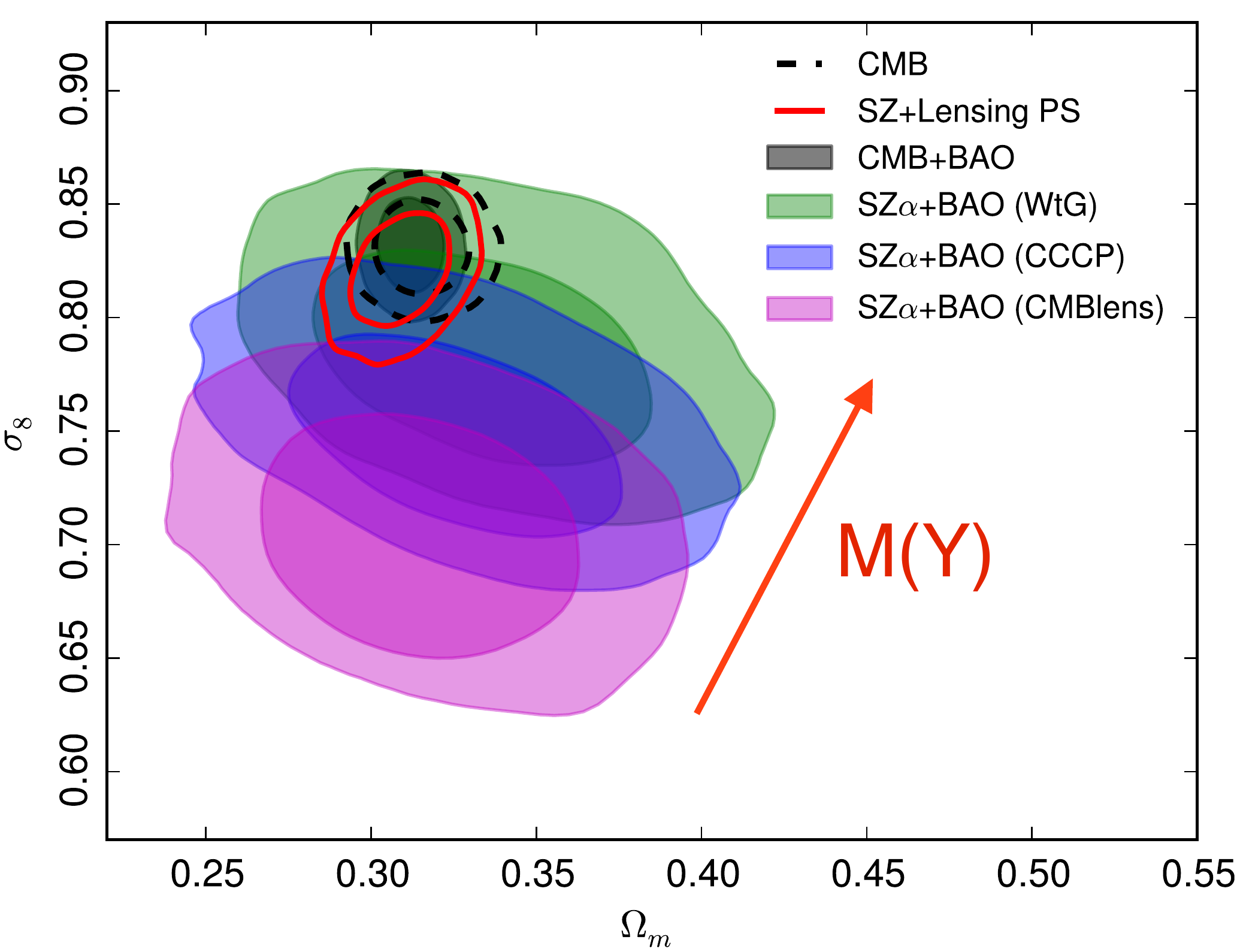}
  \end{minipage}\hfill
  \begin{minipage}[c]{0.35\textwidth}
   \vspace{0pt}
    \caption{Dependence of the constraints on cosmological parameters $\sigma_8$ and $\Omega_{\rm m}$ on the normalisation of the $M_{500}$--$\YSZ$ relation. Each coloured contour represents the cosmological constraints for a given prior on the mass normalisation. The red arrow indicates the effect of an increasing normalisation of the $M_{500}$--$Y_{\rm SZ}$ scaling relation. Reproduced from \citet{planckSZ15,arn17}.
     } \label{fig:mydependence}
  \end{minipage}
\end{figure*}

\paragraph{\citet{mam16}:}
The authors study the mass distribution of 40 relaxed clusters from the Weighing the Giants sample. They find consistency with a simple NFW profile in the central parts, but also significant scatter.

\paragraph{\citet{b17reflex7}} determine the cluster mass function at 10\% level of uncertainty over the mass range $3\times 10^{12} - 5 \times 10^{14} M_{\odot}$ by fitting the observed cluster X-ray luminosity distribution from the REFLEX~II galaxy cluster survey. They conclude that about 14\% (4.4\%) of the matter in the Universe is bound in clusters with a mass larger than $10^{13} (10^{14}) M_{\odot}$, and that it is unlikely that any cluster with a mass $\Mv \gtrsim 10^{15} M_{\odot}$ is present at $z>1$.

\paragraph{\citet{b17noras1}} present the NORAS II galaxy cluster survey, based on X-ray data from the northern part of the RASS down to a flux limit of $1.8 \times 10^{-12}$ erg s$^{-1}$ cm$^{-2}$ (0.1-2.4 keV), containing 860 objects with a median redshift of 0.1. They constrain $\sigma_8$ and $\Omega_{\rm m}$ using the X-ray luminosity, finding results that are in agreement with their previous findings  \citep[][]{b14reflex4}.  \\

\item {\bf SZ samples:}
\label{impact:results:SZ}

\paragraph{\citet{planckSZ15}:}
The cluster sample consists of 439 clusters selected from \planck\ data. Internal mass calibrations based on gravitational lensing are applied, including as baseline the CCCP \citep{hhm15} constraints.
For their baseline assumptions (CCCP+BAO+BBN) they find $\sigma_8(\Omega_{\rm M}/0.31)^{0.3} = 0.774 \pm 0.034$. \citet{planckSZ15} explored several different mass normalisations, including for the first time a calibration based on CMB lensing \citep{mel15}. Figure~\ref{fig:mydependence} illustrates the dependence of the cosmological constraints on the mass normalisation. 
These results refine and confirm the previous results of \citet{PCXX2014}, for which the (internal) mass calibration came from an $M_{500}$--$\YSZ$ relation calibrated from X-ray data, with a bias estimated from comparison to numerical simulations, as discussed extensively in the Appendix of \citet{PCXX2014}. 

\paragraph{\citet{deh16}:} For the mass function constraints 377 clusters with $z>0.25$ are selected from the South Pole Telescope (SPT) survey\footnote{\url{https://pole.uchicago.edu/spt/}}. An external WL mass calibration is applied as well as an additional constraint from \chandra\ data for 82 clusters. 
Note that the central value and uncertainties in the arXiv version (arXiv:1603.06522v1) differ from those in the published version. Here, we use the results from the published version:
$\sigma_8(\Omega_{\rm M}/0.27)^{0.3} = 0.797 \pm 0.031$.
This analysis has been expanded recently to internal WL mass calibration in \citet{bds18}, who find $\sigma_8(\Omega_{\rm M}/0.3)^{0.2} = 0.766 \pm 0.025$.

\paragraph{\citet{has13}:} The mass function is determined using 15 Atacama Cosmology Telescope (ACT)\footnote{\url{https://act.princeton.edu/}}  clusters, using external mass calibration.
We show the BBN+H0+ACTcl(B12) constraints from their Tab.~3, which assume a \emph{fixed} scaling relation:
$\sigma_8(\Omega_{\rm M}/0.27)^{0.3} = 0.848 \pm 0.032$.\\

\item {\bf Optical samples:}
\label{impact:results:opt}

\paragraph{\citet{crs18}:} Clusters selected from a redMaPPer \citep{rrb14} search of the Sloan Digital Sky Survey (DR8) are used. Weak lensing mass profiles stacked in richness bins from the same data are employed for internal mass calibration. They find $\sigma_8(\Omega_{\rm M}/0.3)^{0.5} = 0.79^{+0.05}_{-0.04}$.\\

\item {\bf Other constraints shown in Fig.~\ref{fig:impact:S8}:}

\paragraph{\citet{hoh18}:}
Cosmic shear constraints from the Subaru Hyper Suprime-Cam first-year data. They find $\sigma_8(\Omega_{\rm M}/0.3)^{0.5} = 0.780^{+0.030}_{-0.033}$.

\paragraph{\citet{hkv18,hvh17,vjj18}:}
Cosmic shear constraints from the VST KiDS survey.
They find, respectively
$\sigma_8(\Omega_{\rm M}/0.3)^{0.5} = 0.737\pm 0.040$,
$0.745\pm 0.039$,
and
$0.800_{-0.027}^{+0.029}$.

\paragraph{\citet{DES18}:} DES Y-1 constraints from
galaxy clustering and WL.
Note that the central value and uncertainties in the arXiv version (arXiv:1708.01530v1) differ from those in the published version. Here, we use the results from the published version:
$\sigma_8(\Omega_{\rm M}/0.3)^{0.5} = 0.773_{-0.020}^{+0.026}$.

\paragraph{\citet{P18VI,P15XIII}:} Planck 2018 (VI) and Planck 2015 (XIII) constraints from the CMB. 
For the former, we show the {TT,TE,EE+lowE+lensing} results from the`Combined' column of their Table~1:
$\sigma_8(\Omega_{\rm M}/0.3)^{0.5} = 0.830 \pm 0.013$.
For the latter, we show the TT,TE,EE+lowP+lensing
results\footnote{Note that because of this choice these constraints are tighter than the ones shown in \citet{sr17b}.}
from their Tab.~4:
$\sigma_8(\Omega_{\rm M})^{0.5} = 0.4553 \pm 0.0068$

\paragraph{\citet{hlk13}:} WMAP9 constraints from the CMB. We show the values quoted in their Section~5:
$\sigma_8(\Omega_{\rm M})^{0.5} = 0.434 \pm 0.029$.

\end{itemize}


\subsection{Summary and Interpretation}
\label{impact:summary}

Figure~\ref{fig:impact:S8} leads us to draw the following conclusions:

\begin{itemize}
 \item All the recent cluster constraints agree surprisingly well with each other within the uncertainties despite the fact that they differ dramatically in selection, mass treatment, and analysis. One apparent slight exception is the ACT result for fixed default scaling relation; however, on the one hand, this is expected statistically given seven independent constraints and, on the other hand, as \citet{has13} describe in their paper (see, in particular, their Fig.~14) choosing a different fixed scaling relation (their `UPP') would in fact bring $\sigma_8$  \emph{below} the mean of all the cluster results.
 \item The standard deviation of the cluster results is comparable to the typical uncertainty of the individual results, which can be viewed as indication that confirmation bias is small among the cluster results.
 \item The cosmic shear/galaxy clustering results agree with the cluster results within their uncertainties.
 \item The CMB constraints from WMAP agree with the cluster and cosmic shear results.
 \item The \planck\ CMB constraints are close to all of the above, but outside the standard error of the mean of the cluster results.
\end{itemize}

The general agreement between different probes seems healthy, and one could argue that within the statistical expectations and the still reasonably small ($\ll$$25$) total number of constraints, there is nothing to be excited about. However, progress in physics (also) comes from measurement disagreements, and, initially, these are typically small. Let us briefly outline possible interpretations of the slight tension between the mean cluster constraints ($\sigma_8=0.789\pm 0.012$) and the \citet{P18VI} CMB constraints ($\sigma_8=0.830\pm 0.013$).

The first suspect is unaccounted-for systematic effects. For both clusters and CMB these might come from, e.g. instrumental calibration or modelling issues. For galaxy clusters, further sources of systematic uncertainty include cluster selection effects and the mass determination, and this review article indeed focusses on the cosmological impact of the latter.

Other, more exotic explanations can also be brought to bear. In terms of physics and cosmology, a summed neutrino mass higher than the minimum mass, $\sim$$0.06$ eV, might help alleviate the tension.
It is also interesting to speculate about possible
more exotic physics that might be causing this tension. For instance, a modification of gravity, a self-interacting dark matter component, warm dar matter, or a dark energy component that interacts with dark matter would all change the predicted mass function.


\section{The future}
\label{forward}

\subsection{New surveys and samples}
\label{forward:new_survs}

\subsubsection{X-ray}
\label{forward:new_survs:X}

\paragraph{X-ray surveys:}

The ROSAT All-Sky Survey (RASS) was performed almost 30 years ago and many cluster cosmology samples have been derived from it (e.g. Section~\ref{impact:results:X}). New opportunities based on this venerable dataset include running more sophisticated source detection algorithms, as well as combining the X-ray data with a wealth of new multiwavelength data, particularly in the optical/IR and sub-mm/mm (SZ) regimes.

As an example for the former, \citet{xrp18} recently showed that galaxy groups and clusters with unusually extended surface brightness profiles were missed in previous RASS cluster surveys. They achieved this  by employing a dedicated source detection pipeline particularly sensitive to extended low surface brightness emission, while many of the previous RASS source catalogues were constructed using detection methods optimized for point sources. If missed clusters are not accounted for, their number density will be underestimated, resulting in underestimated values for $\Omega_{\rm M}$ and/or $\sigma_8$ (e.g., Fig.~10 in \citealt{sr17b}); i.e. a qualitatively similar effect as a systematic underestimate of cluster masses. Whether or not the unaccounted-for missed fraction is large enough to significantly affect cosmological parameter constraints derived from previous RASS cluster samples is not yet clear.

Examples of the multiwavelength studies include \citet{sbv04} and \citet{tma18} who combined RASS data in a matched filter approach for joint detection with SDSS and \planck\ data, respectively. They showed that detection probability, purity, and source identification can be improved in such an approach.

With the still-functioning satellites \xmm\ and \chandra, progress is being made particularly in surveys for new clusters serendipitously detected using archival \xmm\ observations. Recent examples  include, e.g. the \xmm\ Cluster Survey  \citep[XCS-DR1][]{mrh12}, the \xmm\ CLuster Archive Super Survey \citep[XCLASS; ][]{csp12,rcs17}, and XXL \citep{pcg16,ppm18}. These surveys have by now detected well over 1\,000 new clusters. Given these surveys are typically much deeper but cover a much smaller area than the RASS, the recovered cluster populations are typically at higher redshifts and/or have lower masses.

\paragraph{Mass and mass proxy calibration}

A large amount of effort continues to be put into follow-up and archival studies of the mass and calibration of the mass proxy relations. Some examples of recent \emph{Chandra} and \xmm\ archival and/or dedicated follow-up of \planck-, SPT-, and ACT-selected clusters, include \citet{PEPXI}, \citet{PIPIII}, \citet{PCXX2014}, \citet{lfj17}, \citet{bcm18}, \citet{msb13}, and of RASS-selected clusters (e.g. eeHIFLUGCS, \citealt{r17}).
The high-quality X-ray data enable many cluster studies but also the determination of hydrostatic masses as well as very precise mass-proxies like the gas mass, temperature, or the $Y_{\rm X}$ parameter. 

One outcome of these studies has been that the comparison of \planck\ SZ-selected clusters with X-ray selected clusters \citep{ros16,ros17,and17,lfj17} has indicated that there may be a tendency in X-ray surveys to preferentially detect clusters with a centrally-peaked morphology, which are more luminous at a given mass, and on average more relaxed. This raises concerns about how representative the X-ray selected samples, used to define our current understanding of cluster physics and to calibrate numerical simulations, have been. In addition, mass comparisons between various samples are hampered by the heterogeneous nature of the data.  A  large effort is now ongoing, in the form of a multi-year heritage project on the \xmm\ satellite, to obtain homogeneous X-ray and lensing mass estimates up to $R_{500}$, of a well-defined \planck\ SZ-selected sample of more than 100 clusters.


\subsubsection{Microwave} 

\label{forward:new_survs:SZ}

The recent arrival of large-area SZ galaxy clusters surveys, by ACT \citep[][]{mar11,has13}, SPT \citep[][]{SPTSZ15} and {\it Planck}\  \citep{esz,psz1,psz2} has resulted in a rapid growth in the number of known, massive galaxy clusters especially at $z>0.5$ -- a regime which was classified as `high redshift' by the galaxy cluster community as recently as a decade ago due to the paucity of known systems at these distances. To date, on the order of 1000s of massive galaxy clusters have been detected through blind surveys using the thermal SZ effect. 
SZ surveys have the advantage over more traditional cluster-detection methods (e.g. X-ray, optical, near-IR) in that they are roughly redshift independent, selecting only on mass. 

There is also a strong complementarity in the coverage of the mass-redshift plane between the new SZ surveys and their X-ray counterparts.  The spatial resolution of ACT and SPT ($\sim 1^\prime$)  allows them to probe the cluster population above a nearly constant mass threshold up to very high redshift, $z\sim1.5$, but their smaller  area limits the number of high mass objects.  The lower spatial resolution of {\it Planck}  ($\sim 5^\prime$) is offset by its being the first all-sky blind survey since the {\it ROSAT} All Sky Survey (RASS). Although less sensitive, it is uniquely suited to finding high mass, high redshift systems. 

In the future, one way to improve the cluster mass calibration is through lensing of the CMB \citep{mel15,planckSZ15,bax18}. CMB-cluster lensing offers a robust and accurate way to constrain galaxy cluster masses, especially at high redshift ($z>1$) where optical lensing measurements are challenging. With CMB lensing we expect to improve mass uncertainty to 3\% for upcoming experiments such as AdVACT, SPT-3G etc., and to 1\% for next generation CMB experiments such as CMB-S4 (discussed below). 


\subsubsection{Lensing and optical/IR}
\label{forward:new_survs:opt}

\begin{table*}
\caption{Wide-field survey properties \citep{HSCWL1styr,DES1stWL,KIDS1stWL}. $^{(a)}$ surveys, $^{(b)}$ bands, $^{(c)}$ planned survey area, $^{(d)}$ limiting magnitude, $^{(e)}$ the number of background galaxies for weak-lensing analysis, and $^{(f)}$ typical seeing-size for shape measurements}\label{table:OIRsurveys}
\begin{center}
 \begin{tabular}{cccccccc}
\hline
Survey$^{(a)}$ & Bands$^{(b)}$ & Area$^{(c)}$ & $m_{\rm lim}^{(d)}$ & $n_{g}^{(e)}$  & Seeing$^{(f)}$ \\
 & & (deg$^2$) & (ABmag) & (arcmin$^{-2}$) &  (arcsec) & \\ 
\hline
HSC-SSP & $grizy$ &  1400 & $r\sim26.4$ ($S/N=5$) & $\sim25$ & $\sim0.58$  \\
DES    & $grizY$  & 5000 & $r\sim23.3$ ($S/N=10$) & $\sim7$  & $\sim0.9$ \\
KiDS   & $ugri$  & 1500 & $r\sim24.9$ ($S/N=5)$ & $\sim7$ & $\sim0.66$ \\ 
 \hline
 \end{tabular}
\end{center}
\end{table*}

On-going imaging surveys (HSC-SSP, DES, and KiDS) enable identification of galaxy clusters using luminous red galaxies and weak-lensing mass maps, and/or enable direct weak-lensing mass measurements of galaxy clusters. The respective survey properties are summarised in Table \ref{table:OIRsurveys}.

The HSC-SSP survey\footnote{\url{https://hsc.mtk.nao.ac.jp/ssp/}} \citep{HSC1styrOverview} is an ongoing wide-field imaging survey
using the HSC \citep{Miyazaki18HSC} which is a new prime focus camera on the 8.2m-aperture Subaru Telescope,  
and is composed of three layers of different depths (Wide, Deep and UltraDeep). 
The Wide layer is designed to obtain five-band ($grizy$) imaging over $1400$~deg$^2$.
The HSC-SSP survey has both excellent imaging quality ($\sim$$0.7^{\prime\prime}$
seeing in $i$-band) and deep observations ($r\simlt26$~AB~mag).
\citet{Oguri18} constructed a CAMIRA cluster catalogue from HSC-SSP S16A
dataset covering $\sim 240$~deg$^2$ using the CAMIRA algorithm \citep{Oguri14b} which is a
red-sequence cluster finder based on the stellar population synthesis model fitting.
The catalogue contains $\sim 1900 $ clusters at $0.1<z<1.1$. \citet{Miyazaki18} have searched galaxy clusters based on weak-lensing analysis of the ˜160 deg$^2$ area and discovered 65 shear-selected clusters of which signal-to-noise ratio is higher than 4.7 in the weak-lensing mass map. 

The DES survey\footnote{\url{https://www.darkenergysurvey.org/}} \citep{DES16Overview} covers a 5000 deg$^2$ area of the southern sky using the new Dark Energy Camera (DECam) mounted on the Blanco 4-meter telescope at the Cerro Tololo Inter-American Observatory. 	
\citet{Rkyoff16} have applied a photometric red-sequence cluster finder (redMaPPer) to 150 deg$^2$ of Science Verification and found $\sim800$ clusters at $0.2<z<0.9$.

The Kilo-Degree Survey\footnote{\url{http://kids.strw.leidenuniv.nl/}} \citep[KiDS;][]{KIDS1stDR} covers 1500 deg$^2$ using the VLT Survey Telescope (VST),located at the ESO Paranal Observatory. \citet{Bellagamba18} have developed Adaptive Matched Identifier of Clustered Objects (AMICO) algorithm and applied to $\sim440\,{\rm deg}^2$ survey data. They found $\sim8000$ candidates of galaxy clusters in the redshift range $0.1<z<0.8$ down to $S/N>3.5$ with a purity approaching 95\% over the entire redshift range.

The optical and weak-lensing cluster finders are complementary to the ICM observations through the SZ and X-ray method. Future multi-wavelength comparison for optical, shear-selected, X-ray and SZ clusters willgive detailed insights into cluster physics and the sample selection functions. 

\begin{figure*}
  \includegraphics[width=\textwidth]{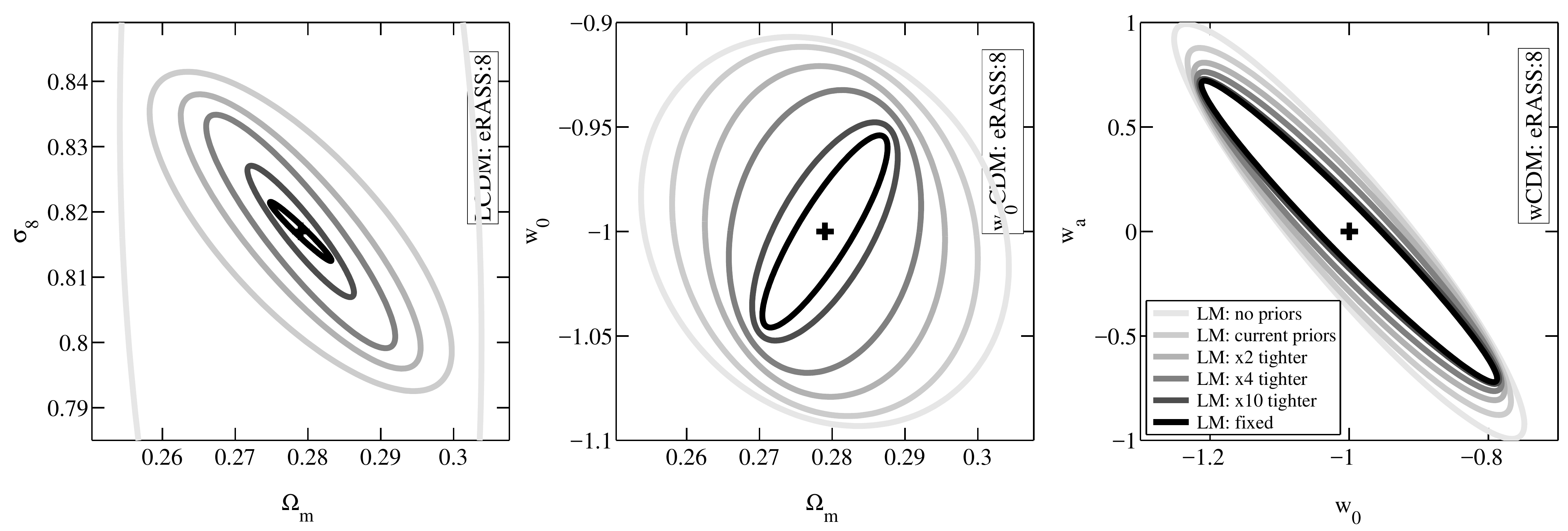}
\caption{Expected constraints on cosmological parameters from {\it eROSITA} (cluster mass function plus clustering) and their dependence on the mass calibration precision. For this purpose, the $L_{\rm X}$--$M$ relation is modelled as a powerlaw with four parameters. The different grey shades illustrate increasing precision coming from direct mass measurements. Light grey means the parameters are completely unconstrained while black shows the constraints when fixing them (see inset on the right figure). As the authors describe, the improvement is strongest for the $\Omega_{\rm M}$--$\sigma_8$ plane and weakest for the $w_0$--$w_a$ plane. Reproduced from \citet{prp18} with permission.}
\label{fig:forward:eROSITA-L-M}       
\end{figure*}
%


\subsection{Next-generation data}
\label{forward:TNG}

\subsubsection{eROSITA} \label{forward:TNG:eROSITA}

\begin{figure*}
  \begin{minipage}[c]{0.65\textwidth}
    \vspace{0pt}
  \includegraphics[bb=270 20 700 450 ,clip,scale=1.,scale=0.55]{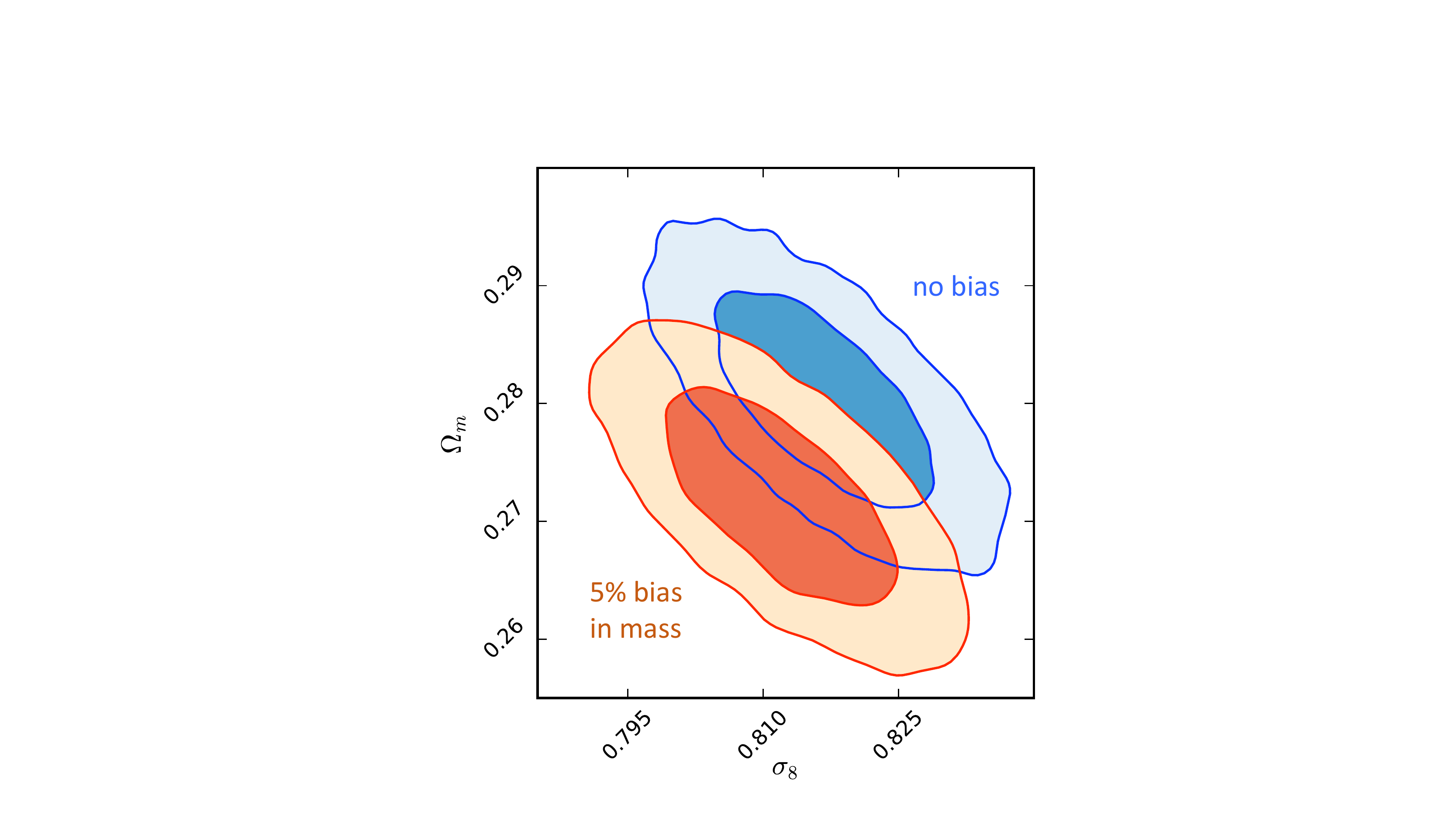}
  \end{minipage}\hfill
  \begin{minipage}[c]{0.35\textwidth}
   \vspace{0pt}
    \caption{Illustrative predicted \emph{eROSITA} constraints on $\Omega_{\rm M}$ and $\sigma_8$ from the cluster mass function and their dependence on the mass calibration accuracy. For the blue contours unbiased mass estimates are assumed while for the red ones a systematic mass bias of 5\% is simulated. Shown are the 68\% and 95\% credibility levels as dark and light shades, respectively. One notes that, for \emph{eROSITA}, this mass bias results in a significant (at the 95\% credibility level) bias for these cosmological parameters. Figure credit: Katharina Borm.
     } \label{fig:forward:L-M_bia}
  \end{minipage}
\end{figure*}

\emph{eROSITA}\footnote{\url{https://www.mpe.mpg.de/eROSITA}} is the main instrument onboard the Spektrum-Roentgen-Gamma satellite to be launched in 2019 \citep{pab14}. It will perform eight X-ray all-sky surveys resulting in at least 20 times higher sensitivity than the RASS \citep{mpb12}. The primary science driver is the study of dark energy with galaxy clusters. \emph{eROSITA} is expected to detect about 100,000 galaxy clusters \citep{ppr12,prp18,crr18}. For a small subsample ($\sim$2\,000 clusters) precise gas temperatures will be measured directly from the survey data \citep{brm14,hsc17}. Competitive constraints on dark energy are expected: e.g., $\Delta w_0 = \pm 0.07$ and $\Delta w_a = \pm 0.25$ in an optimistic scenario with accurate mass calibration down to the low-mass galaxy group regime \citep[][see Fig.~\ref{fig:forward:eROSITA-L-M}]{prp18}, making \emph{eROSITA} one of the first Stage IV dark energy experiments. Mock light cones based on cosmological simulations and including, e.g., expected \emph{eROSITA} photon count rates, are publically available \citep{zfp18}. The dependence of {\it eROSITA} cosmological constraints on the mass calibration accuracy are shown in Fig.~\ref{fig:forward:L-M_bia}. 
%

As discussed in Section~\ref{impact:results}, 
precise and accurate internal mass calibration is important. For \emph{eROSITA}, the plan is to rely particularly on weak gravitational lensing mass calibration using data from, e.g., the VST, DECam, and HSC surveys, and later from Euclid and LSST \citep{mpb12,gmd18}.


\subsubsection{Euclid and LSST}

The LSST\footnote{\url{https://www.lsst.org/}} \citep{LSST08} will cover $20,000\,{\rm deg}^2$ south of $+15$ deg, with a limiting magnitude of $26.9$ during ten years of operation. 
The LSST will have first light in 2020, and start the operations phase in 2022. LSST will construct a large catalogue of clusters detected through their member galaxy population to redshift $z\sim1.2$. Mass calibration using WL mass measurements will measure the cluster mass function. \citet{LSST18} have reported observational requirements for a dark energy analysis consistent with the Dark Energy Task Force definition of a Stage IV dark energy experiment, using conservative assumptions based on current observational resources. 

\begin{figure*}[t]
\begin{center}
\includegraphics[width=0.495\textwidth]{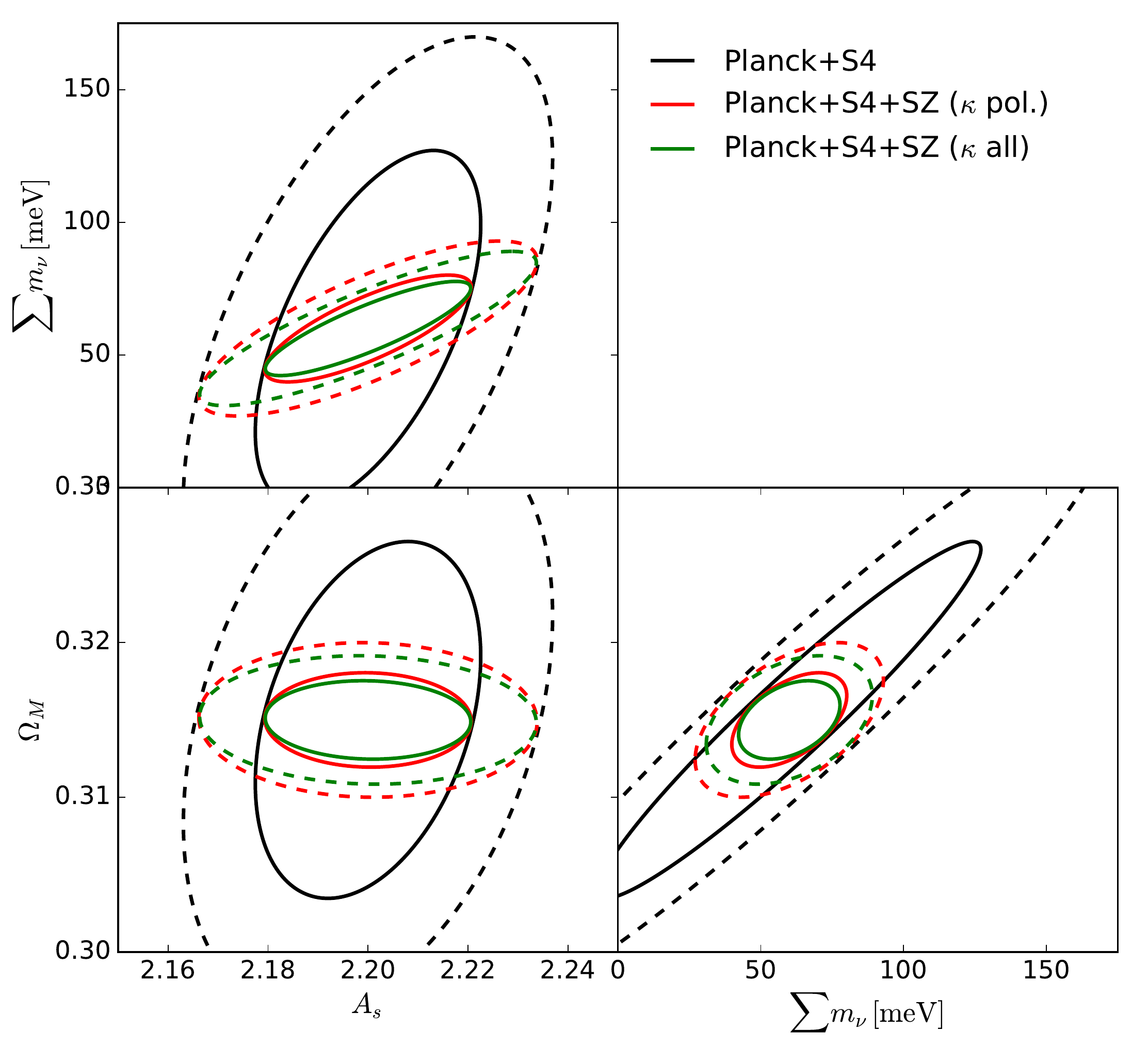}
\hfill
\includegraphics[width=0.495\textwidth]{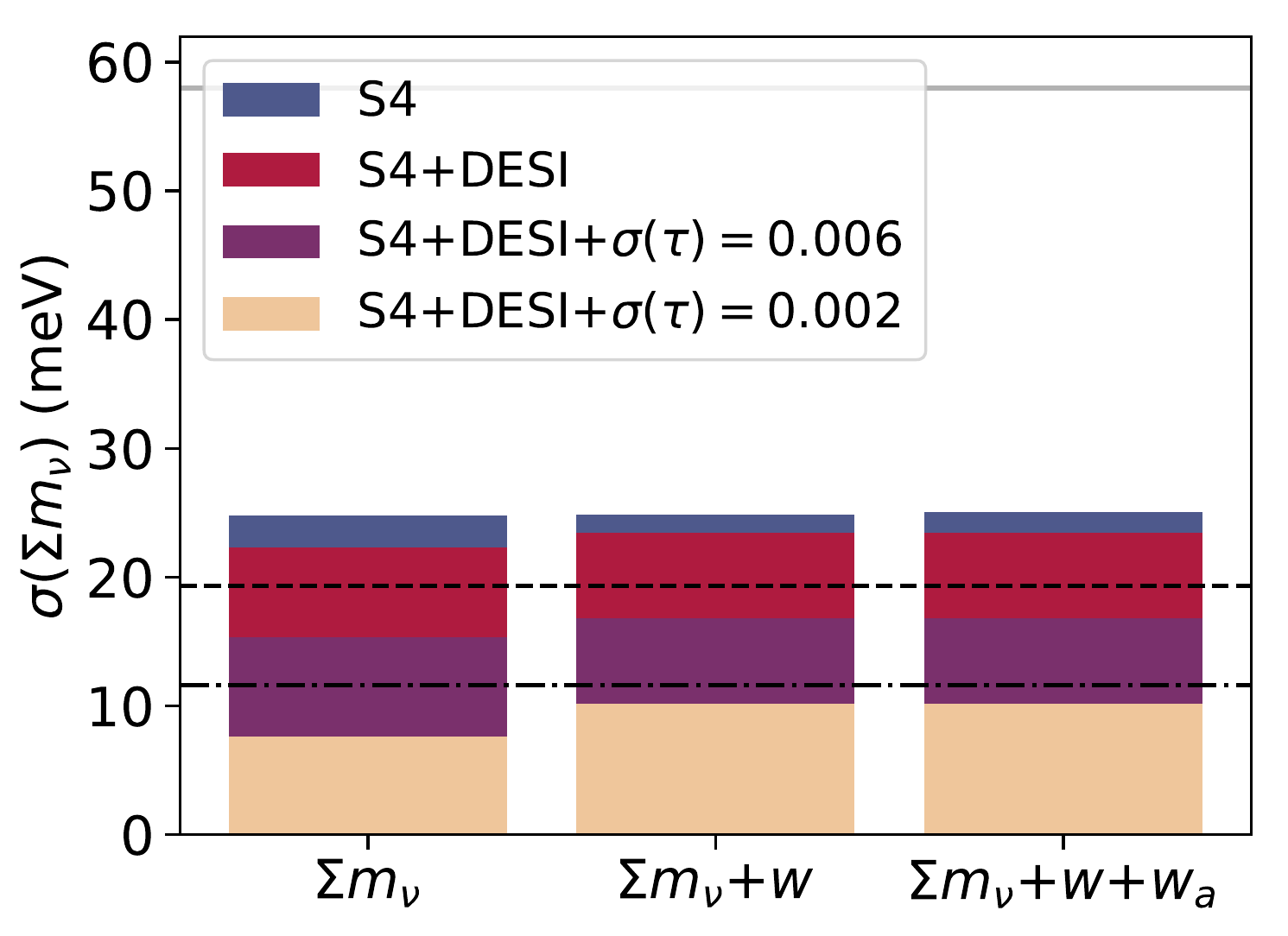}
\caption{
{\it Left:} Uncertainties $\Omega_M$, $A_s$ and the sum of neutrino masses from an SZ catalogue carried out with CMB-S4 in combination with constraints from S4 primary and lensing power spectra, as well as \planck\ temperature and polarisation on $\ell<30$. Results are shown in the absence of lensing mass estimates (black ellipses), and for lensing masses computed using only polarisation (red ellipses) and temperature and polarisation (green ellipses). The results are marginalised over all other cosmological parameters as well as the cluster nuisance parameters. Figure taken from \citet{Louis17}.
{\it Right:} The $1-\sigma$ uncertainty on neutrino mass obtained when marginalizing over $\Lambda$CDM, $\Lambda$CDM+$w_0$ and $\Lambda$CDM$+w_0+w_a$, from tSZ clusters detected using a CMB Stage-4 telescope with $1'$ beam FWHM at 150 GHz. Constraints when the mass calibration is a combination of internal CMB and optical WL. Figure taken from \citet{Madhavacheril17}.
}
\label{fig:CMB-S4}
\end{center}
\end{figure*}

The Euclid survey\footnote{\url{https://www.euclid-ec.org/}} \citep{Euclid11,Euclid16} is a space-based optical/near-infrared survey mission that will operate at L2. A Euclid wide survey will cover $15,000\,{\rm deg}^2$ with limiting magnitude $24-24.5$ during six-years of operations; the deep survey will cover $40\,{\rm deg}^2$ at about two magnitudes deeper. The Euclid survey will collect shape and photo-z for $1.5\times10^{9}$ background galaxies ($\sim30$ per arcmin$^{-2}$) available for WL mass measurements and spectra for $5\times10^7$ galaxies.  The Euclid surveys will show the three-dimensional distribution of dark and luminous matter up to $z\sim2$. 
Euclid will find $\sim 2 \times10^5$ clusters with a S/N greater than 5 at $0.2<z<2$ \citep{Sartoris16}, especially at $z>1$ thanks to the near-infrared bands. 

Ground-based telescopes are helpful to collect photometry in optical bands for an accurate estimation of photometric redshifts in a combination with Euclid bands. The on-going survey data from the DES, HSC-SSP, and KiDS will be used for the purpose.  The CIFS survey using the CFHT will cover $\sim5000$ deg$^2$ with the r-band and $\sim10000$ deg$^2$ with the u-band by Jan. 31st 2020.  The depth will be 24.1 mag and 23.6 mag, which is defined by $10\sigma$ aperture magnitudes for point sources within $2''$ diameter, for the r-band and the u-band, respectively. 
The Javalambre-Euclid Deep Imaging Survey (JEDIS-g) using JST/T250 (Javalambre Suvery Telescope) will collect g-band data of $\sim5000$ deg$^2$ of the northern sky in common with the CIFS survery.

In both Euclid and LSST, cluster mass measurements will be statistically and dramatically improved by the increase in the number of background galaxies. As a result, systematic biases on shape measurements and photometric redshifts will largely dominate over statistical errors. The main challenge for cluster mass measurements is control and minimisation of these systematic uncertainties.  The resulting cosmological constraints from the cluster mass function will be complementary to cosmic-shear cosmology. 


\subsubsection{Simons Observatory and CMB-S4}
\label{sec:futureCMB}

The next decade of CMB survey instruments will continue to progress to large detector counts and additional bands from $\sim$30-300 GHz, promising to dramatically increase the statistical sample of SZ observations of galaxy clusters as well as facilitate the separation of the tSZ, kSZ and rSZ contributions in hundreds of high-mass systems\footnote{For more details, see the review by \citep{mro19} in this volume.}.

Advanced ACTpol, which is the current generation instrument on ACT, and SPT-3G, the current generation camera on SPT, will also see some upgrades that will allow them to detect on the order of thousands of SZ selected clusters. 
Both are in the field and operating.  
SPT-3G for example is predicted to find $\sim5000$ clusters at a signal-to-noise $\geq$ 4.5 \citep{Benson14}. The ACT 6-meter will join the Simons Observatory, and both SPT and ACT will likely become part of CMB-S4.

The Simons Observatory\footnote{\url{https://simonsobservatory.org/}} \citep[S.O.;][]{simons} will combine several existing CMB experiments in the Atacama desert, and add a new 6-metre telescope with a similar optical design to CCAT-prime with an anticipated first light in 2021. 
Looking further ahead, CMB-S4\footnote{\url{https://cmb-s4.org/}} will likely add up to three 6-metre antennas of similar design as the S.O. and CCAT-prime 6-m, and several more lower resolution 1-metre class antennas \citep{CMB-S4} with an anticipated first light in 2028. 

S.O. and CMB-S4 will find on the order of $10^4$ and $10^5$ galaxy clusters respectively, including 10s to 1000s of high-z clusters \citep{Louis17}, through the thermal SZ effect. Figure~\ref{fig:CMB-S4} shows that CMB-S4 will place competitive and independent constraints on cosmological parameters \citep{Louis17}, including e.g. the sum of neutrino masses \citep{Madhavacheril17}.

\begin{figure*}[tbp]
  \includegraphics[clip=true,trim=0.0cm 0.0cm 0.0cm 0.0cm,width=0.45\textwidth]{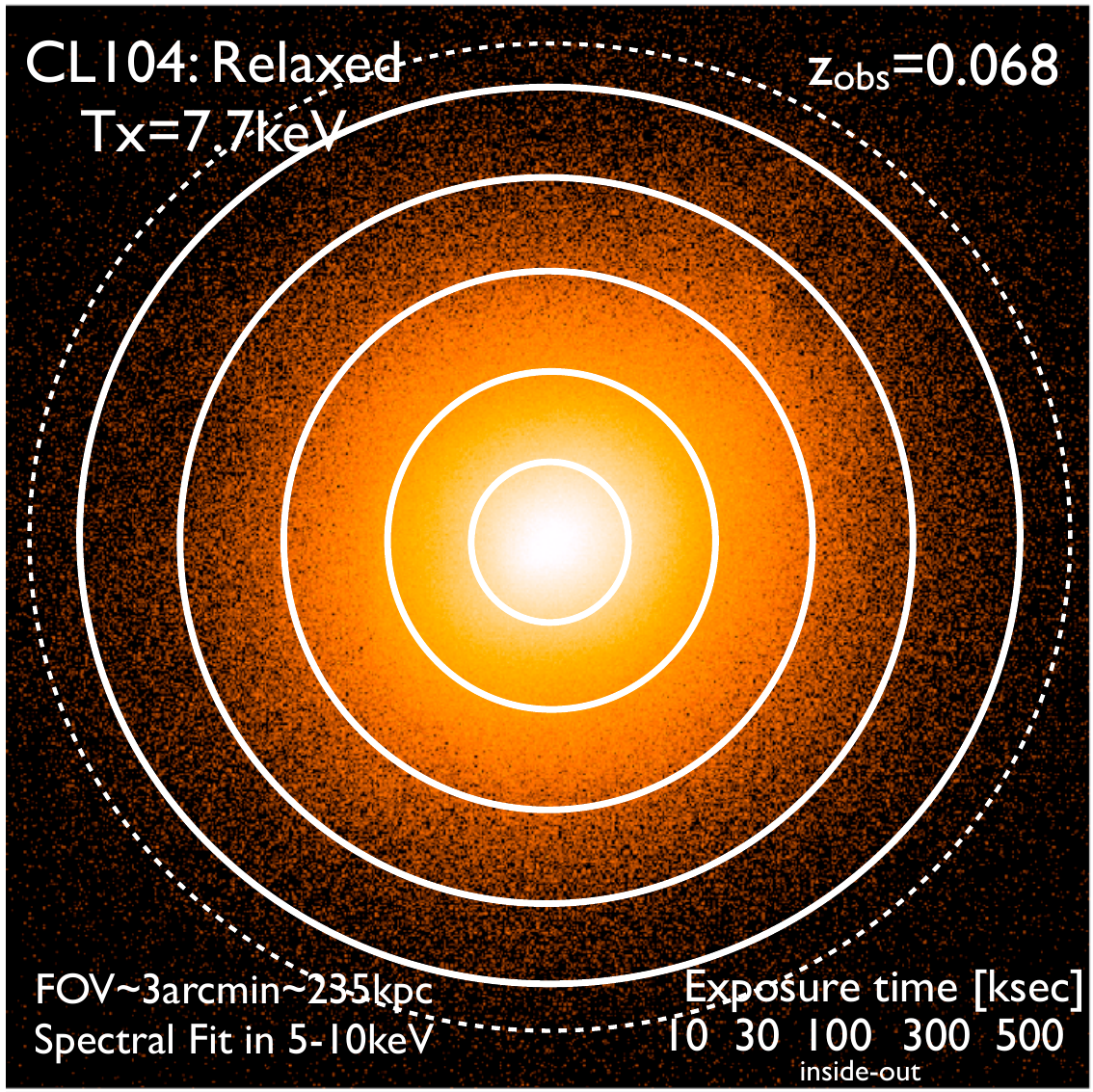}
\hfill
  \includegraphics[clip=true,trim=1.cm 1.0cm 0.0cm 0.0cm,width=0.47\textwidth]{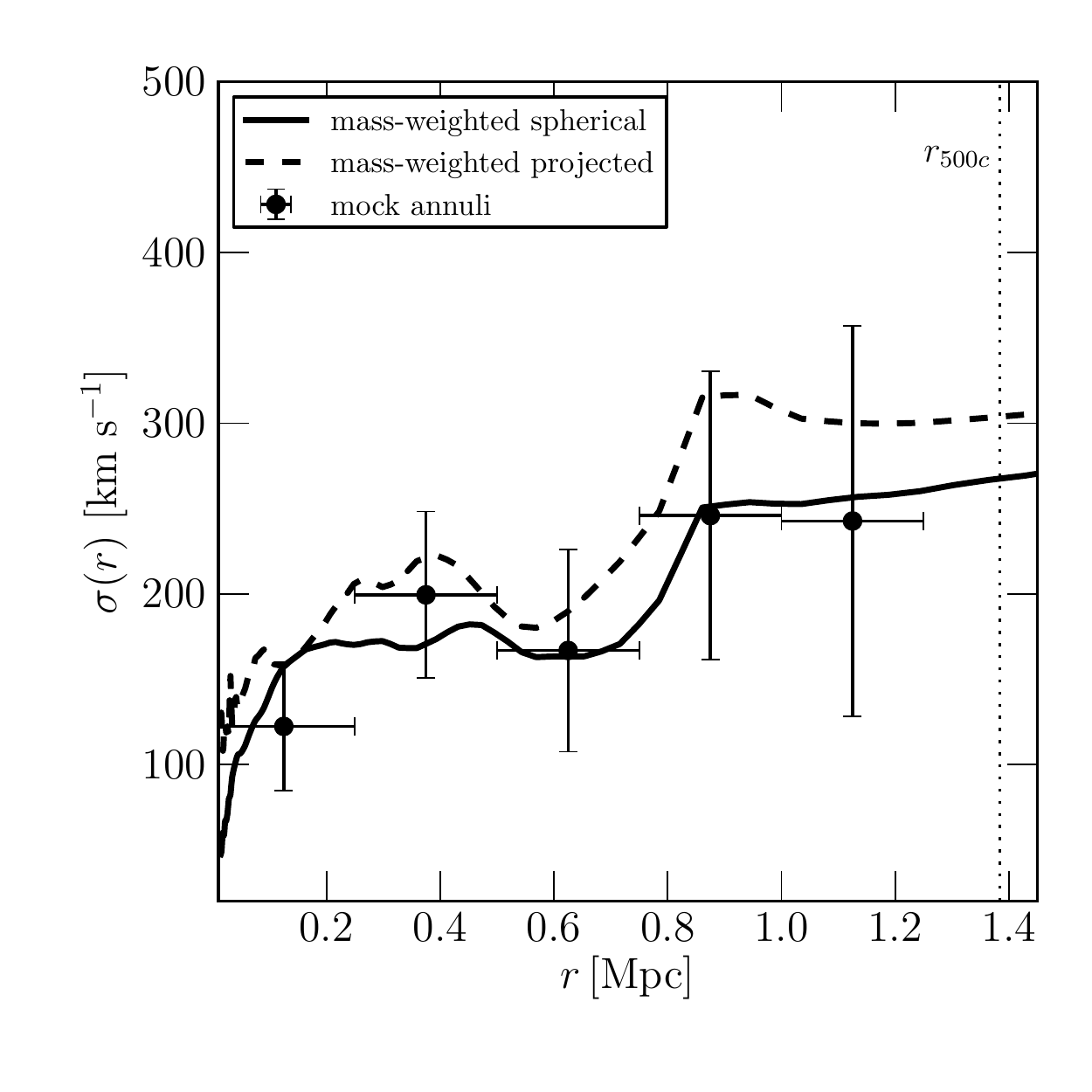}
\caption{Mock SXS / XRISM analysis of a relaxed galaxy cluster extracted from hydrodynamical simulations. {\it Left:} Mock XRISM image in a $5-10$~keV band. The region shown is about $2.6$~Mpc across, and the dotted circle indicates $R_{500}$. {\it Right panel:} XRISM/Resolve measurements of the gas velocity dispersion as a function of radius with $10$, $30$, $100$, $300$, $500$~ksec exposures. The black lines represent the true mass-weighted 3-D (solid) and projected (dashed) gas velocity dispersion profiles. From \citet{nagai13}.}
\label{fig:mock-xarm}
\end{figure*}

\subsubsection{XRISM and Athena}

Gas motions in galaxy clusters play an important role in determining the properties of the ICM and in cosmological parameter estimation through X-ray and SZ effect observations of galaxy clusters. Recently, the Hitomi X-ray satellite has provided the first direct measurements of gas motions in the core of the Perseus Cluster \citep{hitomi16}. Upcoming X-ray missions equipped with high-spectral resolution X-ray micro-calorimeters, such as XRISM and Athena, will continue Hitomi's legacy by measuring ICM motions through Doppler shifting and broadening of emission lines in a larger number of galaxy clusters, and at larger radii.\footnote{Measurements will also be possible in the outskirts of high-redshift clusters using high-resolution SZ spectral imaging observations through tSZ pressure fluctuations and direct kSZ measurements of internal gas motions \citep[see the review by][in this volume]{mro19}.}

To assess the feasibility and future prospects of directly measuring the random and bulk gas motions at large cluster-centric radii in X-ray observations, \citet{nagai13} performed an analysis of mock {\it Hitomi} Soft X-ray Spectrometer (SXS; analogous to Resolve) spectra of a relaxed galaxy cluster extracted from cosmological numerical simulations, and find that a detailed characterization of the gas velocity profile out to beyond $r\approx R_{2500}$ will require of order 500~ksec of XRISM time, with a significant exposure spent on the outermost radial bins.
However, if the significant investment of observing time is made, the gas velocity is recovered in good agreement with the 3D (deprojected) mass-weighted velocity dispersion profile up to $r\approx R_{500}$ (see Figure \ref{fig:mock-xarm}) and enables us to correct for the hydrostatic mass bias of galaxy clusters. On the other hand, perhaps counterintuitively at first, the XRISM mock measurement is slightly ($\sim 30-50$~~km~s$^{-1}$) smaller than the \textit{projected} mass-weighted gas velocity dispersion in a given radial bin, although the mock observations should probe the integrated motions along the line-of-sight. This difference occurs  because the measured velocity is spectral-weighted, and hence the inner regions where the gas density is higher but the gas velocity is smaller carry a higher weight. 
Also, \citet{Ota18} showed that XRISM is capable of measuring non-thermal pressure provided by bulk and random motions and correcting for the hydrostatic mass bias.
Going forward, this synergy between numerical simulations, mock observations, and real data will need to be employed frequently in order to correctly interpret the measurements and uncover any potential biases or significant projection effects.

Some details of the upcoming missions include the following.
\begin{itemize}
    \item The X-Ray Imaging and Spectroscopy Mission (XRISM), formerly XARM (the X-ray Astronomy Recovery Mission,
    \citealt{tmt18})
    is planned as a successor of the Hitomi satellite, and will carry a high spectral resolution X-ray microcalorimeter (Resolve), which is identical to the SXS.
    \item Athena\footnote{\url{https://www.the-athena-x-ray-observatory.eu/}} \citep{nbb13} is ESA's second Large Mission after the planetary mission Juice, and before the gravitational wave mission LISA. Details of cluster science expectations are provided in \citet{pra13,epd13,csh13}. The expected launch is in the early 2030s. The major increase in collecting area together with the superb spectral resolution of the micro-calorimeter instrument (X-IFU) will allow us to map turbulence and bulk motions in the inner regions \citep{ron18}. Together with measurements of these quantities in the outer parts of clusters, it is expected that we will be able to set tight limits on hydrostatic mass bias from non-thermal pressure support resulting from gas velocities. Furthermore, the wide field-of-view of the active pixel sensor (WFI) results in outstanding cluster survey capabilities.
\end{itemize}

\subsubsection{Studies}

X-ray satellite missions under study which would have an impact on cluster science include:

\begin{itemize}

\item AXIS, the Advanced X-ray Imaging Satellite\footnote{\url{http://axis.astro.umd.edu/}} \citep{axis}, is a Probe-class mission under study for the 2020 Decadal. The concept combines $\sim 0.4^{\prime \prime}$ imaging across a $24^\prime \times 24^\prime$ field of view, with CCD-type spectral resolution. 

 \item Lynx\footnote{\url{https://www.lynxobservatory.com/}} \citep{lynx} is a flagship mission concept under study for the 2020 NASA Decadal, for launch around the middle of the 2030s. A large collecting area ($\sim 2$ m$^2$ at 1~keV), high resolution ($\lesssim 1^{\prime\prime}$) imaging, and high-resolution spectroscopy ($\lesssim 2$ eV), combined with survey capability at CCD-type spectral resolution, will push spatially-resolved measurements of clusters and proto-clusters out to $z\sim3$.

\end{itemize}


\section{Summary}

Galaxy cluster mass measurements are fundamental to our understanding of structure formation, and to the use of the cluster population and its evolution to constrain cosmological parameters. As illustrated in this review, there are a number of different methods with which to measure the mass of galaxy clusters, making use of galaxy velocity dispersions, WL, X-ray, and X-ray/SZ observations. In recent years, great progress has been made in all of these methods. Better observations have allowed more precise masses to be obtained; larger samples have allowed statistical analysis; inter-comparison between methods has enabled us to better understand systematic effects. A large parallel effort has been undertaken by the theoretical community, using numerical simulations to explore the assumptions and biases inherent in each mass estimation method. 

In individual systems, mass profile measurements from {\it all} methods now indicate no significant deviation from the cusped dark matter profile shape predicted from numerical simulations, from local systems up to the most distant observations currently possible, at $z\sim1$. Furthermore, the systematic effects inherent in each method are now well-identified: lack of dynamical equilibrium and triaxiality in optical; the hydrostatic bias in X-rays; the purity of background catalogues and triaxiality in lensing. 

A fully consistent picture of the mass of galaxy clusters is necessary for both understanding their formation and evolution, and for their use as cosmological probes. Critically, the methods are fully independent, so that inter-comparison of their results can help to better understand the assumptions and biases inherent in each one. Further progress will require mass measurements of a large sample of clusters, selected in as unbiased a manner as possible, using all available methods. Extension to lower masses is also necessary, both to better understand cluster formation and evolution, and also to exploit the rich data sets that will be available from new surveys in optical (HSC, LSST, DES, Euclid), X-ray ({\it eROSITA}), and SZ (Simons Observatory, CMB-S4). The combination of methods, such as using X-ray and SZ observations of similar angular resolution, will allow extension of mass measurements to higher redshifts. In the future, measurement of bulk motions and turbulence  in the inner regions of nearby systems will be possible with XRISM and Athena, and in the outskirts and in high-redshift systems with high-resolution SZ imaging.

These new surveys in optical, X-ray, SZ, and lensing will yield new samples and allow us to probe selection effects and reveal the properties of the true underlying population. Further progress  is expected given the wealth of current and forthcoming data.


\begin{acknowledgements}

This work was initiated during a visit to the International Space Science Institute (ISSI) in Bern and we acknowledge ISSI’s hospitality. GWP and MA acknowledge funding from the European Research Council under the European Union's Seventh Framework Programme (FP7/2007-2013)/ERC grant agreement No. 340519.
SE acknowledges financial contribution from the contracts NARO15 ASI-INAF I/037/12/0, ASI 2015-046-R.0, ASI-INAF n.2017-14-H.0 and funding from the European Union's Horizon 2020 Programme under the AHEAD project (grant agreement n. 654215).
DN acknowledges Yale University for granting a triennial leave and the Max-Planck-Institut f\"ur Astrophysik for hospitality when this work was carried out.
THR acknowledges support from the German Aerospace Agency (DLR) with funds from the Ministry of Economy and Technology (BMWi) through grant 50 OR 1514.

\end{acknowledgements}

\bibliographystyle{apj}
\bibliography{mass_all}                

\end{document}